\documentclass[a4paper,fleqn,usenatbib]{mnras}

\usepackage{mathptmx}

\usepackage[T1]{fontenc}
\usepackage{ae,aecompl}

\usepackage{graphicx}	
\usepackage{amsmath}	
\usepackage{amssymb}	

\newcommand{\deltac}{\delta_{\rm c}}
\newcommand{\deltal}{\delta_{\rm l}}
\newcommand{\deltam}{\delta_{\rm m}}

\title[Testing the CMF against simulations]{Testing the conditional mass
  function of dark matter halos against numerical N-body simulations}

\author[D. Tramonte et al.]{
D. Tramonte,$^{1,2}$\thanks{E-mail: dtramonte@iac.es}
J.~A. Rubi{\~n}o-Mart{\'{\i}}n,$^{1,2}$
J. Betancort-Rijo$^{1,2}$
C. Dalla Vecchia$^{1,2}$
\\
$^{1}$Instituto de Astrof{\'{\i}}sica de Canarias, C/V\'ia L\'actea s/n, E-38205, La Laguna, Tenerife, Spain\\
$^{2}$Universidad de La Laguna, Dpto. Astrof{\'{\i}}sica, C/Astrof\'isico Francisco S\'anchez s/n, E-38206, La Laguna, Tenerife, Spain
}

\date{Accepted 2017 February 3. Received 2017 January 1; in original form 2016 October 6.}

\pubyear{2017}

\begin{document}
\label{eq:firstpage}
\pagerange{\pageref{firstpage}--\pageref{lastpage}}
\maketitle

\begin{abstract}
We compare the predicted conditional mass function (CMF) of dark matter halos
from two theoretical prescriptions against numerical N-body simulations, both in
overdense and underdense regions and at different Eulerian scales ranging from
$5$ to $30\,h^{-1}$\,Mpc.  In particular, we consider in detail a
locally-implemented rescaling of the unconditional mass function (UMF) already
discussed in the literature, and also a generalization of the standard rescaling
method described in the extended Press-Schechter formalism.  First, we test the
consistency of these two rescalings by verifying the normalization of the CMF at
different scales, and showing that none of the proposed cases provides a
normalized CMF. In order to satisfy the normalization condition, we include a
modification in the rescaling procedure. After this modification, the resulting
CMF generally provides a better description of numerical results. We
finally present an analytical fit to the ratio between the CMF and the UMF (also
known as the matter-to-halo bias function) in underdense regions, which could be
of special interest to speed-up the computation of the halo abundance when
studying void statistics. In this case, the CMF prescription based on the
locally-implemented rescaling provides a slightly better description of the
numerical results when compared to the standard rescaling.
\end{abstract}

\begin{keywords}
methods: statistical -- cosmology: theory -- dark matter -- large-scale structure of Universe
\end{keywords} 


\section{Introduction}
\label{sec:introduction}

The abundance of collapsed dark matter halos as a function of mass, and its dependence on redshift, is one of the basic tools in modern cosmology.
The halo mass function was first estimated analytically, as described in \citet{PS74}. This formalism constitutes the basis for any later analytic development; it is based on the assumption that initial fluctuations are drawn from a Gaussian distribution, and on the physics of spherical collapse. \citet{BA91} developed the so-called `excursion set formalism', in whose framework the Press-Schechter mass function can be derived from the barrier crossing statistics of many independent, uncorrelated random walks. However, this recipe underestimates the abundance of high mass collapsed objects \citep{Efstathiou88} due to the fact that in deriving the analytical expression, all mass elements are assumed to be at the centre of the object they belong to \citep{BRMDa}. As a consequence of this, the accumulated mass fraction is not a universal function of the linear variance of the density contrast but depends on the shape of the power spectrum \citep{BRMDb}. 

The treatment of the ellipsoidal collapse was incorporated into the formalism in several works \citep{Bond96,ST01,ST02} by using a moving barrier. Those studies provided an approximated analytical expression for the distribution of first crossings of the moving barrier, which was used in \citet{ST99} to provide a new parametrization of the mass function. Comparisons with N-body simulations show that this new formalism significantly improves the Press-Schechter mass function in predicting the halo mass distribution. The Sheth-Tormen parametrization has been tested against N-body simulations in \citet{JA01}, providing a very good agreement. In the same reference an alternative fitting formula was proposed: this functional form for the mass function is capable of achieving agreement below $\sim$ 10--30\,\% at all masses, and is said to be universal, in the sense that it can be applied to different cosmological models and different redshifts. However, the fit cannot be extrapolated beyond the mass range employed in that work. \citet{WA06} improved the fit providing a mass function parametrization which reproduce the halo abundance with an error $\sim 5$\,\% at redshift $z=0$ and a fixed cosmology; the work in \citet{Reed07}, instead, provides a good fit for the mass function at redshifts 10--30. Notice that in the works cited so far, the fit is done against catalogues obtained using friends-of-friends (FoF) defined halos \citep{Davis85}, which is computationally efficient and has the advantage that it does not assume any a-priori shape for the halos.

A major issue concernig masses obtained with FoF finders is that they not easily accommodated within the theoretical formalism, which is usually based on spherical overdensity (SO) mass. The corresponding SO algorithm \citep{LC94} identifies objects as spherical regions enclosing a certain overdensity around density peaks. This task has been addressed in \citet{TA08}: the proposed fit for the mass function is capable of reproducing the halo mass function at $z=0$ and in the mass range $10^{11}\,h^{-1}\,\text{M}_{\odot} < M < 10^{15}\,h^{-1}\,\text{M}_{\odot}$ with an accuracy $\lesssim 5$\,\%. They claim the halo mass function cannot be universal at this level of accuracy; the dependence on redshift is explored up to $z=2.5$, and also on the value of the overdensity used in the halo definition, in the range $200$ to $3200$ (computed with respect to the mean background density). The fitting parameters depend on both the redshift and the overdensity, in order to account for the non-universality of the mass function, improving the agreement with N-body simulations. This mass function is the standard reference used in most cluster cosmology analyses, and is the one we shall take as reference in the present work as well. 

There have been subsequent attempts at improving Tinker's fit. The impact of the initial conditions for the simulations used to fit the mass function can be found in \citet{Crocce06}. In \citet{Crocce10}, the halo abundance is fitted over five order of magnitude in mass, and its redshift evolution explored up to $z=2$; the fitting formula is the same as in \citet{WA06}, but with different parameters in order to correct for the underestimation of the most massive halo abundance, resulting in an overall $2\,\%$ fit accuracy. The cosmology and redshift dependence of the mass function is also explored in \citet{Courtin11}, showing the universality is not verified in absolute terms, holding true only for $z=0$ and a limited set of cosmologies; this is due to some cosmology-dependent features in the mass function, like the value of the threshold density for collapse, whose value depends on cosmological parameters. In \citet{Angulo12} an accurate fit with errors within $5\,\%$ is provided over eight orders of magnitude for the halo mass, extending Tinker's fitting range; a discussion on how different halo definitions affect the mass function is also provided. \citet{Watson13} explore in detail the difference between FoF and SO-defined mass functions; they find agreement below $5\,\%$ error with Tinker's fit, with major differences in the high mass end, and extend the result to higher mass values. Finally, \citet{Despali16} provide a fit with $5-8\,\%$ error for different redshifts and cosmologies, provided the mass function is defined in terms of the virial overdensity halo mass; in the same work, scaling relations for the fitting parameters are provided in order to compute the mass function for different redshifts and halo mass definitions.

The inclusion of baryon physics in the simulations may also impact the fits of the halo mass function. For instance, the work in \citet{Bocquet16} shows how the inclusion of baryonic feedback leads to a decrease in the halo masses, which in turn leads to a decrease of the halo abundance at a given mass; the effect is more important at low redshift and low masses, where differences with respect to dark matter-only fits can grow up to $\sim 15\,\%$. The study of baryonic effects on the mass function in other halo mass ranges has been tackled in \citet{Velliscig14} and \citet{Martizzi14}.

All the works on the mass function described so far consider the abundance of halos originated from density perturbations that grow during their evolution, eventually decoupling from the background expansion and finally ending up into collapsed objects. These perturbations are defined with respect to a background which is considered homogeneous. Such a mass function is usually referred to as `unconditional'. We address in the present work the study of the `conditional' mass function, meaning the function describing the abundance of objects inside an environment (the condition) with a definite density contrast with respect to the background. Locally underdense or overdense regions do alter the mass distribution of collapsed halos \citep[e.g.,][]{ShethLemson99}; this applies, for instance, when studying the halo abundance inside a supercluster or a cosmic void. 

Despite all the effort that has been undertaken to provide an accurate parametrization of the unconditional mass function (UMF), the conditional mass function (CMF) has not been so thoroughly studied and not a definite recipe can be found in the literature. Existing work on the subject usually take the UMF as a starting point and try to adapt it to the conditioned environment. 
In \citet{Cole89} the abundance of collapsed halos in perturbed regions is computed starting from the Press-Schecter UMF and using the `peak-background split' formalism; their work is particularly focused on the application of the halo bias to astronomical objects like clusters of galaxies, quasars or the study of the diffuse X-ray background.
The spatial distribution of dark matter halos and its relation to the underlying mass distribution
 has also been studied in \citet{Mo96}: the authors develop an analytical model, based on the extended 
Press-Schechter formalism and the spherical collapse assumption, to describe the halo bias function defined as the ratio
 between the overdensity of halos in a conditioning region and its matter density contrast. Comparison with N-body
 simulations confirms the accuracy of this model in the case of an Einstein-de-Sitter universe. Other attempts worked in the framework of the spherical collapse, and extended the Press-Schechter mass function, in the excursion set formalism \citep{BA91,LC93}. This approach, however, is not adequate to reproduce results from N-body simulations \citep[e.g.,][]{Tormen98,BRMDa,Neyrinck14}. The work in \citet{ST02} extends the excursion set formalism by including the physics of ellipsoidal collapse. Their results reasonably agree with N-body simulations; this formalism, however, is used to quantify the difference in halo abundance when considering the mass function evolution from one redshift to another, and cannot be used for computing the CMF at fixed redshift $z$, which is the problem we address in this work. An extension of the work on the CMF can be found in \citet[][hereafter RBP08]{RBP08}: among the major changes with respect to previous studies, it introduces the dependence of the CMF from the radial position inside the condition (assumed spherical), and explicitly implements the UMF scaling in such a way that the correspondent CMF is properly normalized over the possible values of the density contrast. The resulting CMF shows good agreement with numerical simulation for the conditions explored in that work. 

The present work aims at comparing with N-body simulations the prescriptions for the CMF described in RBP08, both at small and large scales. The results enable us to pinpoint the best recipe for computing the CMF starting from the UMF, depending on the condition scale or the value of the density contrast. We also study the normalization of the CMF at different scales, explicitly modifying the UMF scaling in order to satisfy the normalization condition by construction. In order to simplify the computation of the CMF, we also provide a fitting formula for the bias between the CMF and the UMF in the case of large scale underdense regions, which is needed in the study of void statistics. 

Whenever cosmological parameters are not specified, it is intended we are using a flat $\Lambda$CDM model with the values for the cosmological parameters listed in table~\ref{tab:cosmology}.

\begin{table}
\centering
\caption{Values of cosmological parameters for our reference model.}
\label{tab:cosmology}
\begin{tabular}{cccccc}
\hline
$\Omega_{\rm m}$ & $\Omega_{\rm b}$ & $\Omega_{\Lambda}$ & $h$ & $n_{\rm s}$ & $\sigma_8$ \\
\hline
0.3 & $4.6\times10^{-2}$ & 0.7 & 0.7 & 0.96 & 0.8 \\
\hline
\end{tabular}
\end{table}


\section{Formalism for the UMF}
\label{sec:UMF}

The unconditional mass function (UMF), which we label $\text{d}n_{\rm u}/\text{d}m$, is the mass distribution of the number density of halos at a given redshift. Its definition is such that $\text{d}n_{\rm u}/\text{d}m(m,z)\,\text{d}m$, is the (comoving) number density of halos with mass in the range $[m,m+\text{d}m]$ at redshift $z$. The UMF can be parametrized as in e.g., \citet{TA08}, in the form: 
\begin{equation}
\label{eq:umf}
\frac{\text{d} n_{\rm u}}{\text{d} m} = \frac{\rho_{\rm m}}{m} \frac{\text{d} \ln{\sigma^{-1}}}{\text{d}m} f(\sigma,z),
\end{equation}
where $\rho_{\rm m}$ is the comoving matter density of the Universe and $\sigma^2(m,z)$ is the mass variance of the linear density field extrapolated to redshift $z$, the mass $m$ also defining the scale over which the variance is computed. As a function of the matter power spectrum $P(k)$, the variance can be computed as: 
\begin{equation}
\label{eq:sigma}
\sigma^2(r,z) = \frac{D^2(z)}{2\pi^2} \int_0^{\infty} \;\text{d}k\, k^2\, P(k)\, w^2(kr),
\end{equation}
in which $w$ is the Fourier transform of the window function and $D(z)$ is the growth factor of linear perturbation normalized to unity at $z=0$. Both the window function and the relation between the scale $r$ and the mass $m$ are determined by the employed smoothing filter. In the case of a real space top-hat filter :
\begin{equation}
\label{eq:tophat}
m = \frac{4}{3}\pi \rho_{\rm m}\, r^3,
\end{equation}
and
\begin{equation}
w(x) = \frac{3}{x^3}(\sin{x} - x\cos{x}).
\end{equation}
The term $f(\sigma,z)$ in equation~(\ref{eq:umf}) represents the mass fraction contained in collapsed objects per unit $\ln{\sigma^{-1}}$. It depends on the value of $\deltac$, that is the critical density for collapse, obtained by extrapolating the linear overdensity $\deltal$ of a spherical perturbation at the time it collapses. There exist different functional form for $f(\sigma,z)$ in the literature \citep{JA01}, coming from analytic or semi-analytic models. The original \citet{PS74} UMF provides an analytical expression based on the physics of spherical collapse; however, this simple recipe actually underestimates the abundance of high mass collapsed objects. Improvements to this UMF employ parametric expressions fitted against N-body simulations, like in \citet{ST99} or in \citet{WA06}. In this paper we will use the UMF parametrization by \citet{TA08}, hereafter T08, which explicitly separates a power-law regime at small scale and a high mass exponential cut-off: 
\begin{equation}
\label{eq:TA}
f_{\text{TA}}(\sigma) = A\left[\left(\frac{\sigma}{b}\right)^{-a}+1\right]\,e^{-c/\sigma^2}.
\end{equation}
The function is fitted against N-body simulations in which halos are identified with a spherical overdensity method \citep{LC94}. The improvement in the fit compared to other UMFs comes from the fitting parameters being allowed to depend on redshift and on the value of overdensity $\Delta$ employed by the halo finder (meaning a certain region of the simulation is considered a collapsed object if its density is greater than the mean matter density by a factor $\Delta$). In T08, tables are provided for computing the fitting parameters at different values of $\Delta$ and $z$. In the present work, whenever we talk about the overdensity $\Delta$, it is understood we are considering a density contrast with respect to the matter background density, and not to the critical density, following the convention adopted in T08.


\section{Formalism for the CMF}
\label{sec:CMF}

The conditional mass function (CMF), which we label $\text{d}n_{\rm c}/\text{d}m$, is the proper halo mass distribution inside regions with a definite density contrast with respect to the average matter density, where the predictions from the UMF are no longer adequate. We will always consider spherical regions with matter density contrast $\deltam$, defined as:
\begin{equation}
\label{eq:deltam}
\deltam = \frac{\rho_{\rm c} - \rho_{\rm m}}{\rho_{\rm m}},
\end{equation}
where $\rho_{\rm m}$ is the average background matter density and $\rho_{\rm c}$ the matter density inside the condition. The correspondent linear density contrast $\deltal$ can be computed as \citep{ST02}:
\begin{align}
\label{eq:deltalin}
\deltal(\deltam) = \dfrac{\deltac}{1.68647}\,\left[  1.68647 - \dfrac{1.35}{(1+\deltam)^{2/3}} \right. \nonumber\\
 \left. - \dfrac{1.12431}{(1+\deltam)^{1/2}} + \dfrac{0.78785}{(1+\deltam)^{0.58661}} \right].
\end{align}
In the framework of the CMF it is useful to employ (comoving) Lagrangian coordinates rather than (comoving) Eulerian coordinates. In fact, an overdense (underdense) region does not follow the Hubble flow and it actually shrinks (dilate) during the evolution of the Universe, meaning its Eulerian size changes (the Lagrangian one is constant by construction). Hereafter, we will use the variables $R$ and $Q$ to denote the condition Eulerian and Lagrangian radius, respectively. The two are related by:
\begin{equation}
\label{eq:lagrad}
Q = R\,(1+\deltam)^{1/3}.
\end{equation}
The use of Lagrangian coordinates ensures we can talk about a unique value of $Q$ which is valid at all redshifts. 

Apart from mass and redshift, the conditional mass function also depends on the condition radius and density contrast, $\text{d}n_{\rm c}/\text{d}m(m,z;Q,\deltal)$. There exist different approaches in the literature to derive analytically an expression for the CMF, usually based in the rescaling of some quantities entering the definition of the UMF. We will concentrate in particular on two methods, namely the standard rescaling and one that is implemented locally. It is worth mentioning that other ways of computing the CMF have been developed, based on the rescaling of the cosmological parameters rather than on the mass function itself; these methods are based on the assumption that the condition can be treated as an independent universe with a locally-defined cosmology. For instance, \cite{Gottlober03} stress the efficiency of computing the halo abundance in voids by rescaling the matter power spectrum via a redefinition of the parameter $\sigma_8$; a prescription for computing the local cosmological parameters in voids is derived also in \cite{Goldberg04}, with the aim of improving the efficiency of specialized void numerical simulations.
A complete description of these rescalings goes beyond the scope of this work; for more information we redirect to \citet{BTR16}, where the local cosmology is derived not only for cosmic voids but also for overdense regions.

\subsection{Standard rescaling}
\label{subsec:stdresc}
The most straightforward way to obtain the conditional mass function is by taking the same analytical expression for the UMF and express it in terms of the rescaled variables: 
\begin{align}
\label{eq:stdresc}
\deltac & \longrightarrow \deltac'=\deltac - \delta_1 \nonumber \\ 
\sigma^2  & \longrightarrow \sigma'^2 = \sigma^2 - \sigma_1^2,
\end{align}
where $\delta_1$ is the condition linear density contrast and $\sigma_1^2 = \sigma^2(R)$ is the variance of the matter field at the condition scale. We refer hereafter to eqs.~(\ref{eq:stdresc}) as the standard rescaling. This recipe is the basis for a number of works on the CMF. For instance, the so called excursion set formalism, also known as extended Press-Schechter \citep{BA91,LC93}, which is based on the hypothesis of spherical collapse. An extension to this work, including the ellipsoidal collapse, can be found in \citep{ST02}. 

This rescaling can be applied to any UMF parametrizations. However, care must be taken when rescaling the T08 mass function, which does not show any explicit dependence on the value of the critical density $\deltac$. The parameter can be introduced taking into account that in the previous parametrizations the function $f$ depends on $\deltac$ only via the combination $\nu = (\deltac/\sigma)^2$ \citep{ST99}. In the case of T08 UMF we can rewrite the variance as $\sigma^2=\deltac^2/\nu$, and apply the scaling~(\ref{eq:stdresc}) to the variable $\nu$. This yields:
\begin{equation}
\label{eq:stdrescsigma}
\sigma'^2 = \left[\sigma^2 - \sigma_1^2 \right]\left[\frac{\deltac}{\deltac-\delta_1 } \right]^2,
\end{equation}
which can be applied to eq.~(\ref{eq:TA}). 

\subsection{Local rescaling}
\label{subsec:locresc}

The local rescaling is an attempt at improving the results obtained with the standard recipe. We shall take as a reference for this work the method described in RBP08, where the CMF is obtained from the UMF by a variable rescaling in the style of eqs.~(\ref{eq:stdresc}), but with the rescaling being implemented locally. This means the CMF also depends on the radial distance from the centre of the condition $q$: $\text{d}n_{\rm c}/\text{d}m(m,z;q,Q_1,\delta_1)$. Let $Q_2$ be the Lagrangian radius corresponding to the mass $m$, and $\sigma_2=\sigma(Q_2)$. In formulae, the rescaling reads: 
\begin{align}
\label{eq:locresc}
\deltac' &= \deltac - D(q,Q_1,Q_2) \delta_1 \nonumber \\
\sigma_2'^2 &= \sigma_2^2 - D(q,Q_1,Q_2)^2 \sigma_1^2,
\end{align}
where
\begin{equation}
\label{eq:Dfunction}
D(q,Q_1,Q_2) = \frac{\sigma_{12}(q,Q_1,Q_2)}{\sigma_1^2(Q_1)},
\end{equation}
and $\sigma_{12}$ is the covariance:
\begin{align}
\sigma_{12} &= \sigma_{12}(q,Q_1,Q_2) = \nonumber \\
      &= \dfrac{D(z)^2}{2\pi} \int_0^{\infty} \text{d}k k^2 P(k) w(kQ_1) w(kQ_2) \dfrac{\sin{(kq)}}{kq},
\end{align}
($D(z)$ denotes the normalized growth factor). The full formalism is described in RBP08. Again, in the case of T08 mass function, the dependence on $\deltac$ must be introduced explicitly, yielding the rescaling equation for the variance: 
\begin{equation}
\label{eq:locrescsigma}
\sigma_2'^2 = \left[\sigma_2^2 - D(q,Q_1,Q_2)^2 \sigma_1^2 \right]\left[\frac{\deltac}{\deltac-D(q,Q_1,Q_2) \delta_1 } \right]^2,
\end{equation}
which can be applied to eq.~(\ref{eq:TA}). In RBP08 it is said that the quantity defined in eq.~(\ref{eq:Dfunction}), for $Q_2 \ll Q_1$, is practically independent of $Q_2$, a property that can be exploited to speed the computation up. In particular, this allowed computing the mass derivative of the variance appearing in eq.~(\ref{eq:umf}) without the need of applying each time the rescaling. However, we found that this approximation is not so good in most of the mass range we employ. In this work we therefore always applied the full computation with no approximations.  

In RBP08 the local rescaling was compared to the standard one, showing differences in the predicted halo abundance, particularly in the high-mass end and in underdense regions. In order to get rid of the extra dependence on the radial distance $q$, they considered the volume average of the locally-rescaled CMF inside the condition, in the form 
\begin{equation}
\frac{\text{d}n_{\rm c}^{\text{avg}}}{\text{d}m}(m,z;Q,\delta_1) = \frac{3}{Q^3}\int_0^Q \,\text{d}q\,q^2\,\frac{\text{d}n_{\rm c}}{\mathrm{d}m}(m,z;q,Q,\delta_1).
\label{eq:locrescavg}
\end{equation}
In the following, we shall employ the same definition of averaged CMF when showing the prediction from the local rescaling. 

In RBP08 the CMF computed with function~(\ref{eq:locrescavg}) was directly tested against the halo abundance extracted from numerical simulations, using two specific regions corresponding to underdense regions, resulting in a reasonable agreement.

\subsection{CMF normalization}
\label{subsec:normalization}
In RBP08 it is stressed that the conditional mass function should satisfy the
normalization condition, which means that by integrating the CMF over the
possible values of the linear overdensity inside a condition $Q$, each weighted
by its probability distribution, the unconditional mass function must be
recovered. In the same reference, it has been shown that the local rescaling as
defined in equation~(\ref{eq:locresc}) does not satisfy such a condition. The
problem was solved in RBP08 by slightly modifying the rescaling procedure,
introducing an additional parameter $\alpha$, which we shall refer to as the
normalization parameter. For the case of the T08 UMF, the new proposed
rescaling reads:
\begin{equation}
\label{eq:locrescnorm}
\sigma_2'^2 = \left[\sigma_2^2 - D(q,Q_1,Q_2)^2 \sigma_1^2 \right]\left[\frac{\alpha}{\alpha-D(q,Q_1,Q_2) \delta_1 } \right]^2,
\end{equation}
which should be used instead of eq.~(\ref{eq:locresc}).

A proper value for $\alpha$ should yield a normalized CMF over all the mass
range we are using. In RBP08 the normalization parameter was estimated for the
mass function parametrizations by \citet{ST99} and \citet{WA06}, for one
particular value of the condition Lagrangian radius. In this work we repeated
the same analysis for the T08 mass function and using different values for the
condition Lagrangian radius, in the range $[5,30]\,h^{-1}\,\text{Mpc}$, in order
to study the dependence of $\alpha$ on the condition size. As shown in RBP08,
smaller conditions are much more effective in constraining the value for the
normalization parameter, which we found to be:
\begin{equation}
\alpha = 1.25
\end{equation}
for the T08 mass function. This parameter is instead unconstrained by larger
conditions; requiring a normalized CMF is thus not enough for constraining the
value of $\alpha$ at scales $Q\gtrsim 20\,h^{-1}\,\text{Mpc}$. In order to test
whether the same normalization parameter can be used also for large scales, we
have to compare the CMF with numerical simulations, as detailed in
Section~\ref{sec:simulations}. We also repeated the same analysis for the
standard rescaling, and found the same results; both the local and the standard
rescalings need the introduction of the $\alpha$ parameter for satisfying the
normalization condition. More details on the computation of the normalization
parameter following the analytical procedure of RBP08 can be found in Appendix
\ref{sec:normpar}. In addition, we also explore a different normalization
condition using simulations in Section~\ref{subsubsec:normalization}, and
consistently, we found the same value for the $\alpha$ parameter.


\section{Comparison with simulations}
\label{sec:simulations}

In this section we test the validity of the CMF prescriptions by comparing the predicted abundance of halos with numerical simulations. We want to extend the work in RBP08 by employing a considerably larger number of conditions, in order to reduce statistical uncertainties associated to the particular condition choice. We will test the CMFs against counts extracted from simulations at different scales, and both in overdense and underdense regions. We will also explore the effect of employing different values for the normalization parameter $\alpha$ entering the computation of the locally rescaled CMF.

We present in the following the simulation set we considered, the algorithm we employed to identify regions with a given density contrast and build the halo mass distribution, and the results we obtained when comparing such a distribution to the theoretical recipe for the conditional mass function.

\begin{figure*}
\includegraphics[trim = 85mm 130mm 20mm 30mm, scale=0.5]{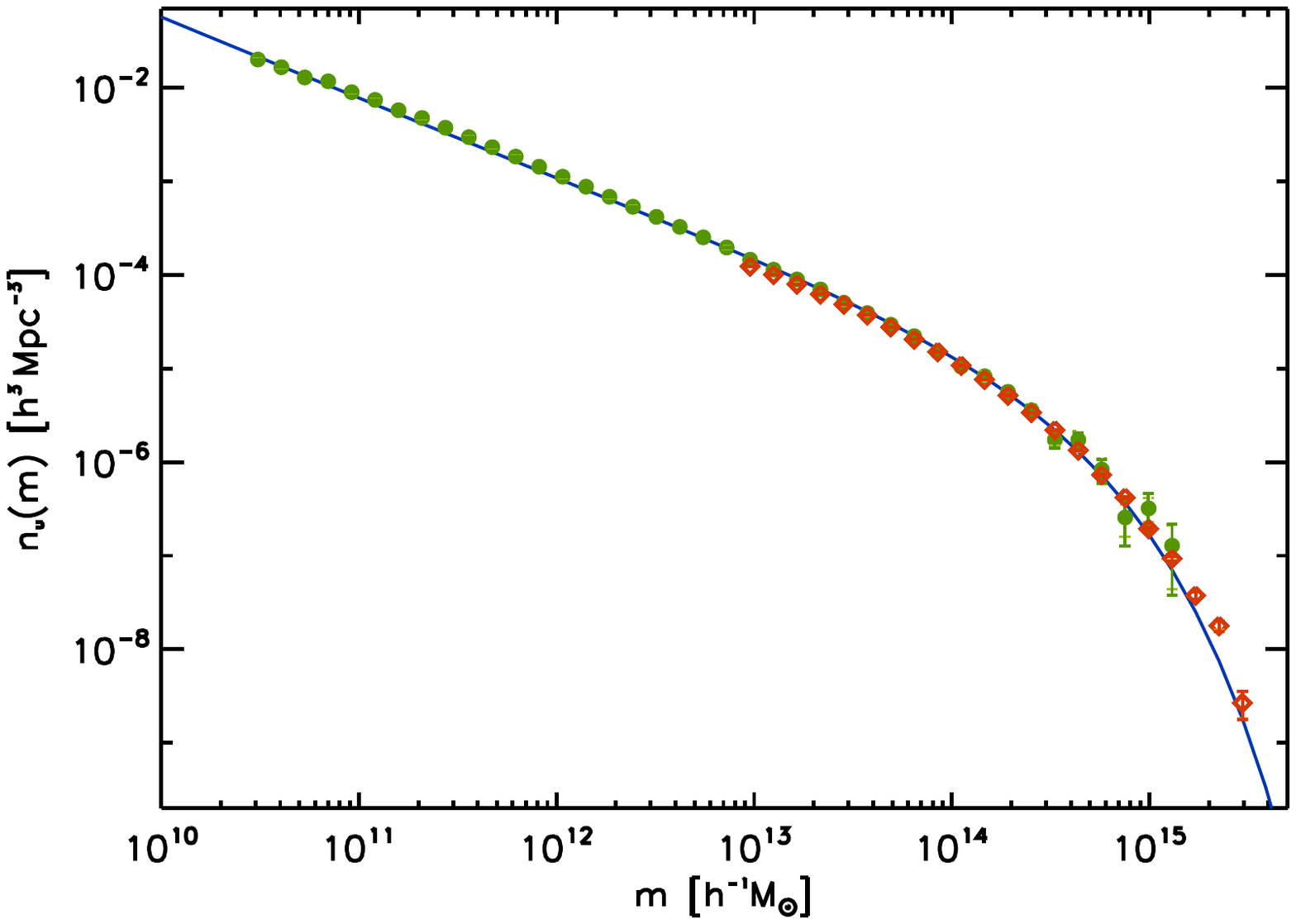} 
\quad
\includegraphics[trim = 30mm 130mm 80mm 30mm, scale=0.5]{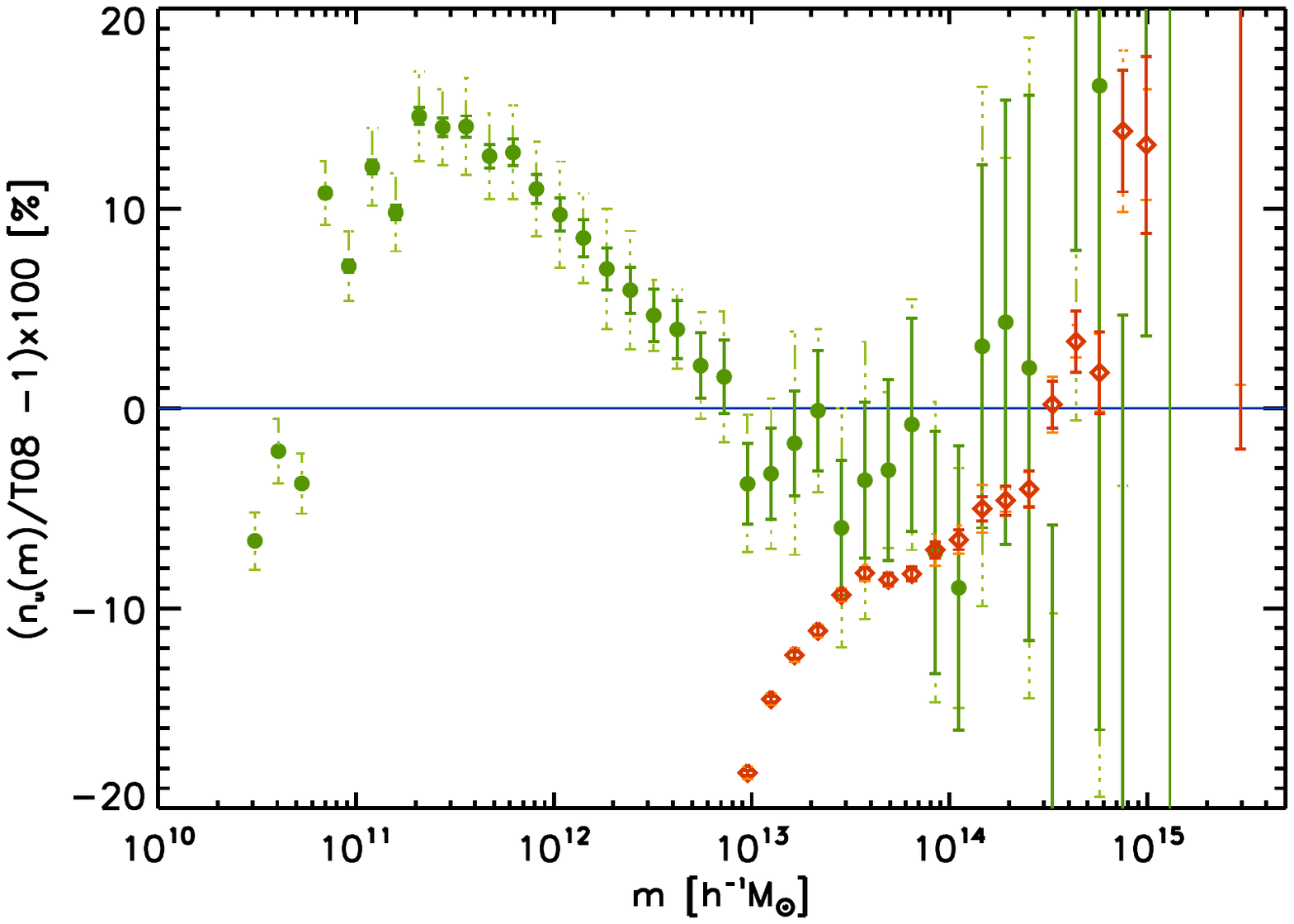} 
\caption{Unconditional mass function test for the simulation set. Left: points mark the halo number density as a function of mass, obtained from the Small Scale Simulation (circles) and the Large Scale Simulation (diamonds); the line shows the abundance predicted by the T08 UMF computed with the same underlying cosmology and an overdensity parameter $\Delta=200$. Right: same as left panel but showing the percent residuals with respect to the theoretical UMF. Poissonian errors are overplotted (thick continuous), together with numerical errors computed dividing each box into eight subsets of equal volume (thin dashed). Although the abundances extracted from simulations reproduce the overall shape of the T08 UMF, they are affected by systematics that can produce deviations from the theoretical prediction at the level of $\sim 10\,\%$. These offsets, which could in principle be strongly reduced by averaging over a large number of simulations, must be accounted for in the subsequent analysis.}
\label{fig:UMFs}
\end{figure*}

\begin{figure}
\includegraphics[trim = 24mm 14mm 30mm 10mm, scale=0.36]{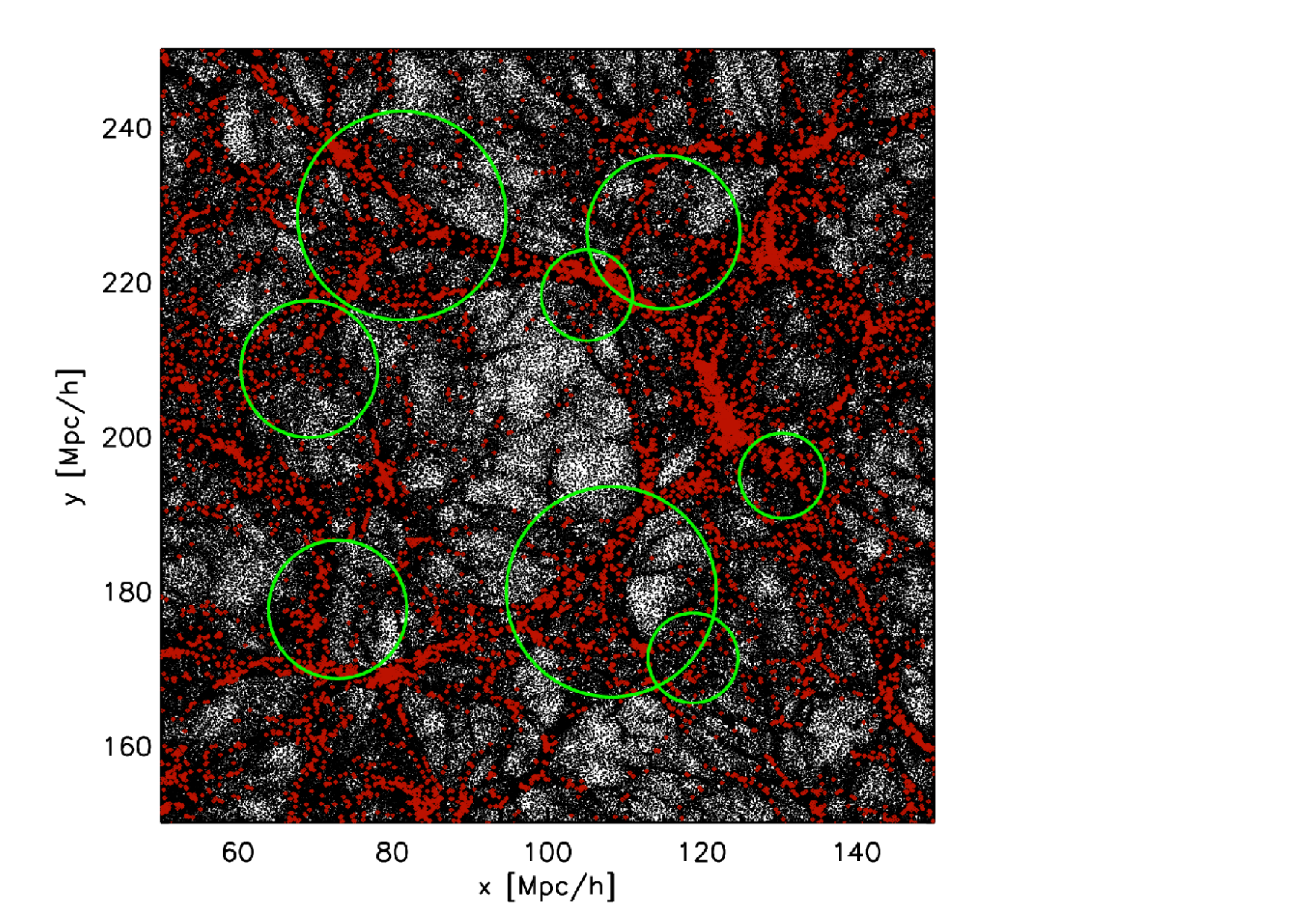} \quad
\caption{Example of a snapshot slice for the SSS simulation, made of a $100\times100\,h^{-2}\,\text{Mpc}^2$ area region with a $5\,h^{-1}\,\text{Mpc}$ depth. We show the particle points (black dots), the collapsed objects extracted by the SO finder (red spots) and some examples of spheres randomly placed over the simulation as conditioning regions (green circles).}
\label{fig:snapshot_cut}
\end{figure}

\subsection{The simulation set}
\label{subsec:simulation_set}
We considered two N-body dark matter-only simulations. Each simulation contains $10^9$ particles in a cubic box with periodic conditions at the boundary, in one case of size $250\,h^{-1}\,\text{Mpc}$ , and in the other of size $1500\,h^{-1}\,\text{Mpc}$. We shall refer to these simulations as the Small Scale Simulation (SSS) and the Large Scale Simulation (LSS), respectively. The basic information on these simulations is summarized in table~\ref{tab:table_sim}. The LSS was selected among the Minerva sample of 100 realisations of the same volume size \citep{Grieb16}; the SSS, instead, was performed in order to extend the lower halo mass range.  Initial conditions were generated starting from an initial glass configuration, and using the 2LPT code implementing the method described in \citet{Crocce06}. We adopted a flat $\Lambda$CDM model with parameters: $h=0.695$, $\Omega_{\rm b}=0$, $\Omega_{\rm m}=0.285$, $n_{\rm s}=0.9632$ and $\sigma_8=0.828$ (\citet{Sanchez13}, Table I), and computed the input linear power spectrum with CAMB \citep{Lewis00}. The initial conditions were evolved from $z=63$ to $z=0$ using GADGET (last described in \citet{Springel05}).  On the simulation snapshot we run a FoF halo finder with linking length $\ell=0.1$. This approximately corresponds to a spherical overdensity $\Delta=1600$, and is used to separate density peaks that would be taken as a single halo by a lower overdensity finder. Then, around each identified peak, halos were searched with a spherical overdensity $\Delta=200$, after processing the FoF output with the SUBFIND code \citep{Springel01}.

In Figure~\ref{fig:UMFs} we show the comparison between the T08 mass function, computed with an overdensity parameter $\Delta=200$, and the halo counts extracted from the simulations. Error bars associated to the simulation data points were computed assuming poissonian statistics inside each mass bin. To avoid boundary effects or resolution effects, the safe mass range we can use for each simulation is $[10^{10.5},10^{14.5}]\,h^{-1}\,\text{Mpc}$ with the SSS, and $[10^{13},10^{15.5}]\,h^{-1}\,\text{Mpc}$ with the LSS. By joining the two simulations we are capable of exploring a range of mass extending over five orders of magnitude. For both simulations, the lower mass limit is a resolution limit: below that scale, the halo finder is no longer able to converge in computing the halo mass with overdensity 200, and all the substructures around the previously identified density peak are lost. When determining the theoretical UMF, we checked that the finite-box effect has a negligible impact on the computation of the variance, remaining below one part in $10^3$ also in the SSS. 

The first plot in Figure~\ref{fig:UMFs} shows that the number density of halos extracted from both simulations is capable of reproducing the overall shape of the T08 mass function, showing the same dependence on mass over the whole range we are exploring. However, the residual plot on the right shows there are still some differences with respect to the theoretical prediction, mostly at the level of a $\sim 10\,\%$ relative error. This is a systematic effect which is due to the use of only one simulation for each box; for instance, the excess of halo abundance from the LSS visible in the high mass end can be removed by averaging many realizations of the same simulation (see Figure 1 of \citet{Grieb16}); this deviation however regards very high masses where errors become large and are not useful for the following analysis.

Besides, a numerical resolution effect is visible in both simulations: it can be recognised as a change in the mass dependence of the residual halo abundance at masses $\sim 2\times10^{11}\,h^{-1}\,\text{Mpc}$ in the SSS and $\sim 5\times10^{13}\,h^{-1}\,\text{Mpc}$ in the LSS. Given the individual particle masses reported in table~\ref{tab:table_sim}, these values correspond in both cases to halos containing around 200 particles. We actually considered such halos in LSS when performing the analysis described in Section~\ref{subsec:results}. In order to test to what extent this numerical issue can affect our results, we explicitly computed the bias as the ratio of the variance of halo number density and the variance of matter, as a function of mass. We found that numerical resolution effects only show up below masses $\sim 6\times10^{12}\,h^{-1}\,\text{Mpc}$, corresponding to halos made of around 20 particles, which are not considered in our analysis. These resolution effects are therefore not a significant issue.

In order to achieve a better agreement with the theoretical mass function, the average of many simulations in each box should be used. This goes beyond the scope of this work. Here, in order to account for possible systematics in the following analyses, when considering comparisons between simulated and predicted halo abundances in conditioned environments, we will always be comparing the ratios of each CMF to the corresponding UMF, i.e. directly comparing the halo bias function.

\begin{table}
\centering
\caption{Table summarizing the properties of the two simulations used in this work.}
\label{tab:table_sim}
\begin{tabular}{cccc}
\hline
 Name & Size & Particle mass & Softening \\
  &  $(h^{-1}\,\text{Mpc})$ &  ($h^{-1}\,\text{M}_{\odot}$) &  ($h^{-1}\text{kpc}$)\\
\hline
 LSS & 1500 & $2.7\times10^{11}$ & 60 \\
 SSS & 250 & $1.2\times10^9$ & 10 \\
\hline
\end{tabular}
\end{table}

\begin{figure*}
\includegraphics[trim = 45mm 175mm 30mm 20mm, scale=0.45]{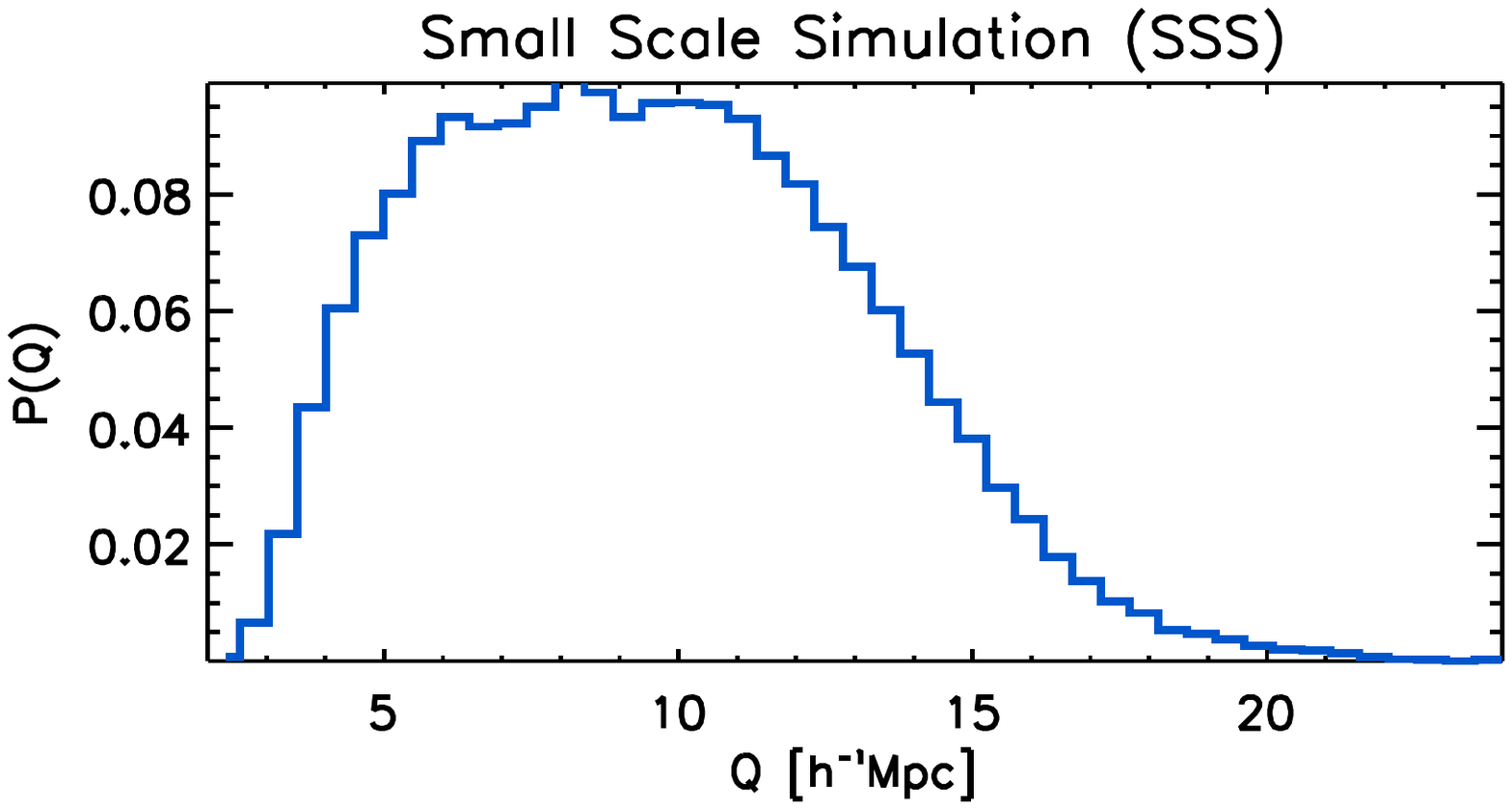} \quad
\includegraphics[trim = 10mm 175mm 30mm 20mm, scale=0.45]{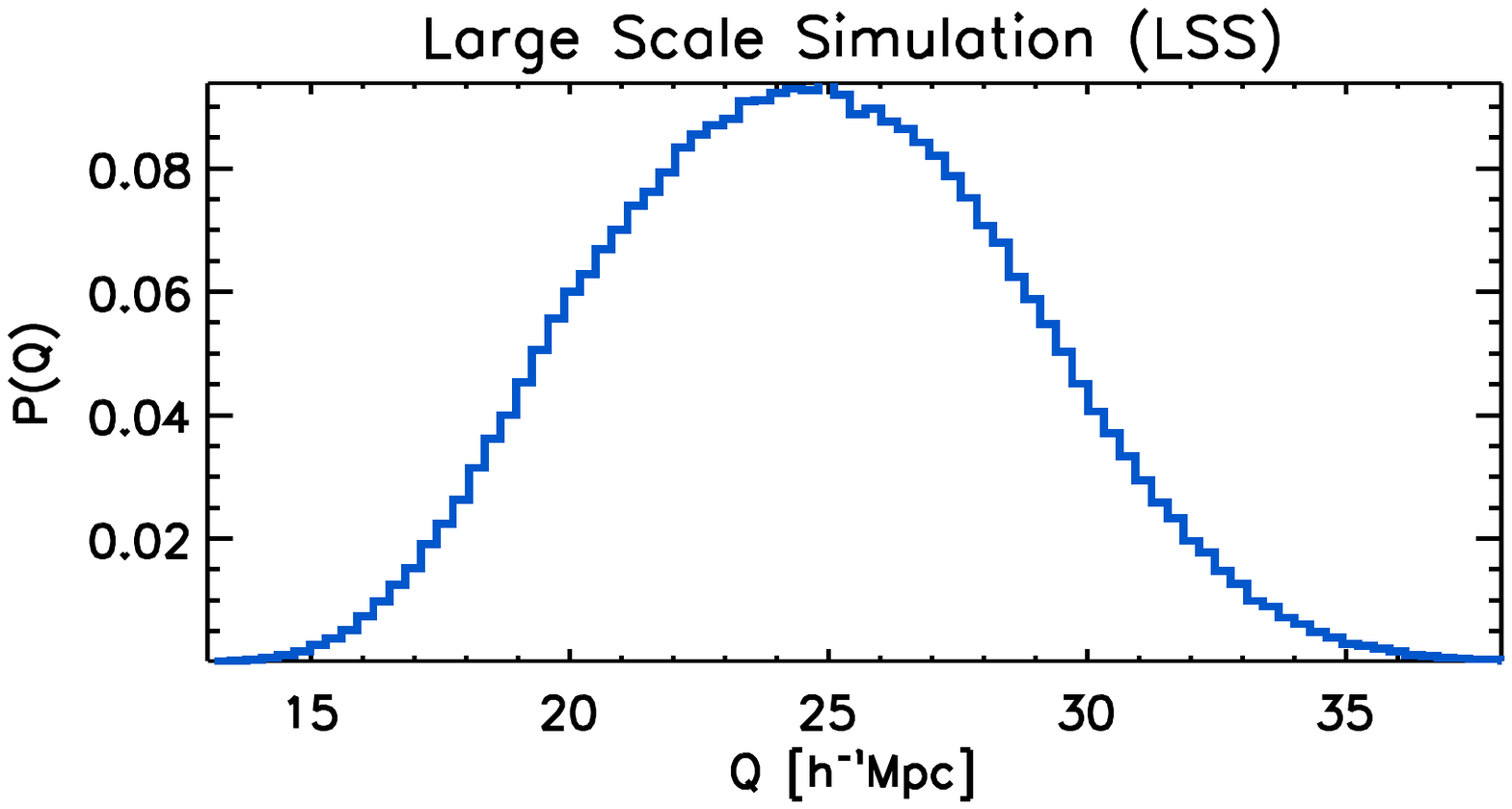} \\
\includegraphics[trim = 45mm 175mm 30mm 20mm, scale=0.45]{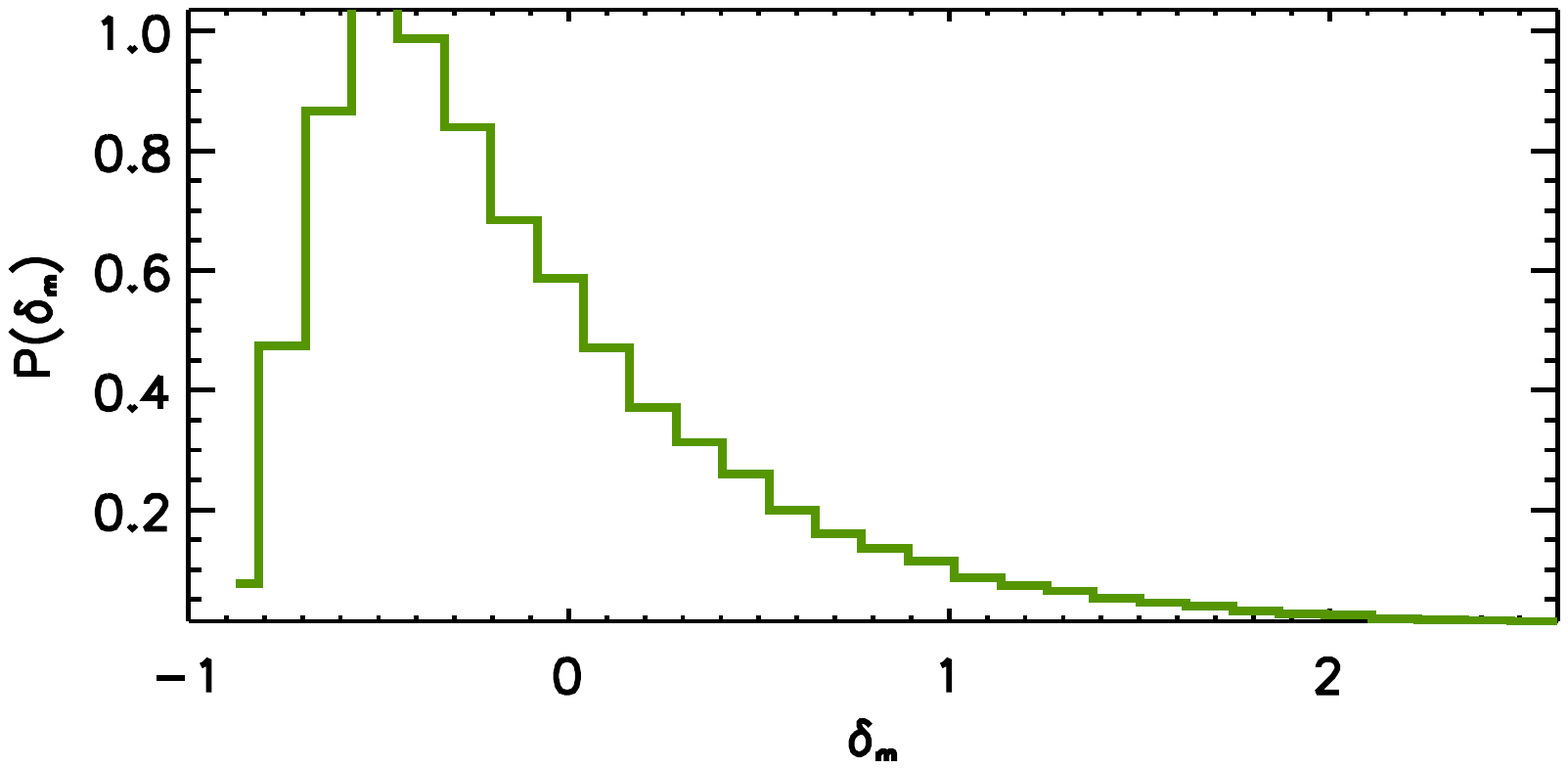} \quad
\includegraphics[trim = 10mm 175mm 30mm 20mm, scale=0.45]{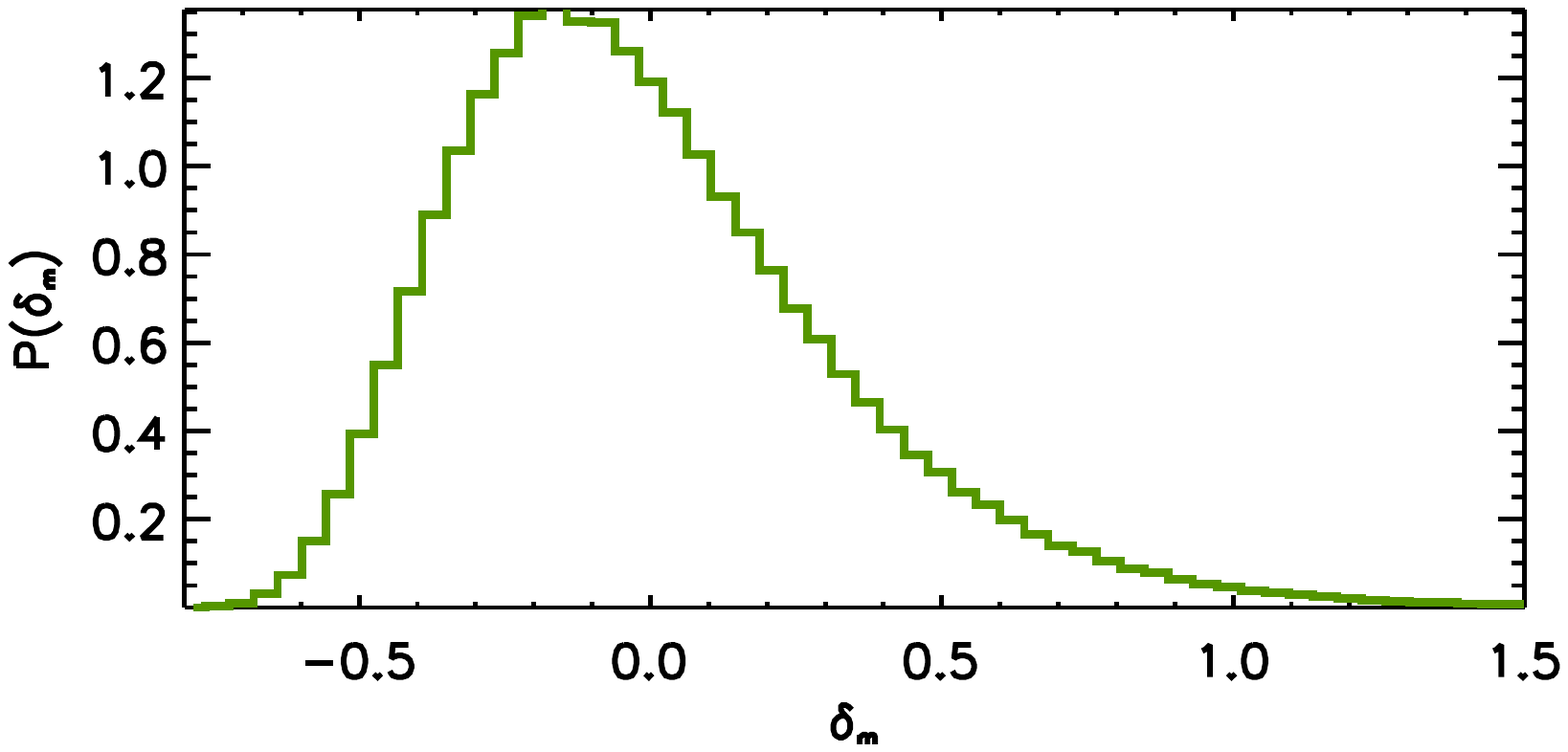} \\
\includegraphics[trim = 45mm 165mm 30mm 20mm, scale=0.45]{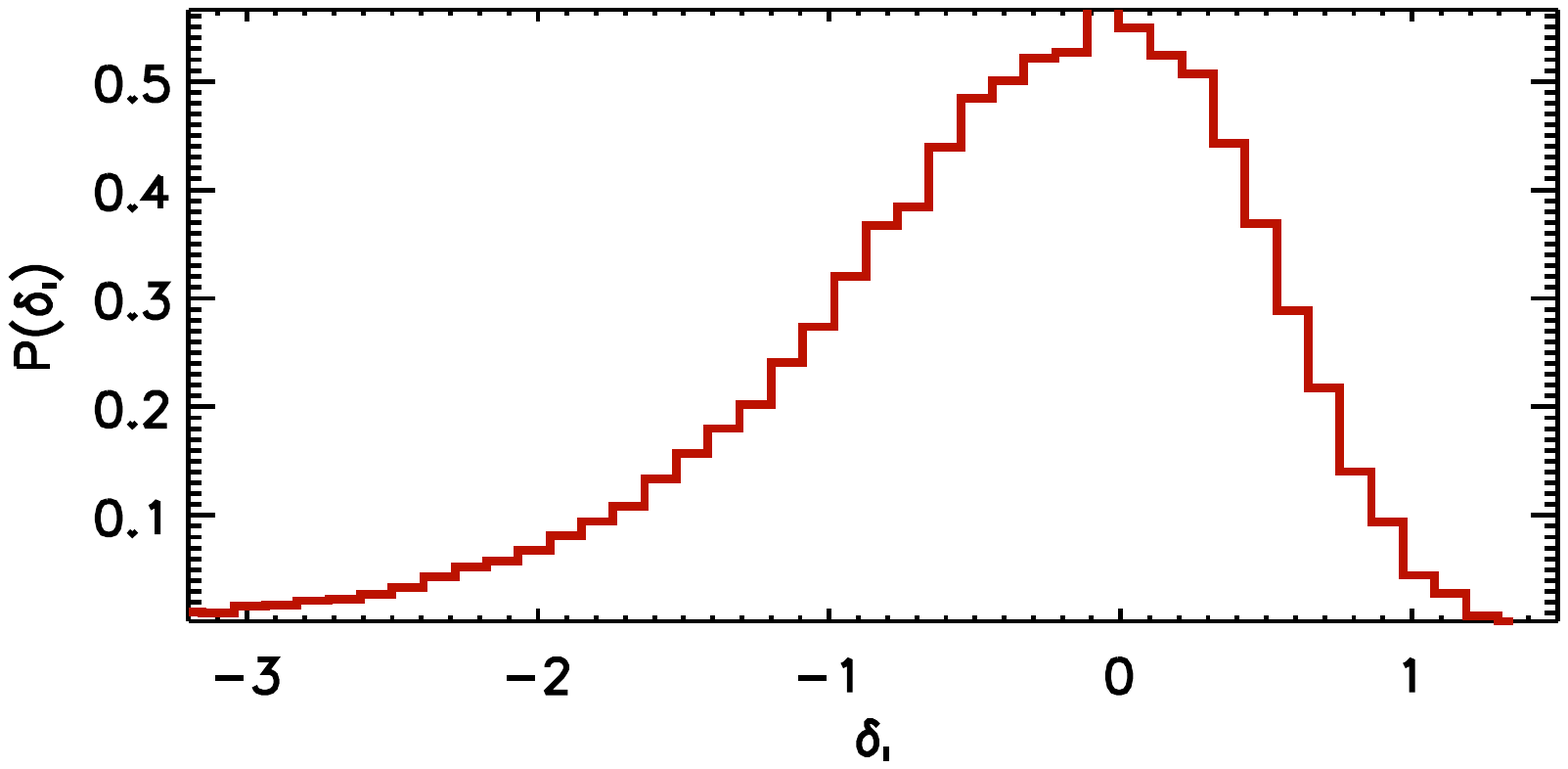} \quad
\includegraphics[trim = 10mm 165mm 30mm 20mm, scale=0.45]{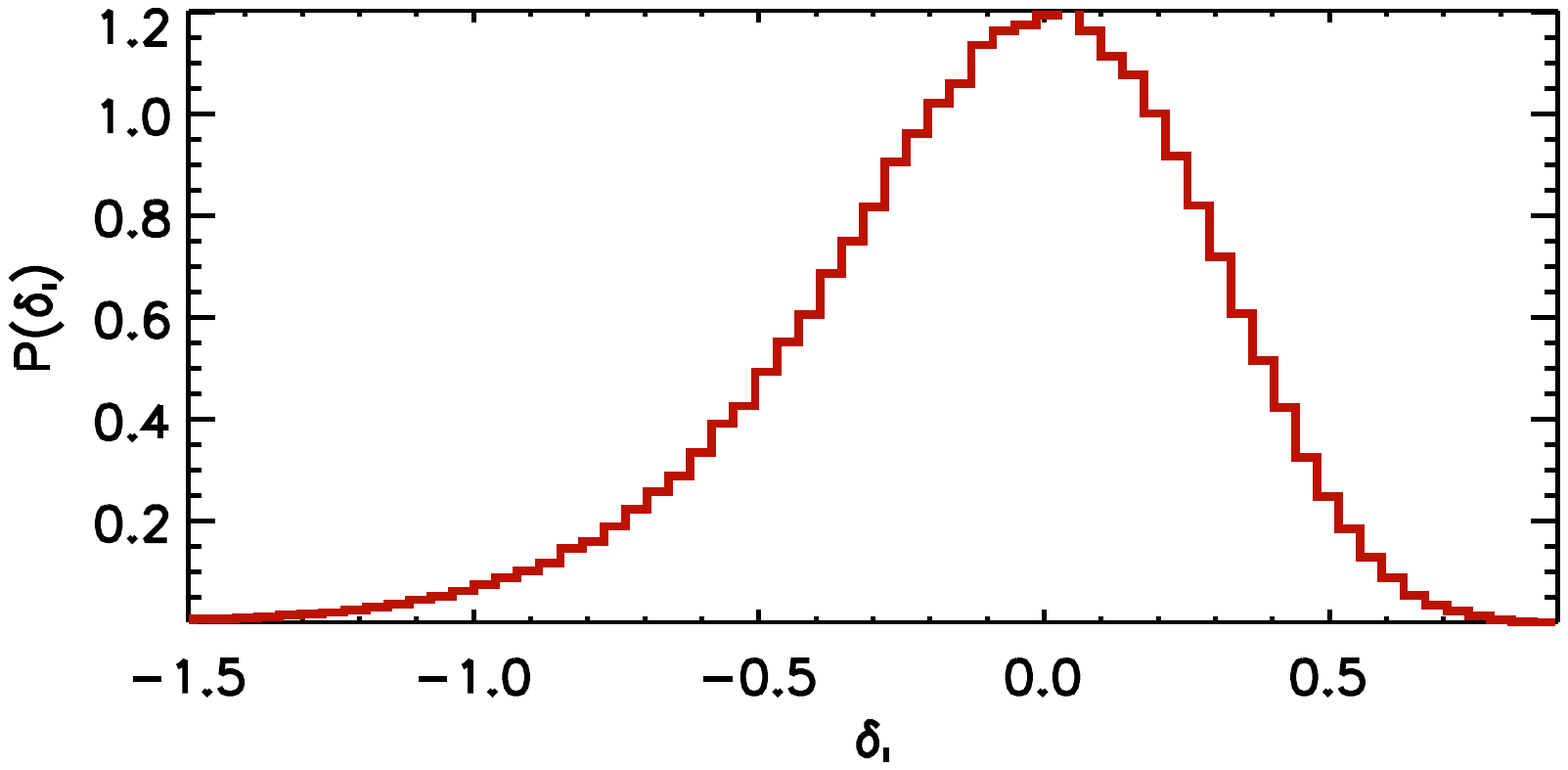} \\
\caption{Posterior distributions for the quantities characterizing the conditions, resulting from placing randomly placed spheres over the simulations with a top-hat prior for their Eulerian radius (see the text for details). From top to bottom: normalized distribution for the Lagrangian radius $Q$, the physical density contrast $\deltam$ and the linear density contrast $\deltal$. Left column: Small Scale Simulation, with a total $6\times10^4$ spheres; right column: Large Scale Simulation, with a total $4.4\times10^5$ spheres.}
\label{fig:posteriors}
\end{figure*}

\subsection{Methodology}

Testing the CMF requires comparing the theoretical abundance of halos to the actual counts extracted from regions of the simulation with a definite density contrast. In order to identify such regions we threw a large number of spheres on the simulation snapshot, storing the total number of particles inside each one of them; this allowed us to determine the number density of particles inside each sphere. By knowing the average particle density, the sphere density contrast could be computed using~(\ref{eq:deltam}). The sphere positions were generated randomly with a uniform distribution inside the simulation box; taking advantage of the periodic boundary conditions, any point of the box was a possible sphere centre. The sphere radii were also generated randomly; in the case of the SSS, which we employed for the smaller conditions, the values for the Eulerian radius $R$ were drawn from a uniform distribution in the interval $[5,15]\,h^{-1}\,\text{Mpc}$. The largest conditions were instead placed over the LSS, with radii drawn from a uniform distribution in the interval $[20,30]\,h^{-1}\,\text{Mpc}$. We employed a total of $4.4\times10^5$ spheres for the LSS, and of $6\times10^4$ spheres for the SSS: the smaller simulation volume ensures the same spatial coverage can be obtained with a lower number of spheres. In both cases the spheres are allowed to overlap; in fact, this is not an issue when building the final halo distributions. In Figure~\ref{fig:snapshot_cut} we show a sample slice of the SSS, showing the simulation particles, the identified halos and a few spheres randomly placed over the simulation.

Note that the aforementioned radii are the Eulerian quantities $R$; once the physical overdensity $\deltam$ inside each sphere is computed, its Lagrangian radius can be determined with~(\ref{eq:lagrad}). The correspondent linear overdensity $\deltal$ can be obtained via eq.~(\ref{eq:deltalin}), and this is the last quantity needed to completely characterize each sphere. In Figure~\ref{fig:posteriors} we show the final normalized distributions for the quantities $Q$, $\deltam$ and $\deltal$, in both simulations. We see that in the case of the SSS, a consequence of using smaller spheres is that the density contrast is peaked at more negative values compared to the LSS case, and as a result the distribution of Lagrangian radii is rather asymmetric. With the LSS spheres the density contrast distribution peak lies closer to zero, and the Lagrangian radius distribution has a more symmetric shape around the interval allowed by the prior Eulerian radius distribution.

This set of characterized conditions can be used to test the theoretical predictions of the CMF. The comparison between theory and simulation was performed at the level of the halo density as a function of mass, computed as a histogram over a specified set of mass bins. In principle, this comparison could be performed for each one of the spheres we threw over the simulations. Typically, a single sphere contains around $\sim 10^2$--$10^3$ halos in the SSS and $\sim 10$--$10^2$ in the LSS. However, given the high number of spheres we employed, it was convenient to group them into bins of Eulerian radius and linear density contrast, and consider the comparison between the simulated and theoretical histograms averaged inside each $(R,\deltal)$ bin. This way we could take advantage of the higher statistics resulting from averaging the halo mass distribution from many conditions with common size and density contrast.

We obtained the simulated histograms as follows. For each sphere, using the information of its position and Eulerian radius, we recorded all collapsed objects retrieved by the halo finder that lie inside the sphere. Let us fix a $(R,\deltal)$ bin and let $N_{\rm s}$ be the total number of spheres whose size and density contrast fall in this bin; we joined in a single set all the halos contained inside theses $N_{\rm s}$ spheres, and binned these halos over a specified set of mass bins $(m)_j$. We call $N_j$ the number of halos whose masses enter the $j$-th mass bin. We defined our halo density histogram extracted from the simulation the quantity:
\begin{equation}
\label{eq:simhist}
n_{\rm c}^{\text{sim}}(m_j;R,\deltal) = \frac{N_j}{N_{\rm s}V_{\rm R}},
\end{equation}
where $V_{\rm R}$ is the average eulerian volume of the spheres belonging to the $(R,\deltal)$ bin, $V_{\rm R}=4\pi R^3/3$. We chose a sufficiently small step $\Delta R = 0.5\,h^{-1}\,\text{Mpc}$ between radius bins, in order to ensure the central value $R$ was representative of the size of all spheres in the bin.

We then computed the theoretical prediction for the halo density in the bin $(R,\deltal)$. For each sphere in this bin, we computed the theoretical CMF $\text{d}n_{\text{c},i}/\text{d}m(m;Q,\deltal)$ using the exact values of $\deltal$ and $Q$ of the sphere; we integrated the CMF over the same mass bins $(m)_j$ considered earlier, to transform the distribution into a density as a function of mass, $n_i(m_j;Q,\deltal)$. Here the subscript $j$ runs over the mass bins, the subscript $i$ over the different spheres contained inside the $(R,\deltal)$ bin, $i=1,...,N_{\rm s}$. The $N_{\rm s}$ histograms obtained this way, one for each sphere, had to be combined to get an average estimate for the $(R,\deltal)$ bin. Note that the density of objects extracted from the simulation is Eulerian, while the formalism for the CMF described so far yields the Lagrangian density. So, the theoretical histograms were combined by first multiplying each one by the sphere Lagrangian volume $V_{\text{Q},i} = 4\pi Q_i^3/3$, by summing the correspondent halo numbers, and by translating to final Eulerian density dividing by the mean Eulerian volume of the bin $V_{\rm R}$. In formulae:
\begin{equation}
\label{eq:theohist}
n_{\rm c}^{\text{theo}}(m_j;R,\deltal) = \frac{1}{N_{\rm s}V_{\rm R}}\,\sum_{i=1}^{N_{\rm s}} n_i(m_j;Q,\deltal) V_{\text{Q},i}
\end{equation}

The theoretical histograms were generated using both the CMF from the standard
rescaling and from the local rescaling. We also explored the effect of the
normalization parameter $\alpha$ in changing the agreement of the CMF with the
simulation; we considered in particular the value $\alpha=1.25$, which is the
one required for normalizing the T08 mass function, and also an intermediate
value $\alpha=1.5$, which lies between 1.25 and $\delta_{\rm c}$. Notice that,
although the CMF is naturally expressed in Lagrangian density, we could not bin
the spheres in $Q$ and $\deltal$ because these quantities are not independent.

\begin{figure*}
\includegraphics[trim = 30mm 130mm 30mm 20mm, scale=0.5]{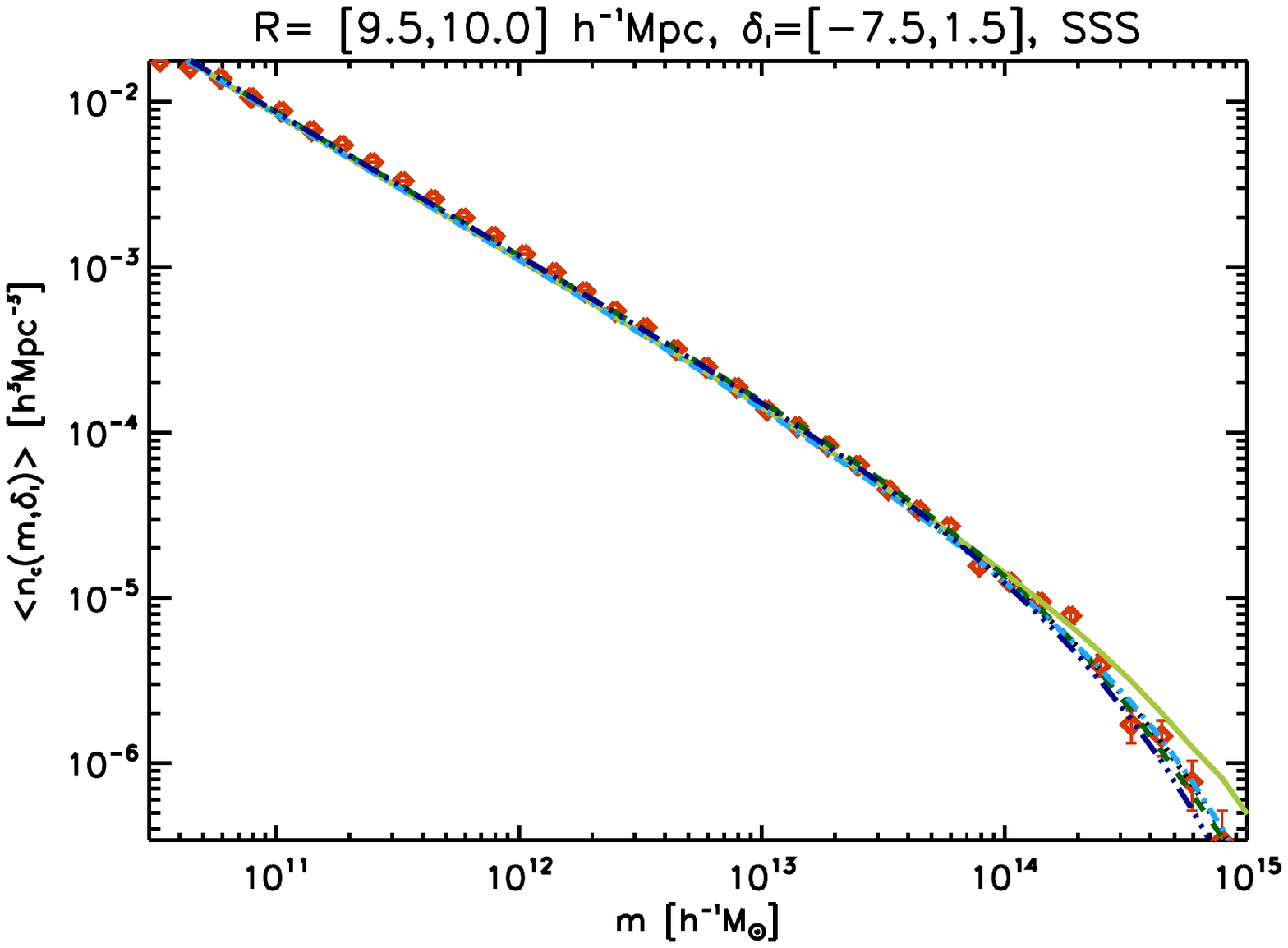} \quad
\includegraphics[trim = 20mm 130mm 30mm 20mm, scale=0.5]{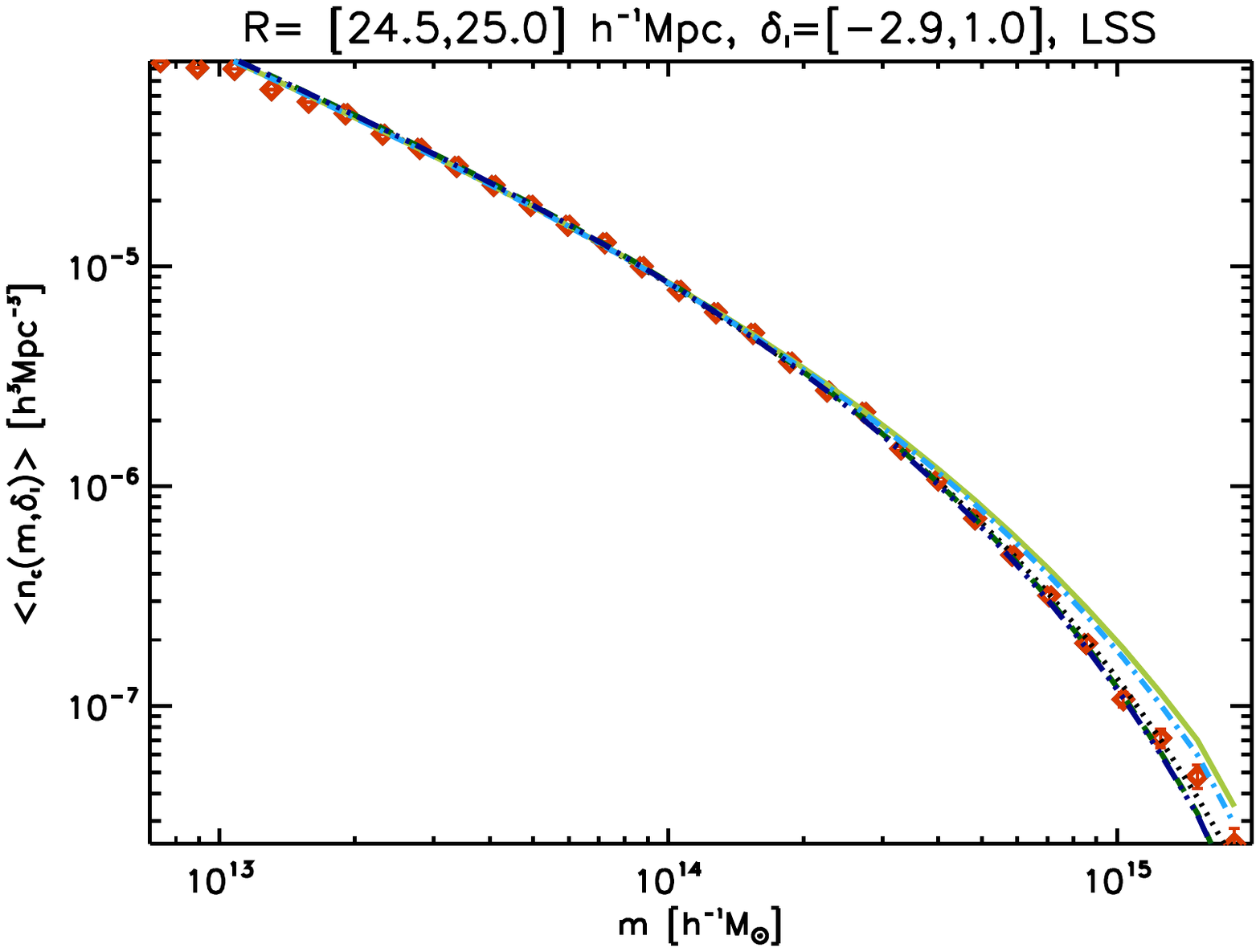} 
\includegraphics[trim = 30mm 130mm 30mm 30mm, scale=0.5]{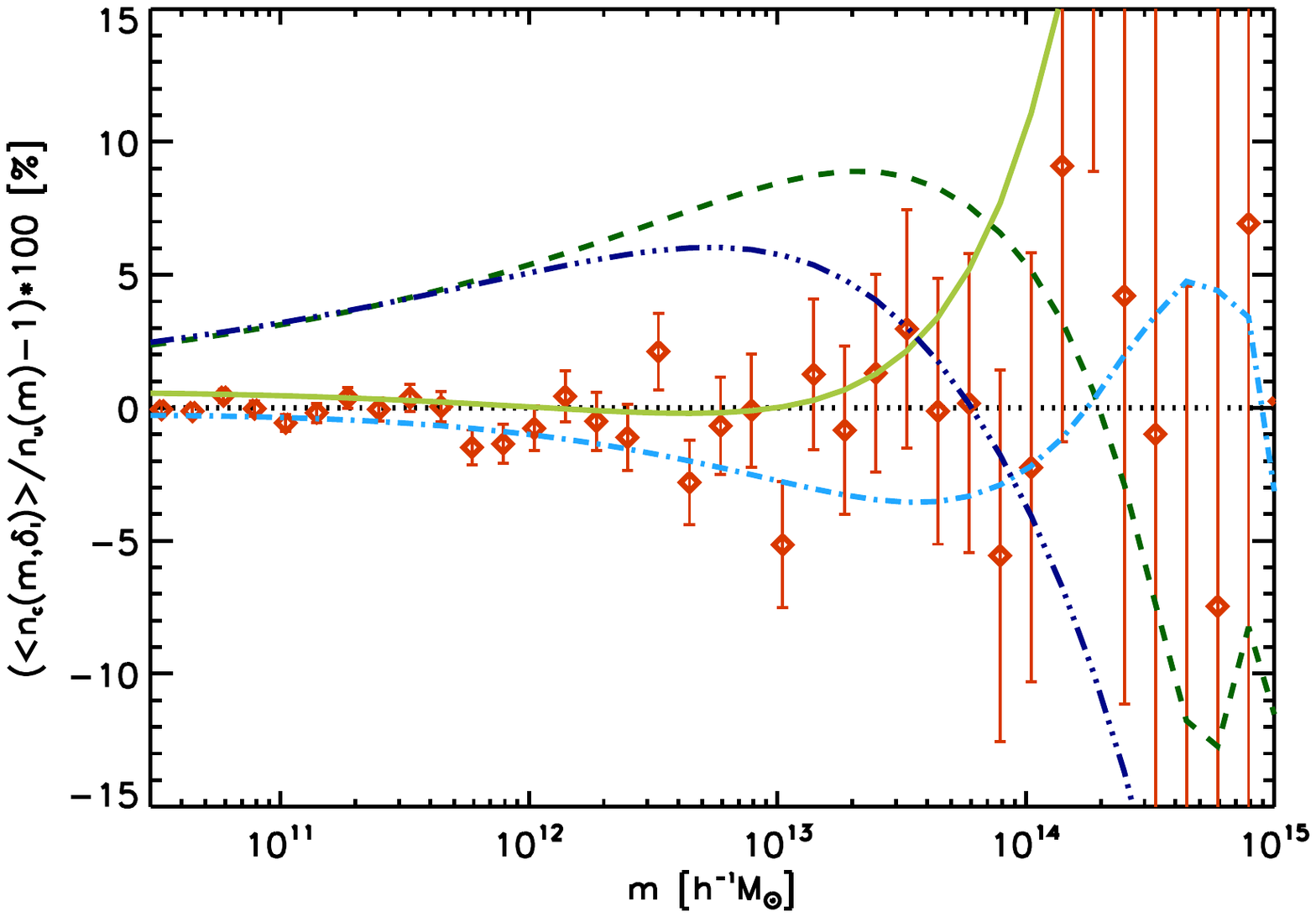} \quad
\includegraphics[trim = 20mm 130mm 30mm 30mm, scale=0.5]{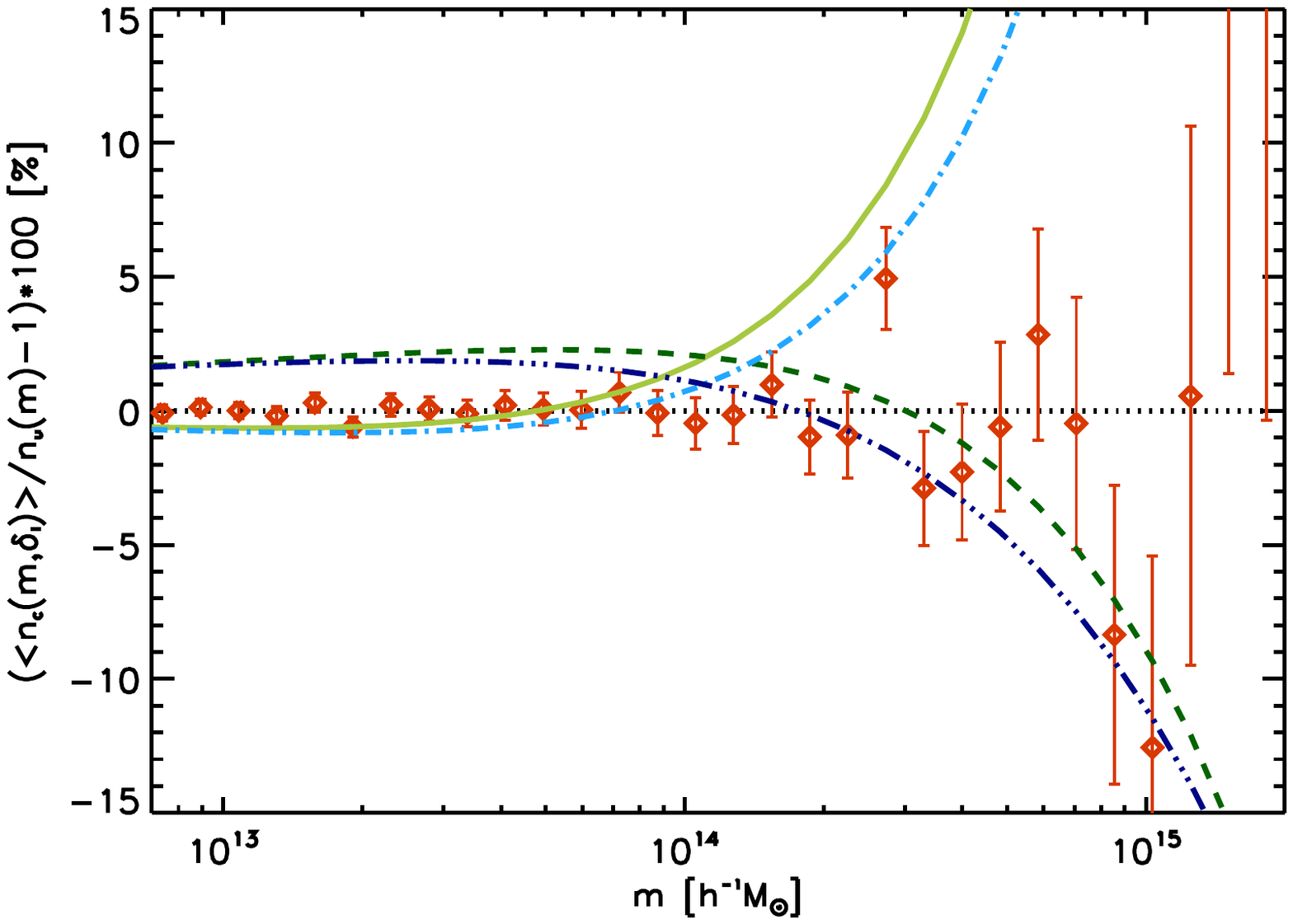} 
\caption{Normalization test for the analytical CMF described in
    Section~\ref{sec:CMF}, for both the SSS (left column) and the LSS (right
    column). Top panels show the average (over all the spheres satisfying the
    condition indicated in the title of each panel) of the analytical CMF for
    each of those conditions. Bottom panels show the corresponding residuals (in
    percent) against the reference UMF, i.e., the T08 mass function. The solid
    line corresponds to the standard rescaling normalized with $\alpha=1.25$;
    the dashed line is the standard rescaling with no explicit normalization;
    the dot-dashed line is the local rescaling normalized with $\alpha=1.25$;
    and the triple dot-dashed line is the local rescaling with no explicit
    normalization. For the analytical predictions, the normalization is clearly
    improved by the inclusion of the $\alpha=1.25$ parameter in the rescaling
    procedure. Finally, as a consistency check, we also include, using data
    points, the average of all the CMFs evaluated numerically from the
    simulation in the same spheres. By construction, this average recovers the
    numerical UMF (i.e., data points in Figure~\ref{fig:UMFs}), with high
    accuracy (see data points in the bottom panel).}
\label{fig:normalization}
\end{figure*}


\subsection{Results}
\label{subsec:results}

\subsubsection{CMF normalization}
\label{subsubsec:normalization}

The first consistency check to be applied to any analytical prescription to
compute the CMF is a normalization condition. The conditional halo mass function
must recover the unconditional one when averaged over all possible values of the
density contrast. In practice, this is done by evaluating the analytical CMF for all the
set of possible conditions $(R,\deltal)$ that we have obtained with the spheres
described in the previous subsection, and averaging over all of them.
Note that this test is independent and complementary to the analysis that we
perform in Appendix~\ref{sec:normpar}. There are two main differences: first, 
we are using here, by construction, the true
probability distribution of $\deltal$ as derived from our simulations (see
Figure~\ref{fig:posteriors}); and second, we are integrating for a fixed
Eulerian $R$ value, instead of a fixed Lagrangian $Q$ value.

The normalization condition must hold true for each value of the radius, that
is, for all bins in $R$. We show in Figure~\ref{fig:normalization} the results
for one radius bin, for both simulations, plotted in terms of the halo
density. We plot the analytical predictions from eq.~\ref{eq:theohist} obtained using the two different
recipes for the CMFs, namely the local and the standard rescalings of the T08 
function, both with no explicit normalization and with normalization parameter
$\alpha=1.25$. In order to better visualize the deviations among the different
predictions, we show in the lower panel of Figure~\ref{fig:normalization} the
corresponding plot of residuals, computed as percent deviations with respect to
the unconditional mass function (T08 in this case).
Residuals from the analytical determinations of the CMF give us information
about the accuracy at which the normalization condition is satisfied. These
residuals are clearly reduced when introducing the normalization parameter
$\alpha=1.25$ in the UMF rescalings, resulting in a final $\sim 1\,\%$ error
with respect to the T08 mass function. This confirms that the modification of
the rescaling introduced to accomplish the normalization condition also works
when using the proper $\deltal$ distribution from numerical simulation, and for
a fixed Eulerian condition radius. These conclusions are the same for both the
CMF computed with the standard rescaling and the CMF computed with the local
rescaling.
Residuals from all theoretical CMFs grow larger than $10\,\%$ above a certain
mass threshold, around $\sim 10^{14}\,h^{-1}\,\text{M}_{\odot}$ for the SSS, and
$\sim 4 \times 10^{14}\,h^{-1}\,\text{M}_{\odot}$ for the LSS. Now, in RBP08 it
is underlined that the formalism for computing the CMF is expected to fail when
approaching the mass of the condition; in the same reference, a fraction of
$1/30$ of the condition mass is chosen as an upper limit setting the mass range
where the CMF formalism is valid. According to this prescription, the mass
values above which our residuals grow very large correspond to the limiting top
masses for conditions of size $Q\sim 20\,h^{-1}\,\text{Mpc}$ for the SSS, and $Q
\sim 35\,h^{-1}\,\text{Mpc}$ for the LSS. By looking at the $Q$ probability
distributions in Figure~\ref{fig:posteriors}, we found these radii to fall in
the very high tail of the distributions. Errors in normalizations therefore
arise in a mass interval dominated by rare massive objects that are found in
highly overdense regions, which we are not going to consider in the following
analysis.

For completeness, we also added in those panels in
Figure~\ref{fig:normalization} the data points corresponding to the average of
the actual CMFs obtained from the simulations for those spheres, computed using eq.~\ref{eq:simhist}. As expected,
the average of all the numerical CMFs recovers the UMF plotted with points in
Figure~\ref{fig:UMFs} with few percent accuracy, for all $R$ bins, and for both
SSS and LSS simulations. This result proves the consistency of our algorithm in
placing random spheres and averaging their content. Note that errors grow larger
in the high mass end where the box finite-size effect becomes important and the
statistics of collapsed objects is poorer.

\begin{figure*}
\includegraphics[trim = 60mm 130mm 30mm 20mm, scale=0.5]{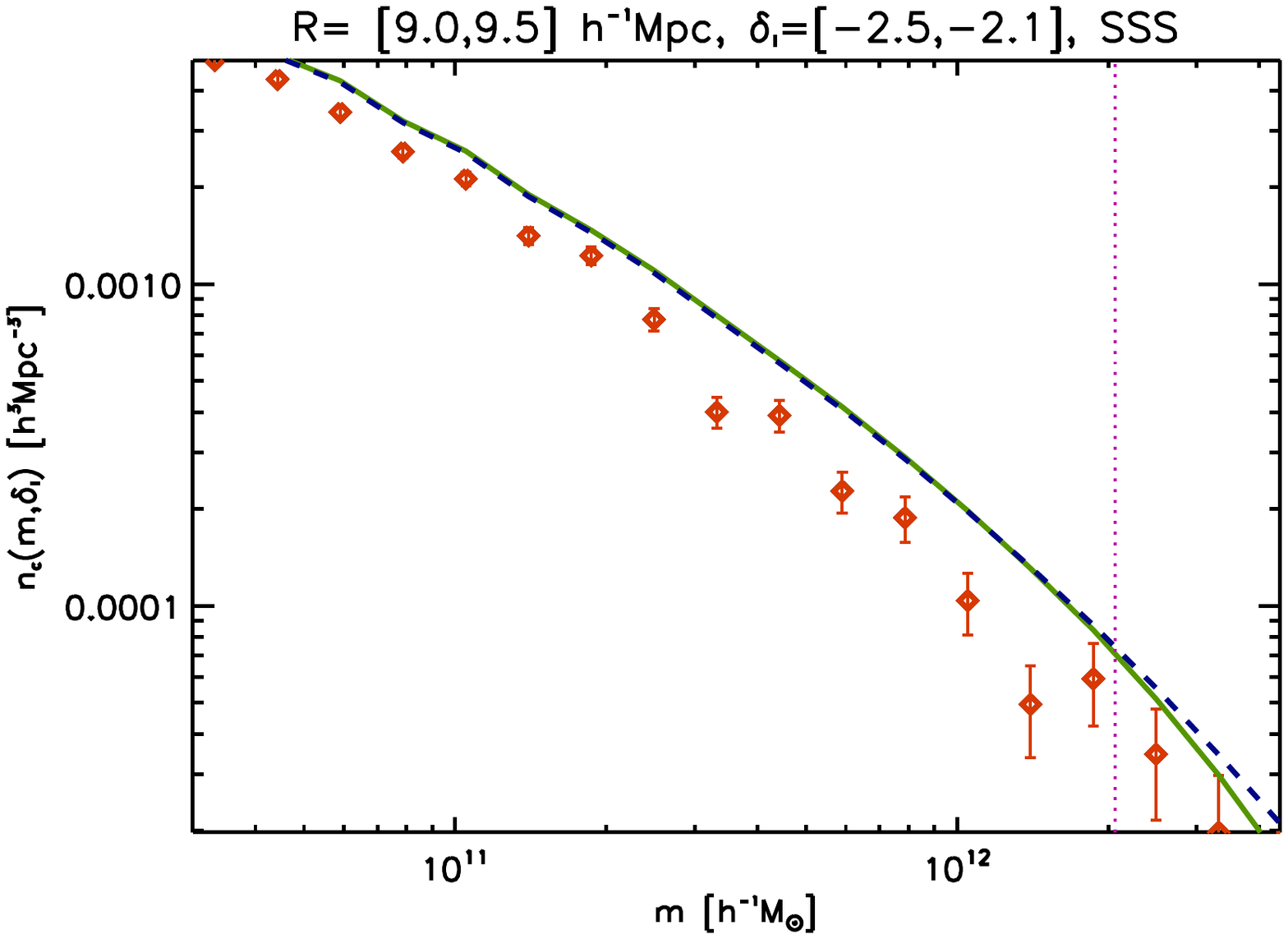} \quad
\includegraphics[trim = 20mm 130mm 30mm 20mm, scale=0.5]{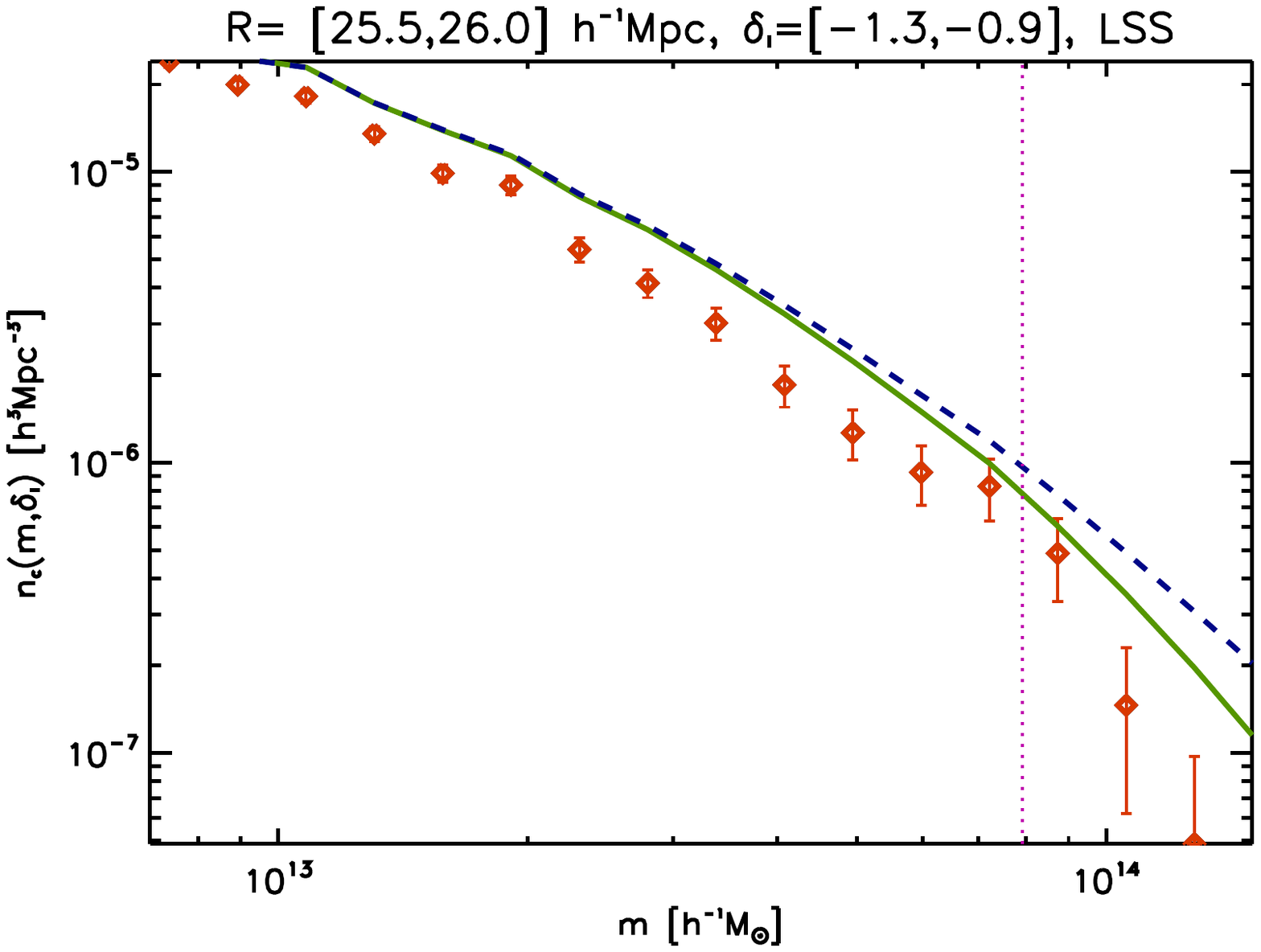} \\
\includegraphics[trim = 60mm 130mm 30mm 20mm, scale=0.5]{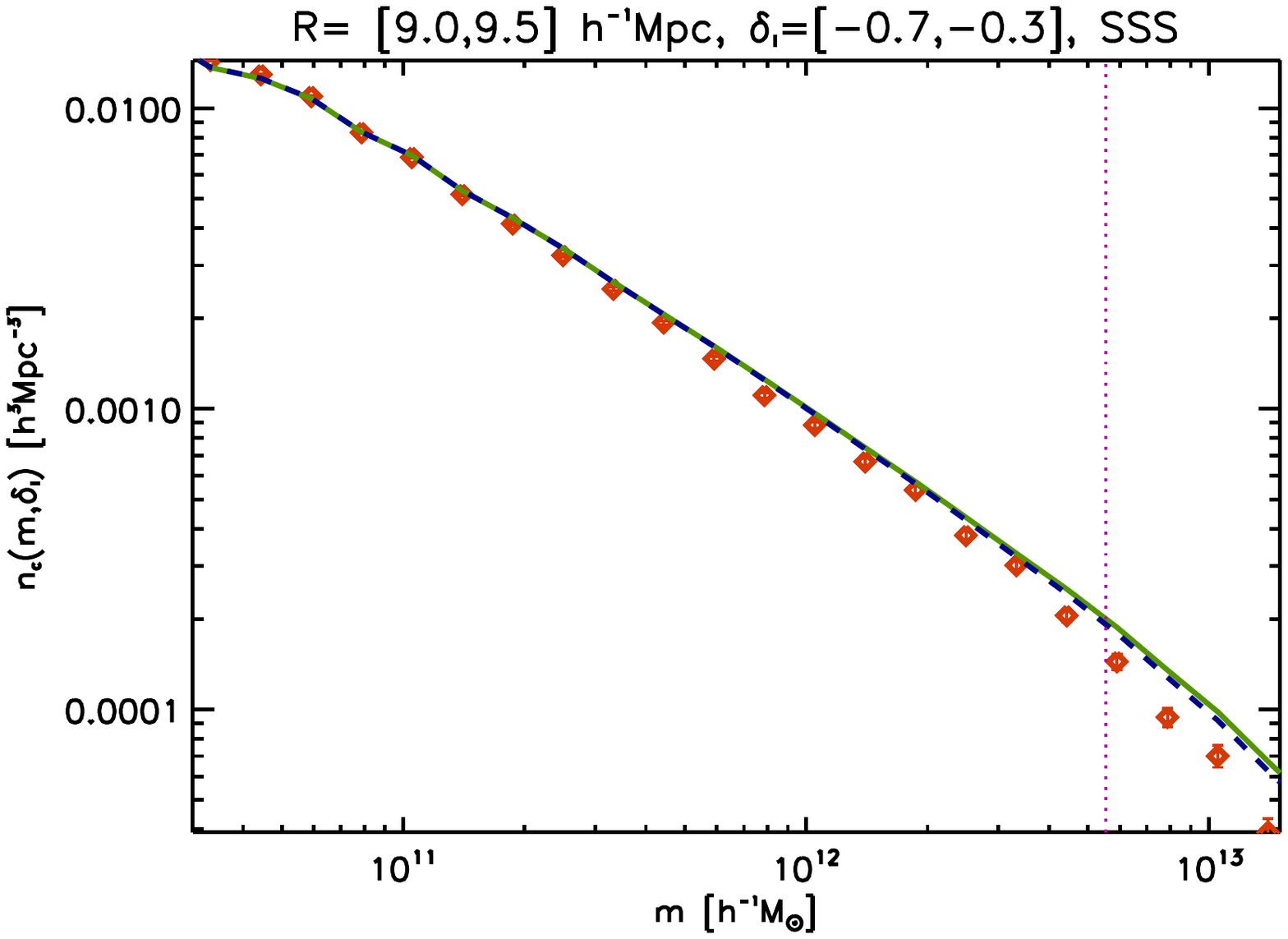} \quad
\includegraphics[trim = 20mm 130mm 30mm 20mm, scale=0.5]{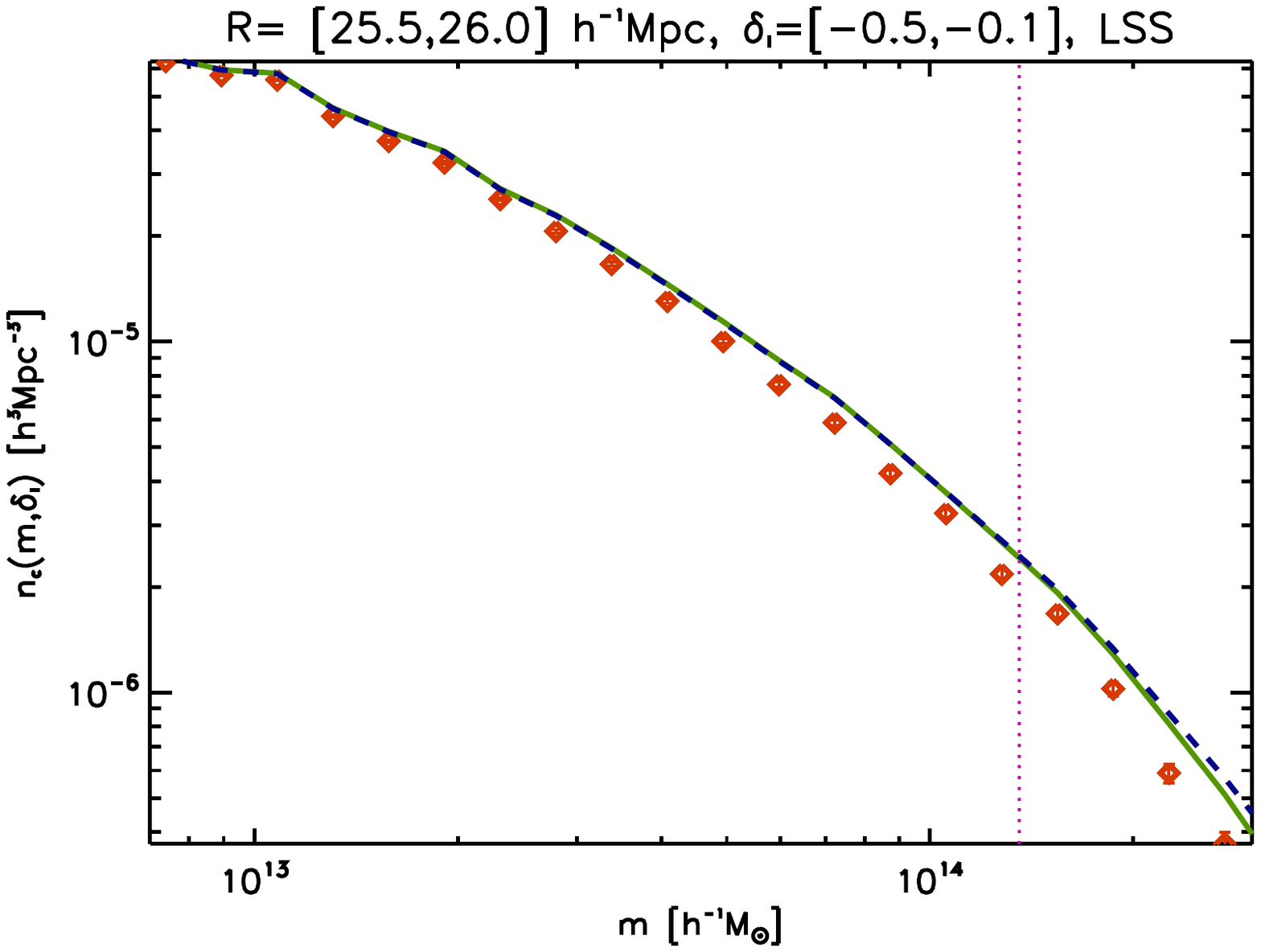} \\
\caption{Comparison between simulation and theoretical CMFs, plotted as the halo
  number density as a function of mass. Left column shows results for a smaller
  condition in the SSS, right column for a larger condition in the LSS. Upper
  panels show the predictions for an underdense region with a higher density
  contrast with respect to the lower panels. Points: simulation; solid line:
  standard rescaling; dashed line: locally rescaled CMF, without
  normalization. The vertical dotted line represents a fraction $1/30$ of the
  average condition mass. The two theoretical CMFs show general agreement except
  in the high mass tail, but they both fail in reproducing results from
  simulations in case of strongly underdense regions.}
\label{fig:comparison_neg_1}
\end{figure*}
\begin{figure*}
\includegraphics[trim = 60mm 130mm 30mm 20mm, scale=0.5]{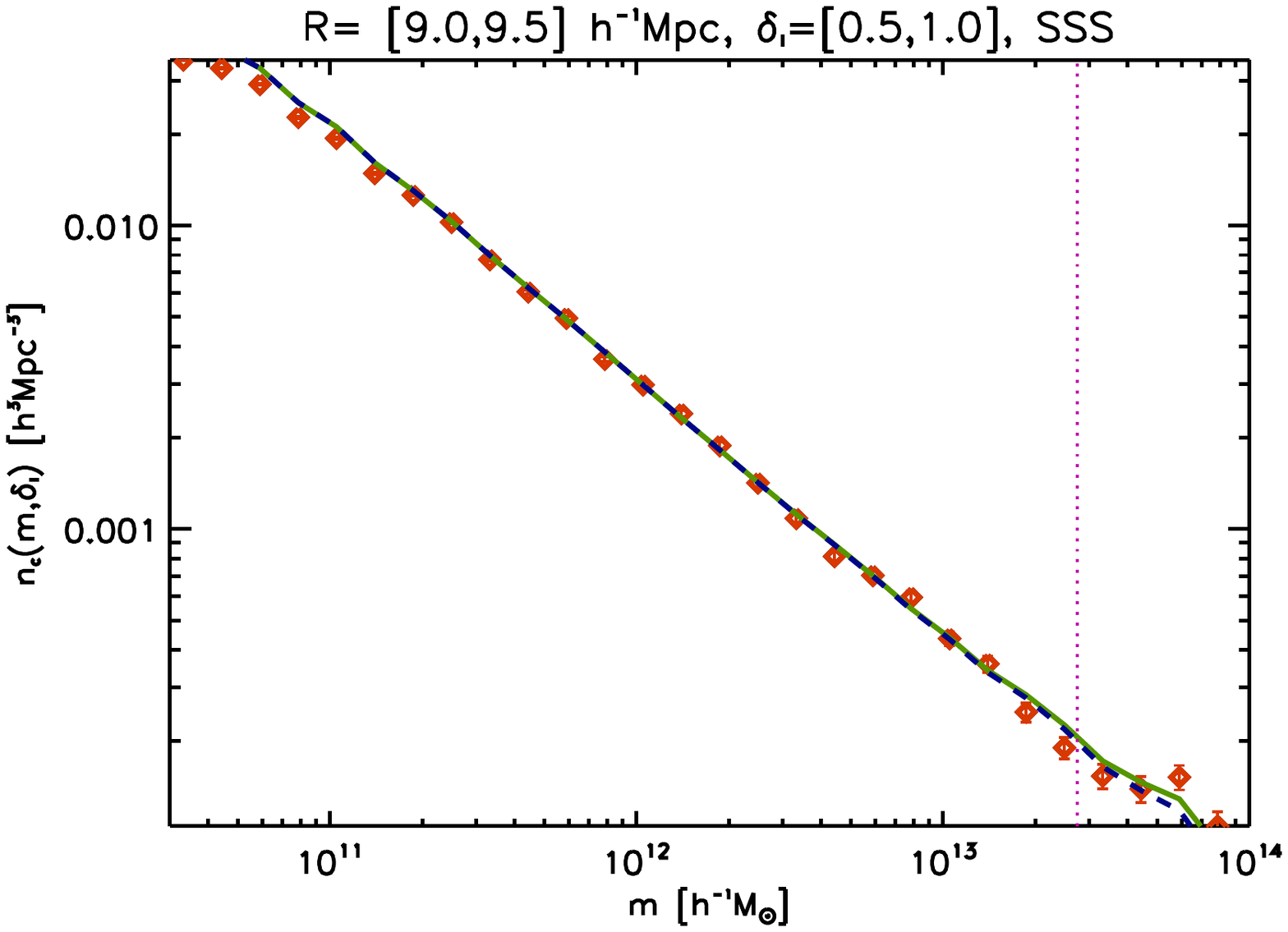} \quad
\includegraphics[trim = 20mm 130mm 30mm 20mm, scale=0.5]{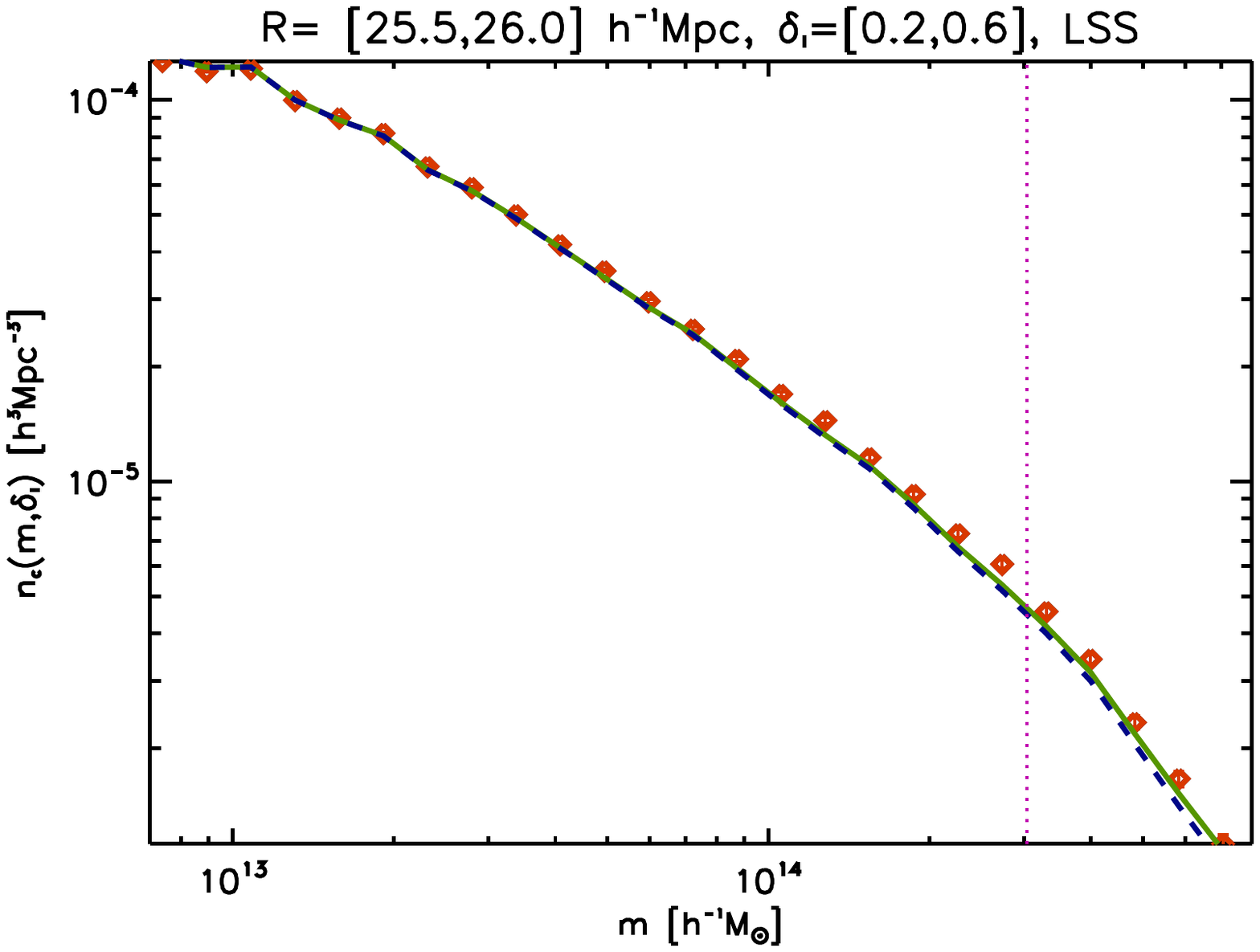} \\
\caption{Same as Figure~\ref{fig:comparison_neg_1}, but in the case of an overdense region. In this case both CMFs show overall agreement with counts from simulations.}
\label{fig:comparison_pos_1}
\end{figure*}

\begin{figure*}
\includegraphics[trim = 60mm 130mm 30mm 20mm, scale=0.5]{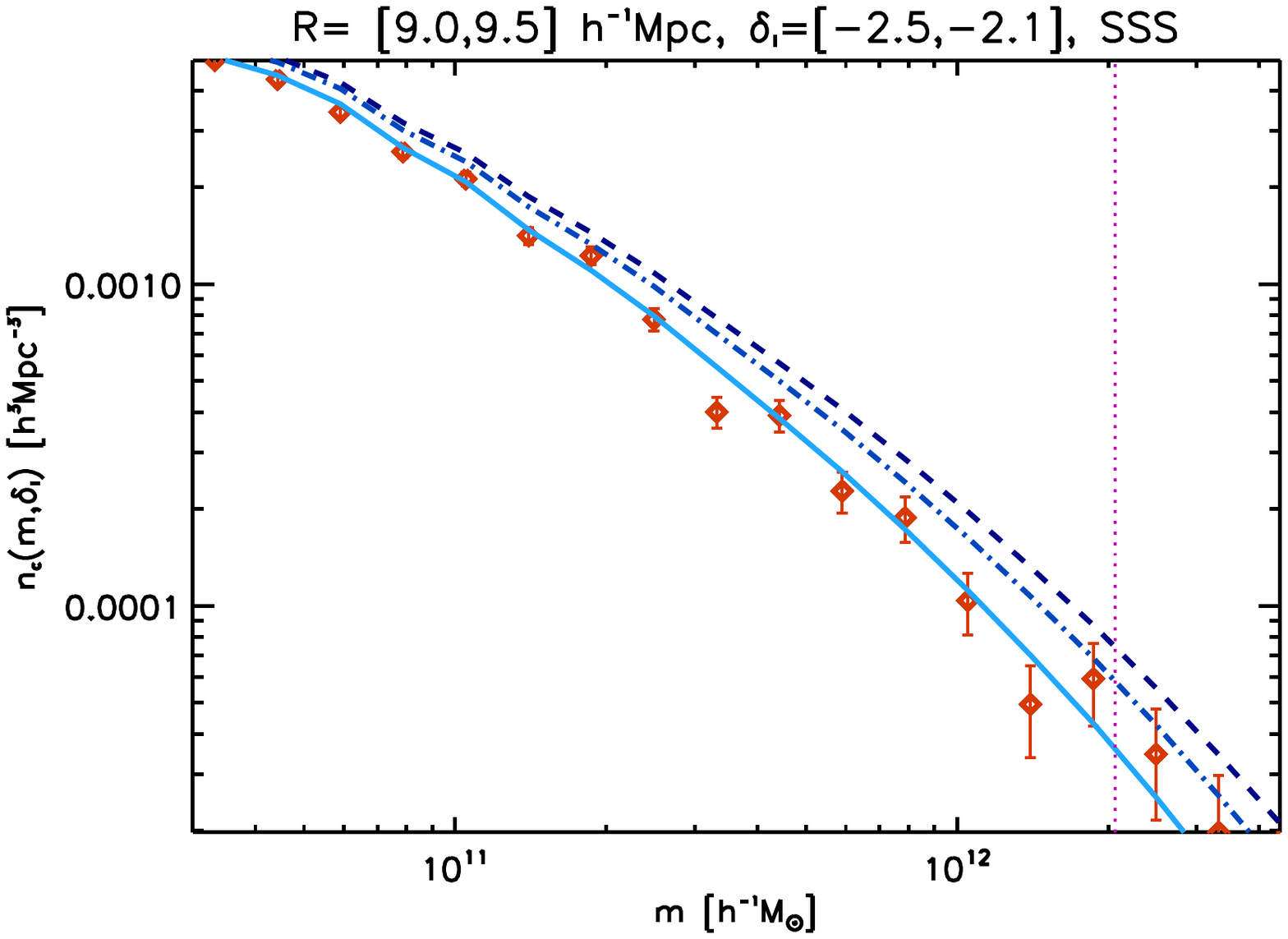} \quad
\includegraphics[trim = 20mm 130mm 30mm 20mm, scale=0.5]{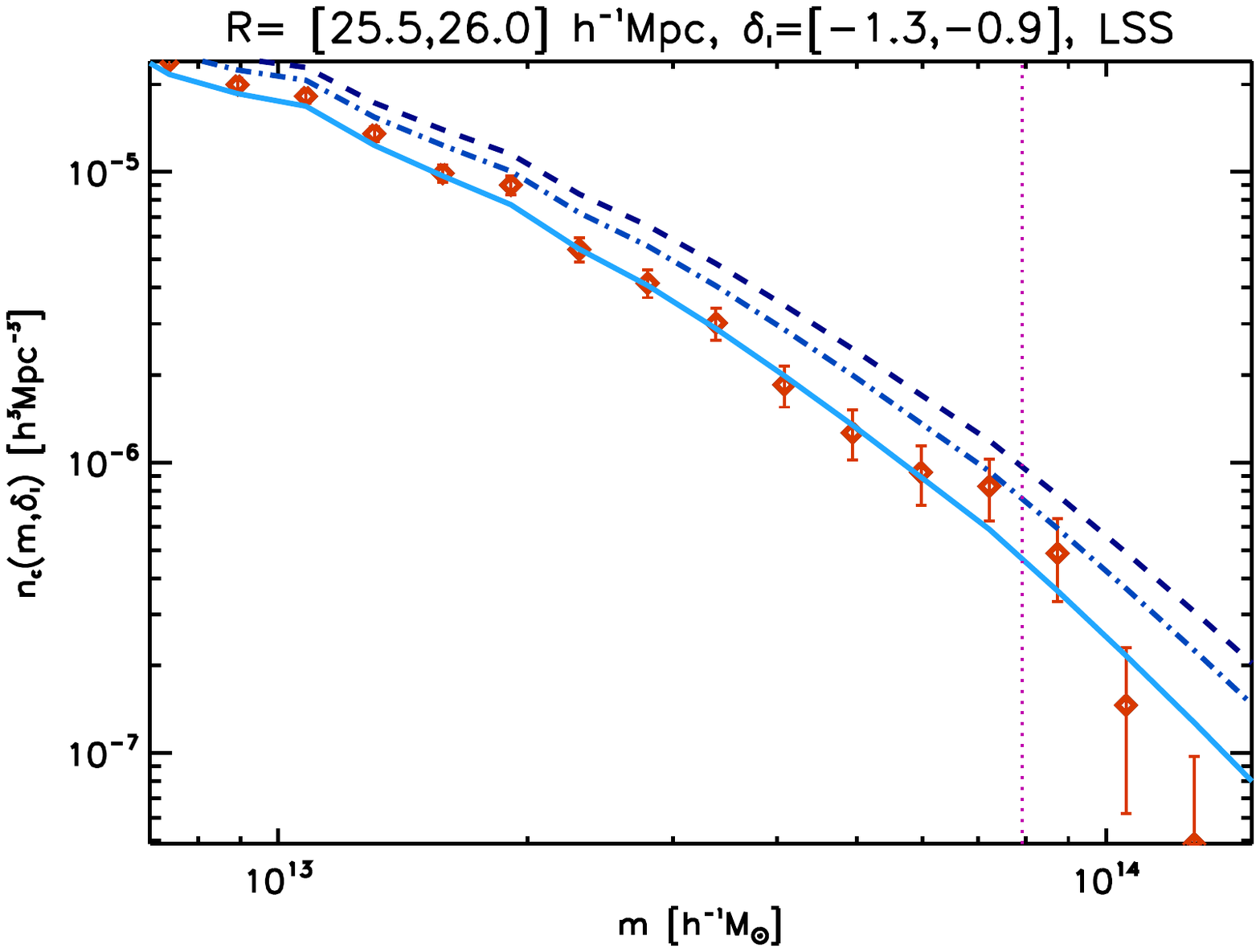} \\
\includegraphics[trim = 60mm 130mm 30mm 20mm, scale=0.5]{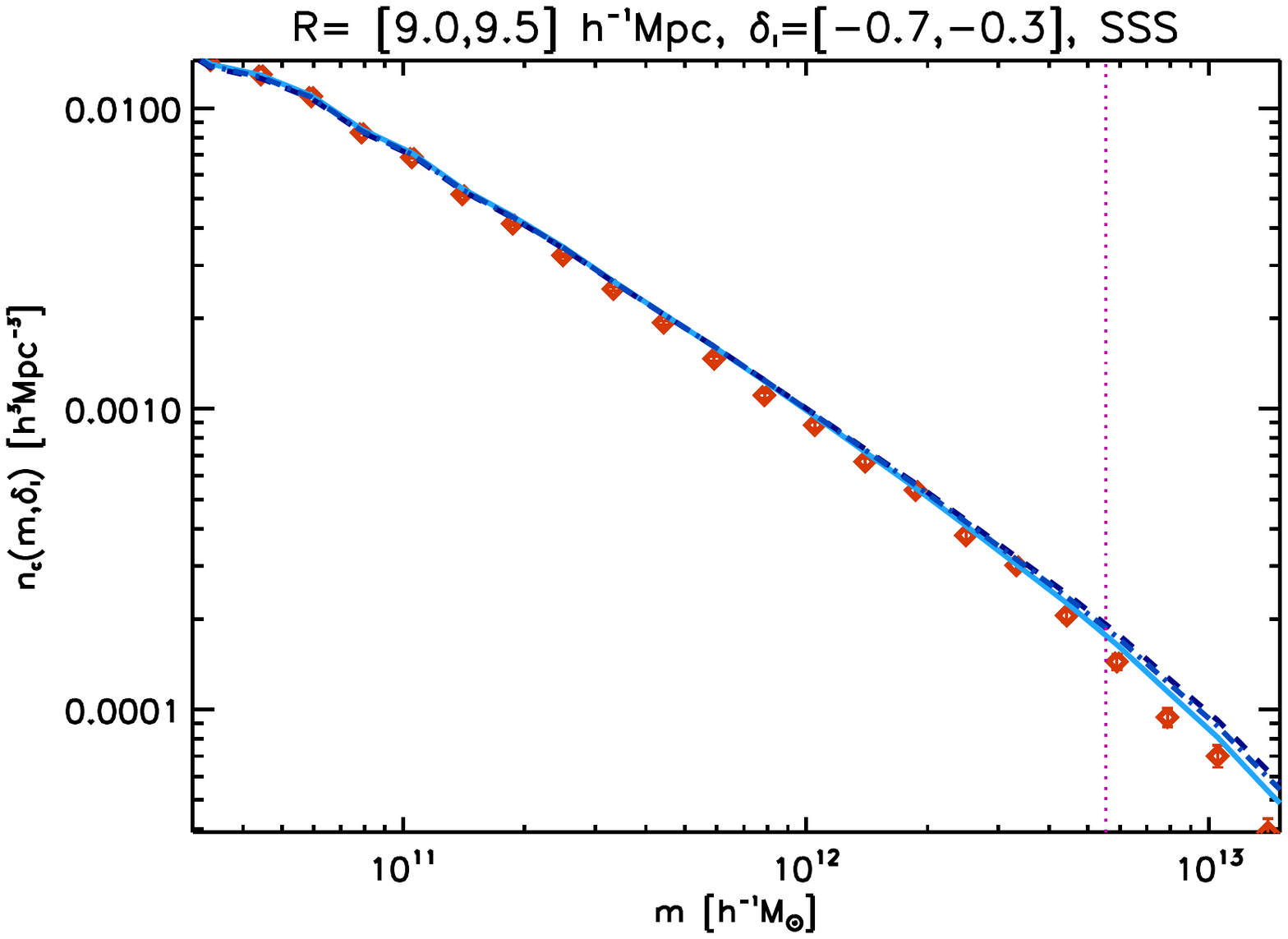} \quad
\includegraphics[trim = 20mm 130mm 30mm 20mm, scale=0.5]{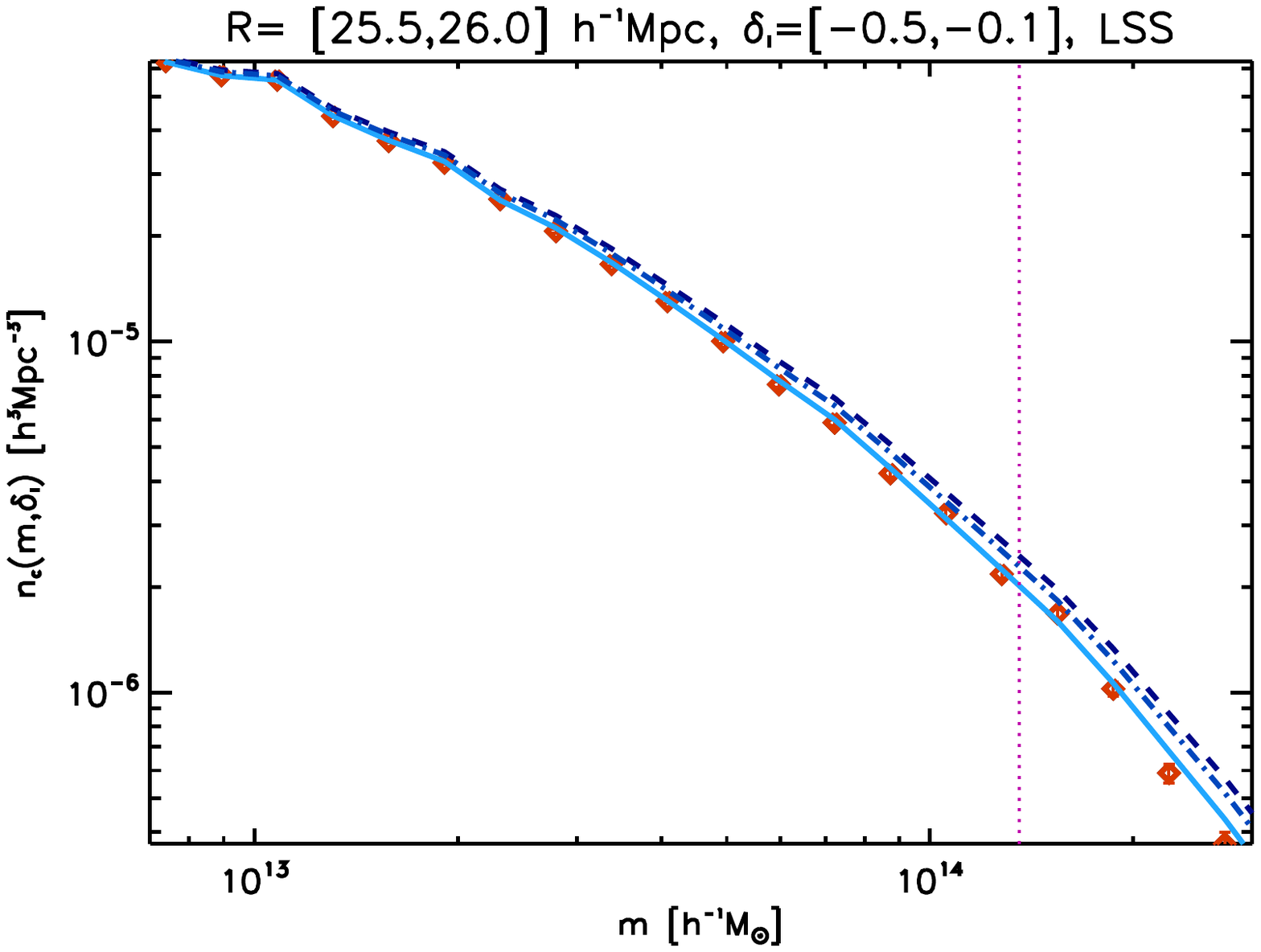} \\
\caption{Comparison between simulation and the locally rescaled CMF, considering the same cases as Figure~\ref{fig:comparison_neg_1}, but showing the effect of the normalization parameter in the rescaling procedure. Points: simulation; solid line: locally rescaled CMF, with normalization parameter $\alpha=1.25$; dot-dashed line: locally rescaled CMF, with a more conservative value $\alpha=1.5$; dashed line: locally rescaled CMF with no explicit normalization. By comparison with Figure~\ref{fig:comparison_neg_1}, it is evident that the inclusion of the normalization parameter improves the agreement of the theoretical prediction with the counts from simulations.}
\label{fig:comparison_neg_2}
\end{figure*}
\begin{figure*}
\includegraphics[trim = 60mm 130mm 30mm 20mm, scale=0.5]{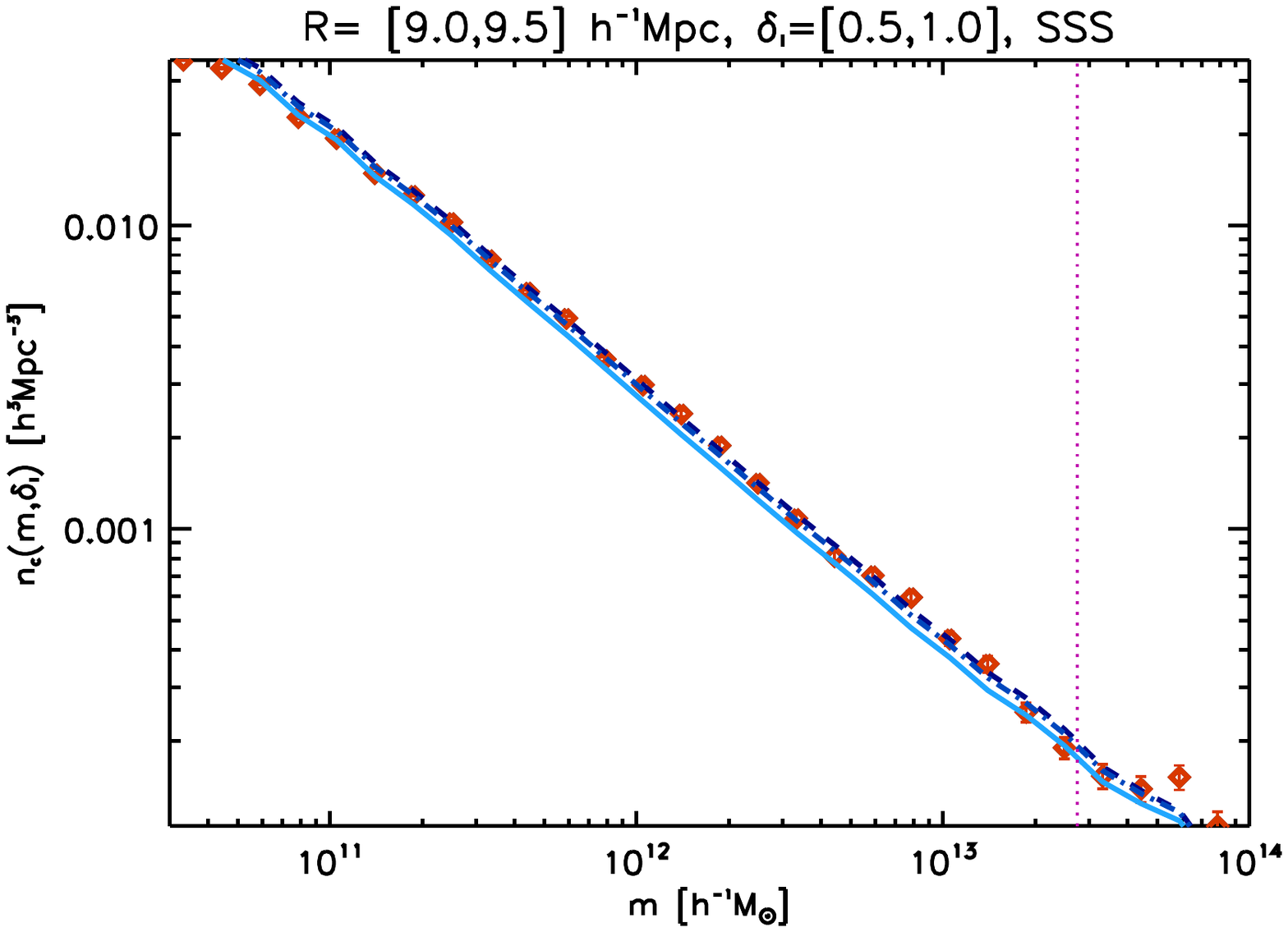} \quad
\includegraphics[trim = 20mm 130mm 30mm 20mm, scale=0.5]{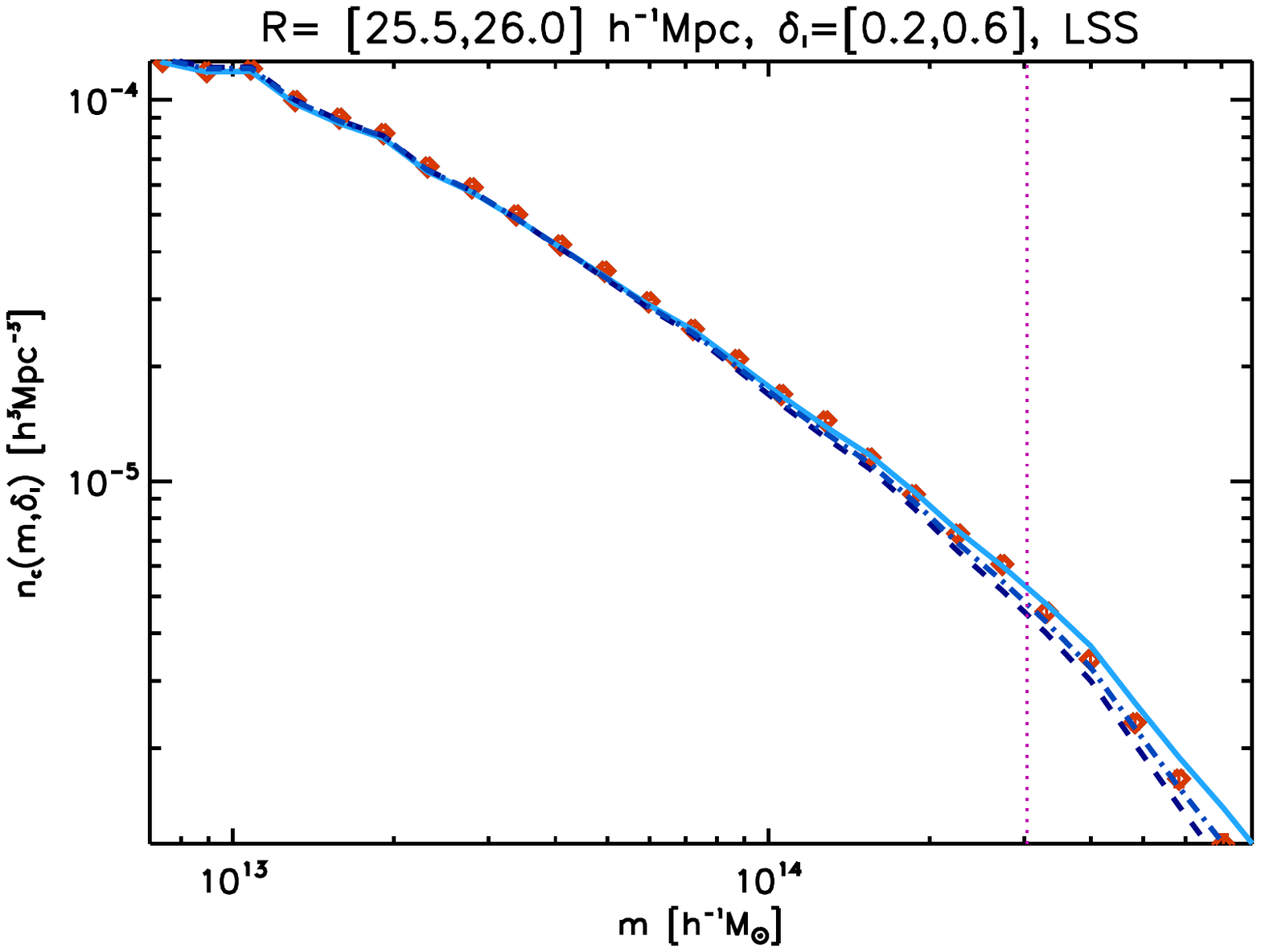} \\
\caption{Same as Figure~\ref{fig:comparison_neg_2}, but in the case of an overdense region. The inclusion of explicit normalization in the CMF rescaling still improves the agreement with simulations, even though in the case of SSS it seems to require a more conservative value for the parameter $\alpha$.}
\label{fig:comparison_pos_2}
\end{figure*}

\begin{figure*}
\includegraphics[trim = 60mm 130mm 30mm 20mm, scale=0.5]{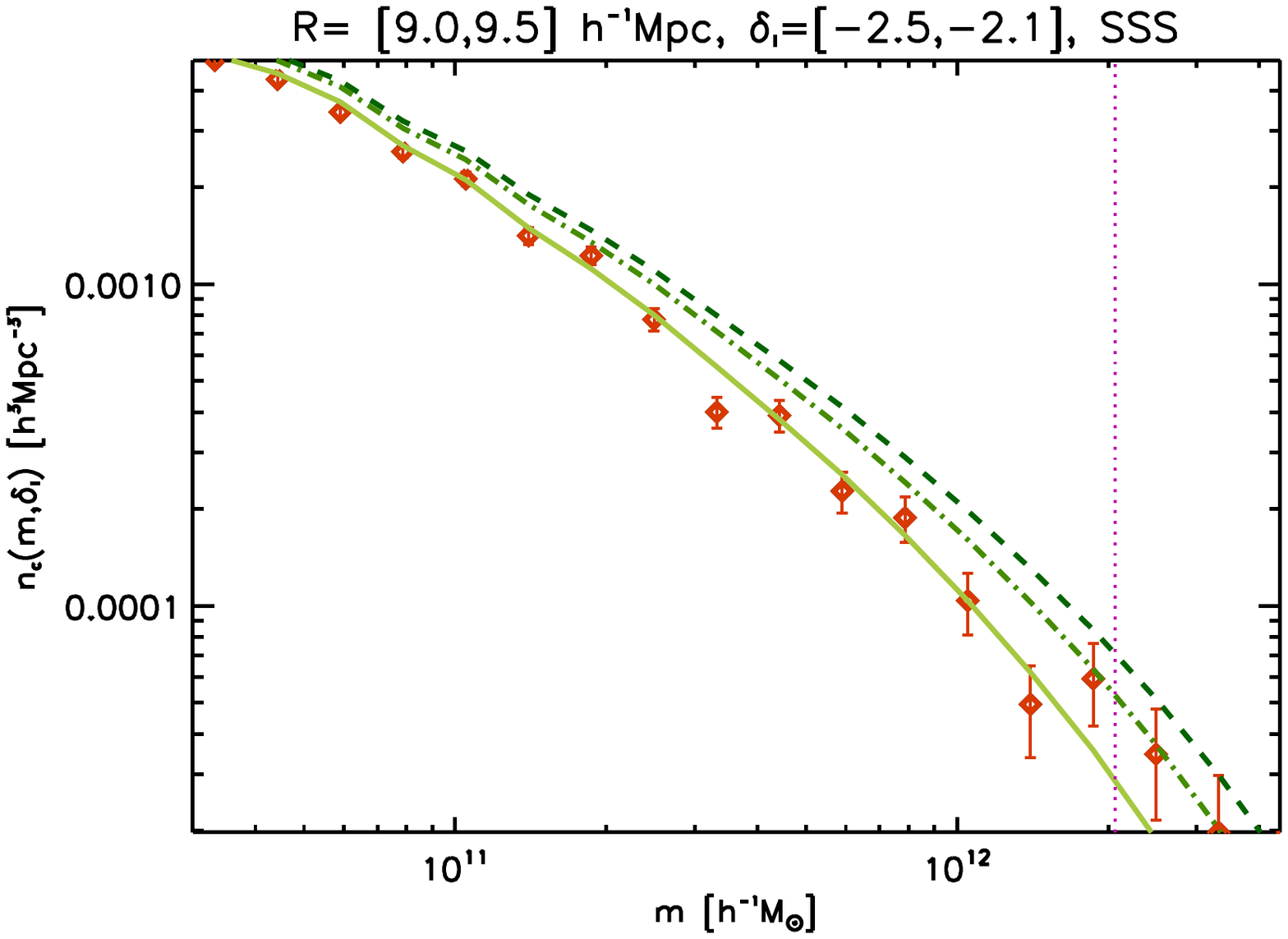} \quad
\includegraphics[trim = 20mm 130mm 30mm 20mm, scale=0.5]{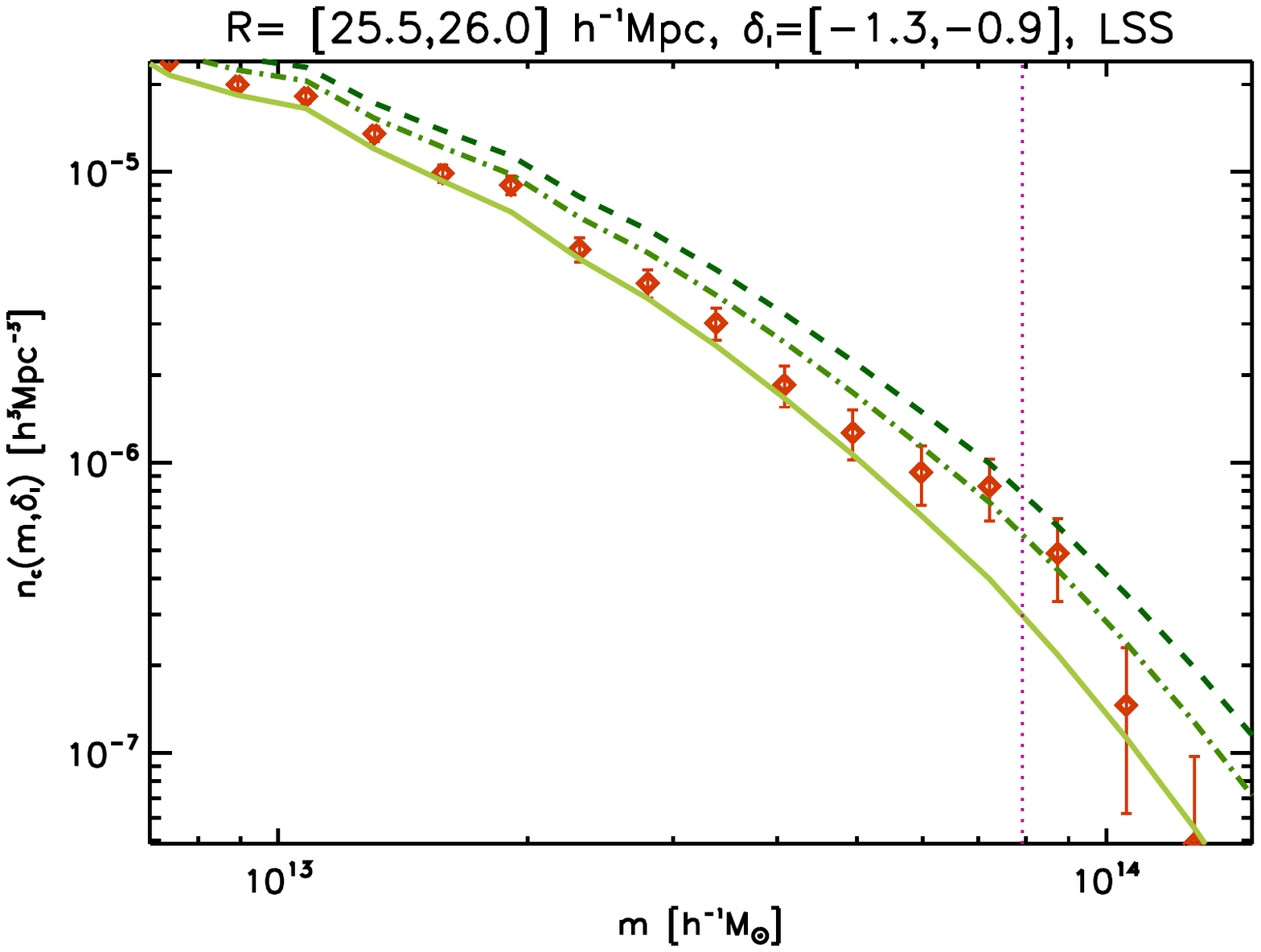} \\
\includegraphics[trim = 60mm 130mm 30mm 20mm, scale=0.5]{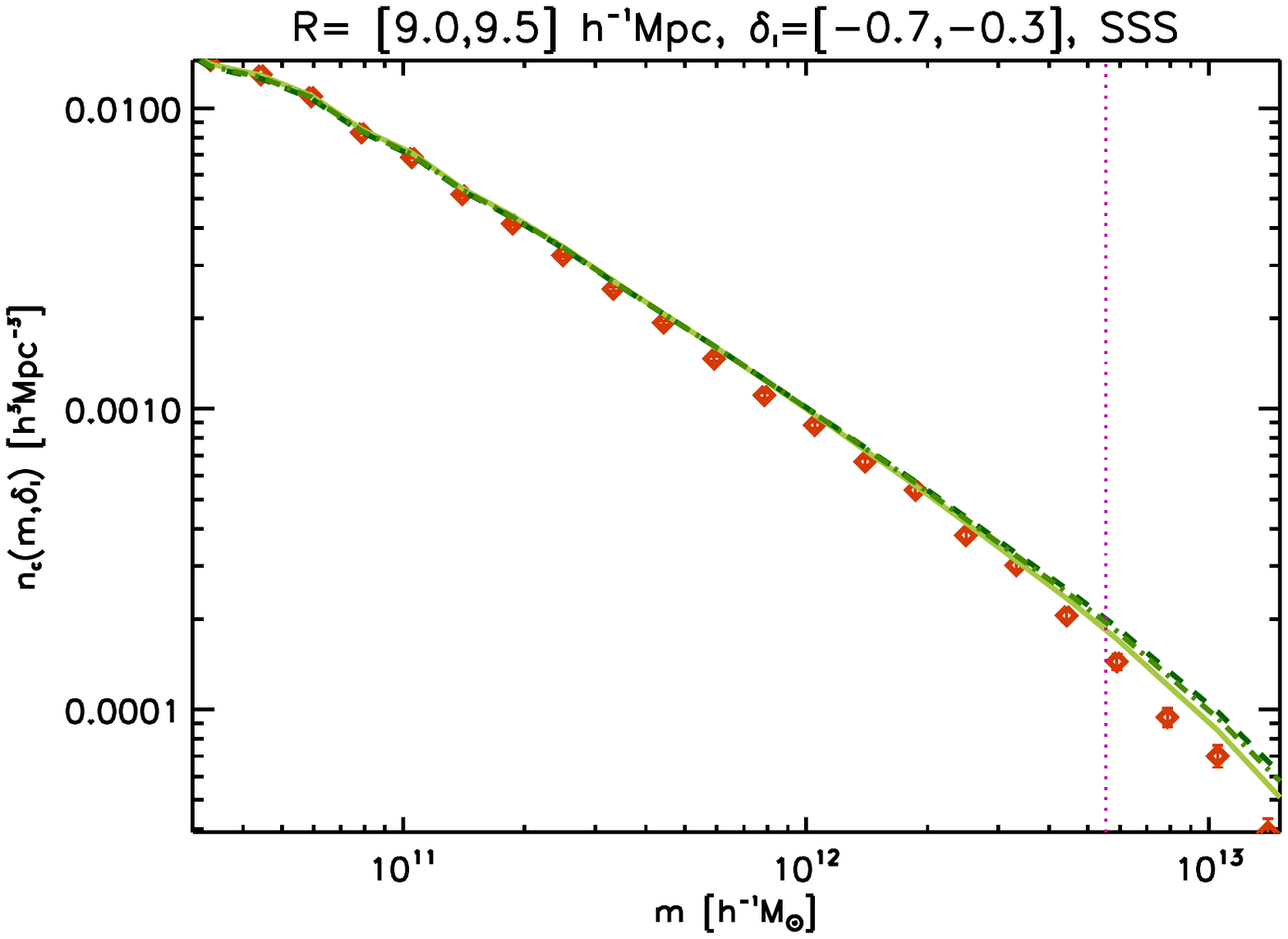} \quad
\includegraphics[trim = 20mm 130mm 30mm 20mm, scale=0.5]{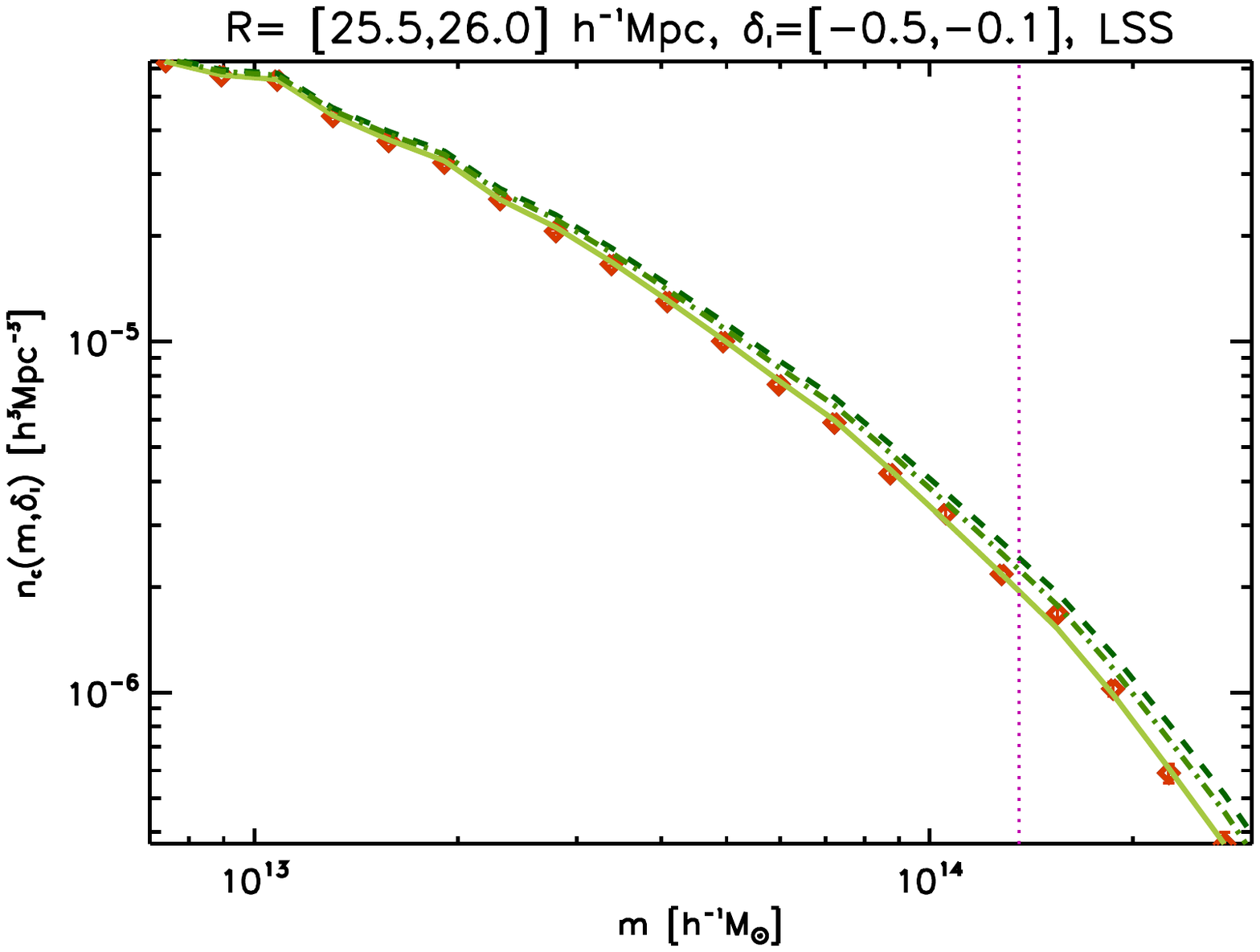} \\
\caption{Comparison between simulation and theoretical CMF, for the same cases as Figure~\ref{fig:comparison_neg_2}, but considering this time the effect of the normalization parameter on the standard rescaling. Points: simulation; solid line: standard CMF with normalization parameter $\alpha=1.25$; dot-dashed line: standard CMF with normalization parameter $\alpha=1.5$; dashed line: standard CMF with no explicit normalization. Comparison with Figure~\ref{fig:comparison_neg_2} shows that normalizing the locally rescaled CMF or the standard CMF yields very similar results when reproducing counts from simulations, the inclusion of explicit normalization achieving a better agreement.}
\label{fig:comparison_neg_3}
\end{figure*}
\begin{figure*}
\includegraphics[trim = 60mm 130mm 30mm 20mm, scale=0.5]{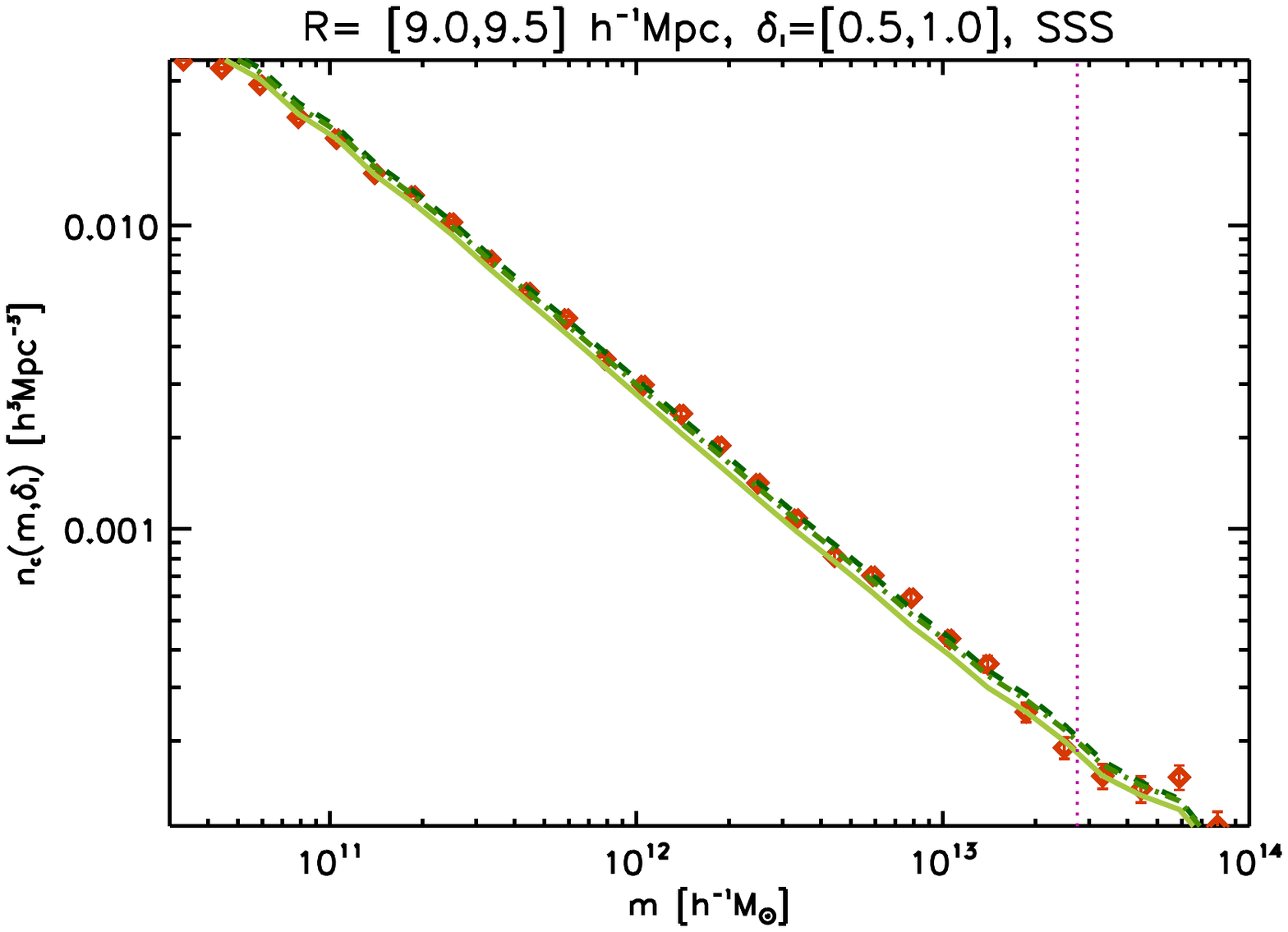} \quad
\includegraphics[trim = 20mm 130mm 30mm 20mm, scale=0.5]{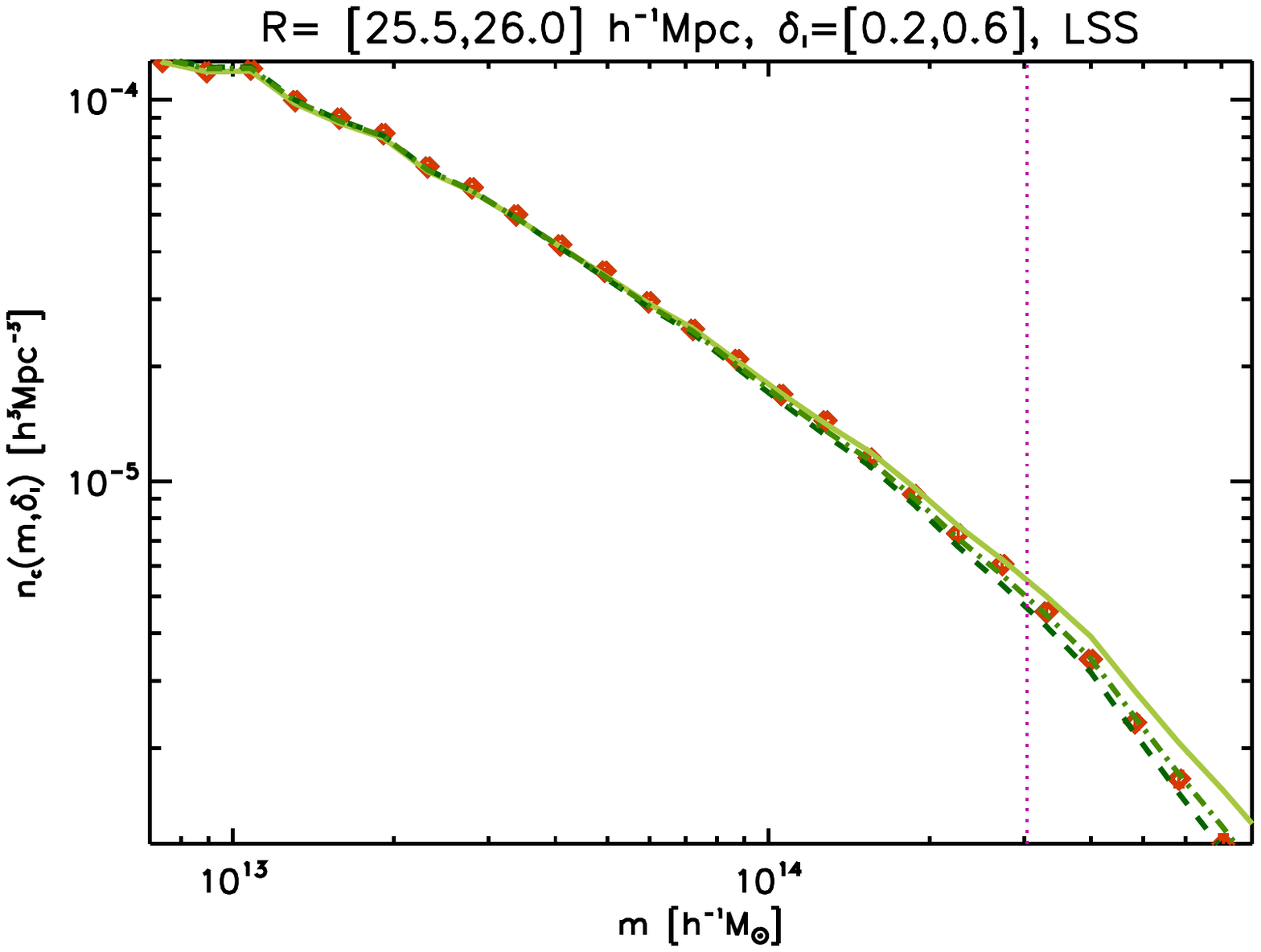} \\
\caption{Same as Figure~\ref{fig:comparison_neg_3}, but in the case of an overdense region. Again, results are very similar to the normalization of the locally rescaled CMF, as can be verified by comparing with Figure~\ref{fig:comparison_pos_2}. Explicit normalization improves the agreement with simulation, the SSS case demanding a higher value of $\alpha$. }
\label{fig:comparison_pos_3}
\end{figure*}

\subsubsection{Halo abundance in overdense and underdense regions}
We then checked the predictions of the CMF recipes for individual density
contrast bins. Note that results from eq.~\ref{eq:theohist}, in this case, are
not the quantity that can be compared to the simulation results
eq.~\ref{eq:simhist}; indeed, according to the discussion in
Section~\ref{subsec:simulation_set}, the abundances extracted from simulations
are affected by systematics due to cosmic variance. So, let $n_{\rm u}^{\text{theo}}$
and $n_{\rm u}^{\text{sim}}$ be the halo abundances computed using the unconditional
mass functions from theory (T08) and simulations, respectively; both are
plotted in Figure~\ref{fig:UMFs}. In order to take the aforementioned
systematics into account, the theoretical prediction can be rescaled as:
\begin{equation}
\label{eq:theohist_scaled}
\tilde{n}_{\rm c}^{\text{theo}}(m_j;R,\deltal) = n_{\rm c}^{\text{theo}}(m_j;R,\deltal)\, \frac{n_{\rm u}^{\text{sim}}(m_j)}{n_{\rm u}^{\text{theo}}(m_j)}.
\end{equation}
The prediction $\tilde{n}^{\text{theo}}$ is the one we finally compared with the simulation counts $n^{\text{sim}}$ for testing the CMFs recipes. In practice, this is equivalent to directly compare the so called matter-to-halo bias function \citep{ShethLemson99}, both from the simulations and the theoretical predictions.

In Figure~\ref{fig:comparison_neg_1} we show the comparison between the simulations and the CMFs computed using the standard rescaling and the local rescaling with no normalization, in the case of underdense regions; we show results for four bins of $\deltal$. Figure~\ref{fig:comparison_pos_1} shows two examples for the case of an overdense region. In all plots, a vertical line shows the mass correspondent to $1/30$ of the condition mass. We see that, in overdense regions, both the standard and the local CMF provide a very similar prediction, and in good agreement with the counts from simulations. In underdense regions, however, the agreement with simulations gets poorer, especially when the density contrast is stronger. Note that the oscillations of the theoretical predictions at low masses is simply a consequence of the normalization performed with eq.~\ref{eq:theohist_scaled}.

In Figures~\ref{fig:comparison_neg_2} and~\ref{fig:comparison_pos_2} we consider the same radius and density contrast bins, this time showing the comparison between counts from simulations and the prediction from the local rescaling with explicit normalization. We employed the value $\alpha=1.25$ as required by the T08 mass function normalization; we also show, as a reference, the locally rescaled CMF without normalizing and normalized with an intermediate choice of the normalization parameter, $\alpha=1.5$. In case of underdense regions, the explicit normalization of the CMF implies a clear improvement when reproducing the counts from simulations, particularly when the density contrast is stronger. For the most underdense bin we show, when changing from the not normalized CMF to the one using $\alpha=1.25$, the reduced $\chi^2$ changes from $9.7$ to $2.4$ in the SSS, and from $22.6$ to $0.7$ in the LSS. These results suggest that normalizing the local CMF with $\alpha=1.25$ provides a more accurate recipe in underdense regions, both at small and large scales; this is in agreement with the results in RBP08, where the same value for $\alpha$ is employed for normalizing the \citet{ST99} mass function. In case of overdense regions, the use of the same value of $\alpha$ does not always produce an improvement in the agreement with simulated counts. When introducing the $1.25$ normalization parameter, the reduced $\chi^2$ changes from $1.8$ to $42.7$ in the SSS, actually worsening the fit, while in the LSS we have again improvement with the reduced $\chi^2$ changing from $16.5$ to $3.3$.

We repeated the same comparisons using the standard CMF, computed with explicit normalization parameters $\alpha=1.25$ and $\alpha=1.5$. The resulting plots comparing the theoretical prediction with the simulation counts, for the same set of radius and overdensity bins considered so far, are shown in Figures~\ref{fig:comparison_neg_3} and~\ref{fig:comparison_pos_3}. The results are much alike the ones obtained with the local rescaling, with the explicit normalization substantially improving the agreement with simulation in underdense regions but not always in overdensities. In this case, for the most underdense bin, when introducing the $\alpha=1.25$ parameter the reduced $\chi^2$ changes from $10.3$ to $2.4$ in the SSS and from $18.0$ to $2.0$ in the LSS. In the positive $\deltal$ bin, instead, the reduced $\chi^2$ changes from $1.7$ to $40.5$ in the SSS, and from $11.4$ to $2.9$ in the LSS.

When comparing these $\chi^2$ with the ones quoted above, and for all possible cases, we find that the agreement with simulation is basically the same when using the local CMF or the standard CMF (see e.g., Figures~\ref{fig:comparison_neg_1} and~\ref{fig:comparison_pos_1}), although the local rescaling provides a better description in underdense and intermediate regions than the standard rescaling. In overdense regions, however, the standard rescaling seems slightly better for this particular simulation. 

In conclusion, when trying to reproduce the abundance of halos inside
conditioning regions, the best recipe for computing the CMF, in either the
standard or the local rescaling, requires the use of the $\alpha$ parameter.
When using $\alpha=1.25$, for underdense regions we clearly improve the agreement with numerical simulations, while for the case of overdensities, although there is still an improvement for large scales, we have an overcorrection for small regions. We checked that these conclusions still hold when changing the mass range over which we computed the $\chi^2$. The large values for the $\chi^2$ obtained with explicit normalisation in the SSS may be due to an underestimation of the errors; what matters here is that in this case the normalised CMF worsen the agreement with simulated abundances. It is not clear whether this issue is due to the particular simulation we are using. The same comparison could be repeated by combining results from many N-body simulations; this goes beyond the scopes of this work. In any case, we conclude that the normalized CMF with $\alpha=1.25$ is the one that works better in the majority of the cases.


\section{A fitting formula for the CMF in underdense regions}
\label{sec:fit}

As discussed in RBP08, one of the main driving goals for developing a proper parametrization of the conditional mass function is the theoretical study of void statistics. A void can be considered, within our framework, as a locally underdense region, and the distribution of collapsed objects within it can be described by the CMF. The formalism for computing the number density of void is discussed in \citet{PBP06}; voids can be used as a tool to constrain cosmological parameters, as it is shown in \citet{BPP09}.

\begin{figure*}
\includegraphics[trim = 30mm 120mm 10mm 20mm, scale=0.45]{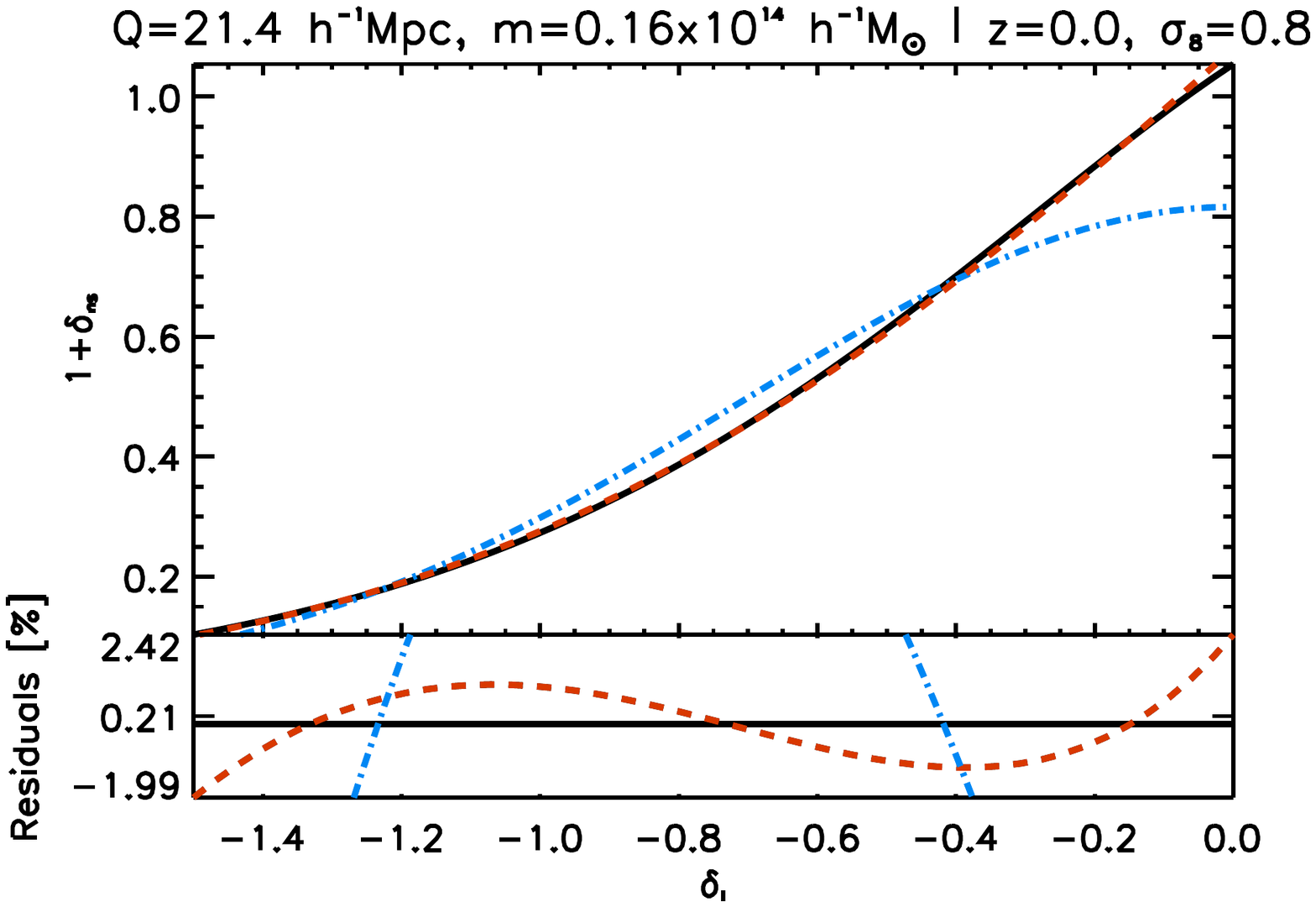} \quad
\includegraphics[trim = 10mm 120mm 30mm 20mm, scale=0.45]{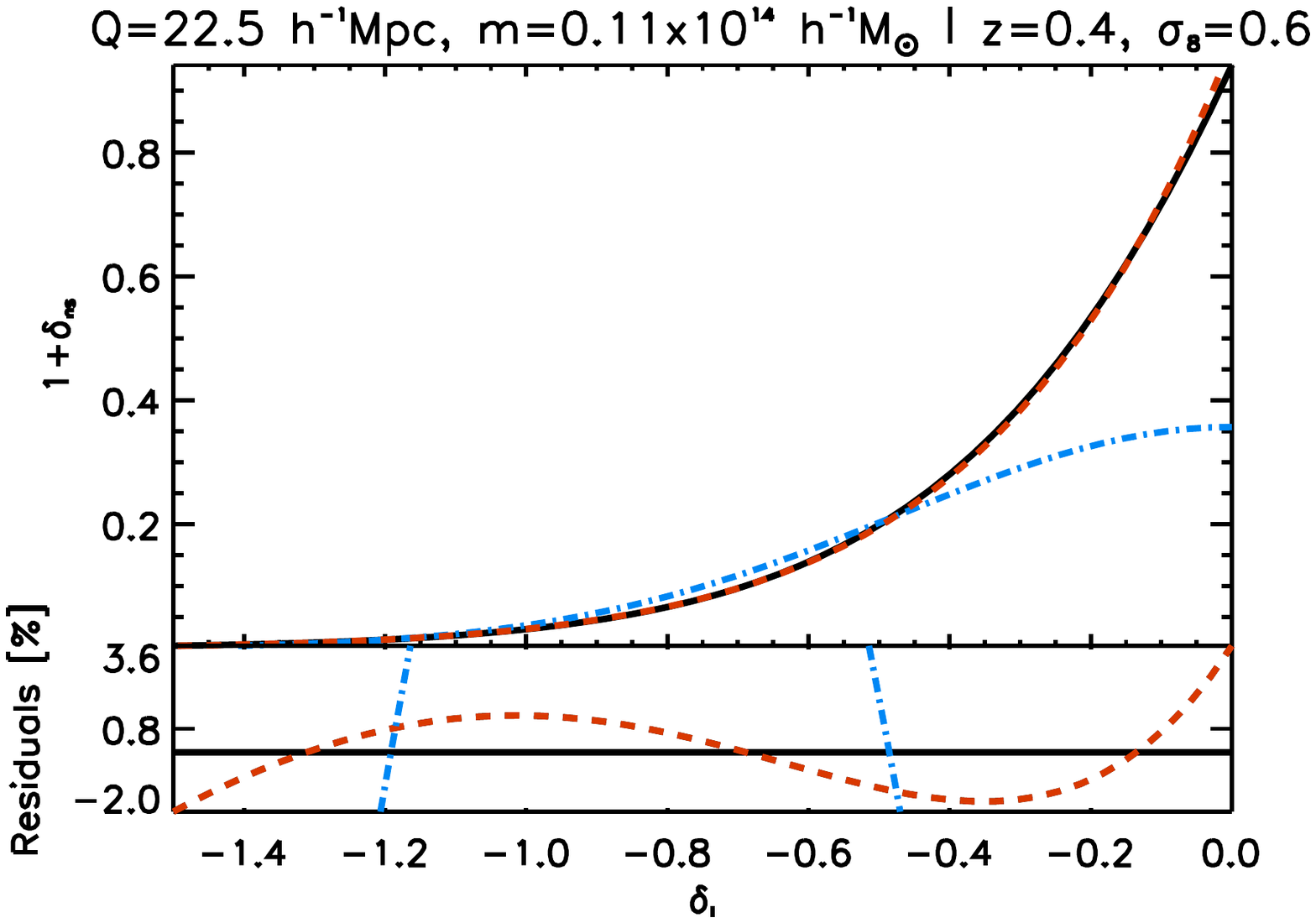} \\
\includegraphics[trim = 30mm 120mm 10mm 20mm, scale=0.45]{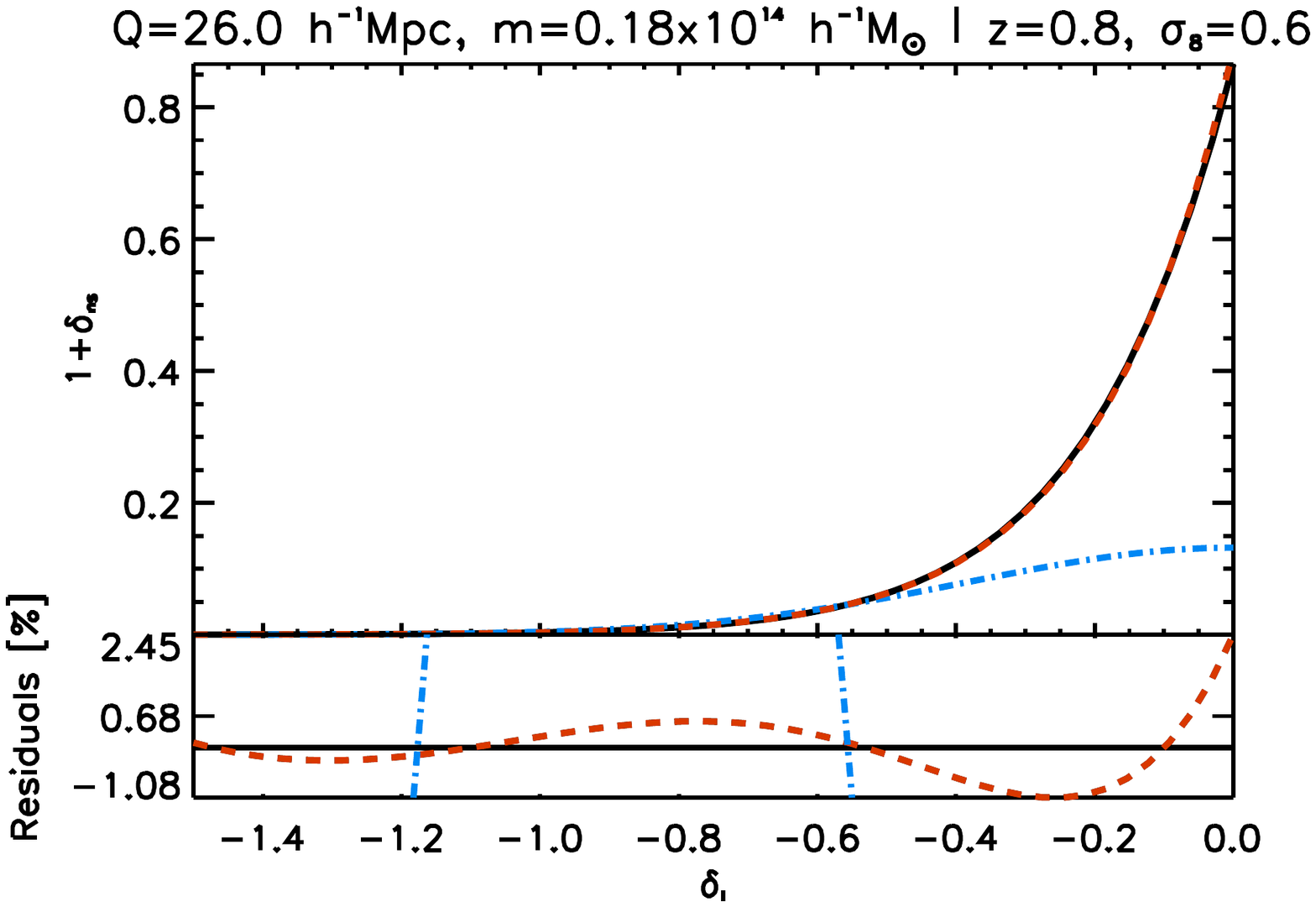} \quad
\includegraphics[trim = 10mm 120mm 30mm 20mm, scale=0.45]{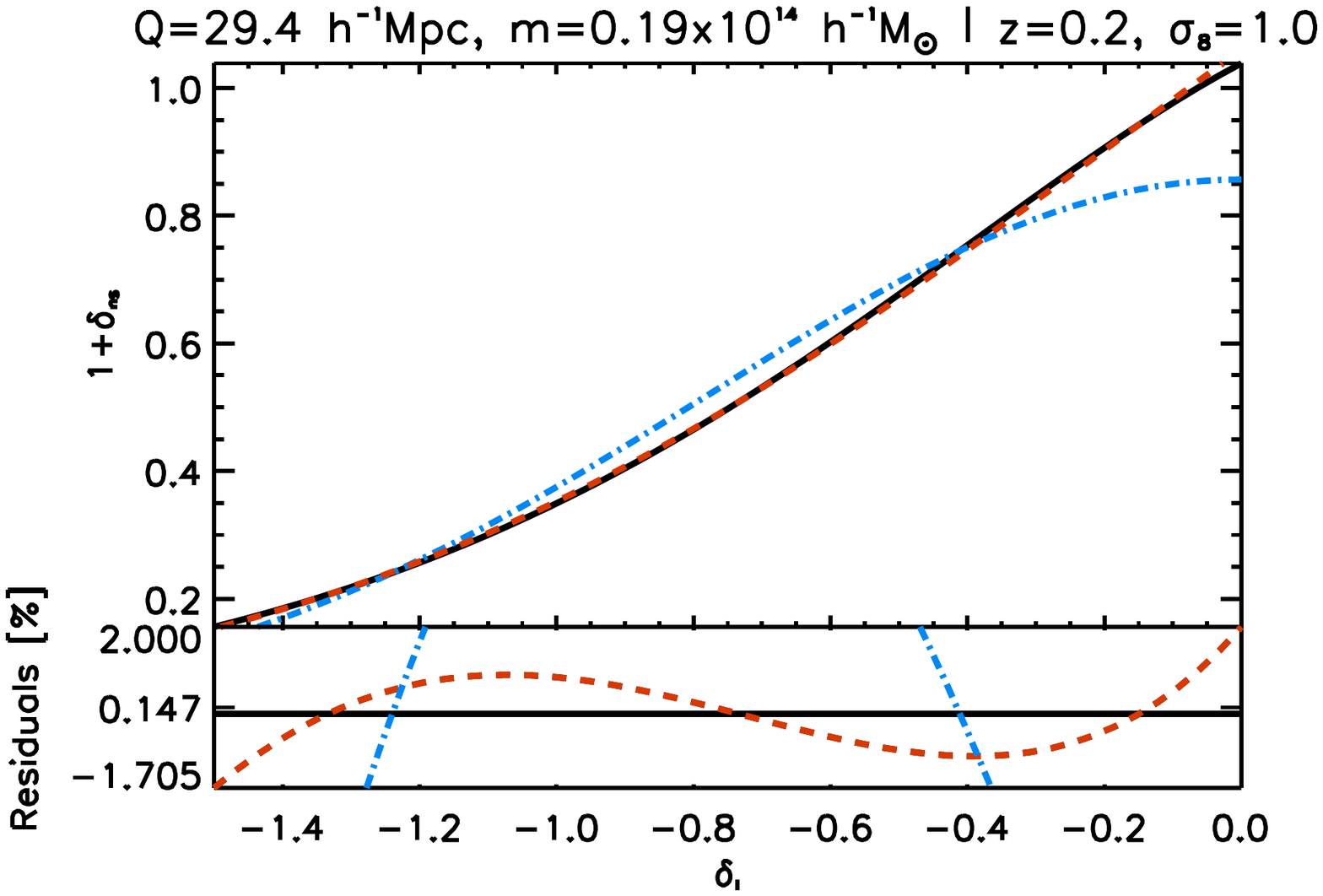} \\
\caption{Bias between the integrated CMF and UMF, as defined in eq.~(\ref{eq:bias}), plotted as a function of the linear density contrast. We compare the prediction from the two-parameter fit~(\ref{eq:fitRBP}) and the three-parameter fit~(\ref{eq:fit}), for different values of condition radius, mass, redshift and $\sigma_8$ (the case is made explicit in the title). Residuals are shown below each plot, in units $[\%]$. Solid line: theoretical bias computed with~(\ref{eq:bias}); dashed line: three-parameter fit; dotted line: two-parameter fit. It is clear how the introduction of the additional parameter $c$ improves the quality of the fit, particularly when $\deltal$ gets closer to zero. Residuals for the three-parameter fit are always below a few percent, but can grow very large with the two-parameter fit.}
\label{fig:fits}
\end{figure*}

In those works the probability of finding a void of a given radius requires the computation of the quantity:
\begin{equation}
\label{eq:totalbias}
1 + \delta_N(m,z;Q,\deltal) = [1+\deltam(\deltal)][1+\delta_{\text{ns}}(m,z;Q,\deltal)],
\end{equation}
where we made the dependence on the objects mass, on redshift and on the void radius explicit. We can call the first factor a `systematic bias', which is due to the different expansion rate of the void compared to the background. The second factor in eq.~(\ref{eq:totalbias}) is the `statistical bias' (or simply the bias), accounting for the clustering of the protohaloes within the condition before it expands in comoving coordinates. This is usually referred in the literature as the (matter-to-)halo bias function \citep{SvdW04, Furlanetto06, Paranjape12, Jennings13}. Again in \citet{PBP06}, it is shown that the relation between the bias and the CMF is given by:
\begin{equation}
\label{eq:bias}
1+\delta_{\text{ns}}(m,z;Q,\deltal) = \dfrac{1}{N_{\rm u}(m,z)} \left[ \frac{3}{Q^3}\int_0^Q \,\text{d}q\,q^2\, N_{\rm c}(m,z;q,Q,\deltal) \right].
\end{equation}
Here $N_{\rm u}(m)$ is simply the cumulative mass function for collapsed objects of mass greater than $m$, in the unconditional case:
\begin{equation}
\label{eq:intumf}
N_{\rm u}(m,z) = \int_m^{+\infty}\,\text{d}m'\, \frac{\text{d}n_{\rm u}}{\text{d}m'}(m',z),
\end{equation}
and $N_{\rm c}(m;q,Q,\deltal)$ is the same for the conditional mass function:
\begin{equation}
\label{eq:intcmf}
N_{\rm c}(m,z;q,Q,\deltal) = \int_m^{M}\,\text{d}m'\,\frac{\text{d}n_{\rm c}}{\text{d}m'}(m',z;q,Q,\deltal).
\end{equation}
In this case the mass range over which the mass function has to be integrated is limited by the mass of the condition; the upper integration extrema should therefore be proper fraction $M$ of the condition mass $m(Q)$. In our computations we always used $M = m(Q)/30$. Notice that the mass function we use in eqs.~(\ref{eq:bias}) and~(\ref{eq:intcmf}) is the locally rescaled CMF with normalization parameter $\alpha=1.25$. Indeed, the voids that are best suited for constraining cosmological models are the largest ones, and according to the discussion in Section~\ref{subsec:results} this is the best CMF recipe in the case of underdensities and large conditions.

Eq.~(\ref{eq:bias}) is the ratio of the total density of collapsed objects above a given mass in the conditioned environment with respect to the unconditional one. It contains all the information on the conditional mass function, and is the starting point for the study of void statistics. Given all the dependences that show up in eq.~(\ref{eq:bias}), the full numerical computation of the bias is very time-consuming. Therefore, efforts have been made to find a numerical fit for the bias, which allows us to avoid heavy numerical computations. In RBP08 a fitting formula in the form:
\begin{equation}
\label{eq:fitRBP}
1 + \delta_{\text{ns}}(m,z;Q,\deltal) = A(m,z,Q)\,e^{-b(m,z,Q)\deltal^2},
\end{equation}
is discussed, and a procedure for computing the fitting parameters $A$ and $b$ is provided. The fit~(\ref{eq:fitRBP}) is a good approximation for the bias at the scales discussed in RBP08 (around $Q=9\,h^{-1}\,\text{Mpc}$) and for strongly underdense regions. However, as already stated, when dealing with void statistics the largest voids are the ones that best constrain cosmological parameters, on scales around $Q=20\,h^{-1}\,\text{Mpc}$. The fit~(\ref{eq:fitRBP}) cannot be applied at those scales, nor when the density contrast approaches zero. In this work, we consider an extension of the fitting formula~(\ref{eq:fitRBP}), which can be applied to a broader range of condition sizes and overdensities; the function we propose has the form:
\begin{equation}
\label{eq:fit}
1 + \delta_{\text{ns}}(m,z;Q,\deltal) = A(m,z,Q)\,e^{-b(m,z,Q)\deltal^2+c(m,z,Q)\deltal},
\end{equation}
which differs from~(\ref{eq:fitRBP}) in the addition of the extra parameter $c$. 

We wanted to check the goodness of this fit in reproducing the bias. To this aim, we implemented the full computation given in eq.~(\ref{eq:bias}), using the reference cosmology from table~\ref{tab:cosmology}, for different values of the condition radius in the range $Q\in[18,28]\,h^{-1}\,\text{Mpc}$. The linear density contrasts we considered span the range $[-1.5,0.0]$. We also considered the dependence on redshift, which we moved in the range $[0,1]$, and on the parameter $\sigma_8$, which we allowed to vary in $[0.6,1.0]$. According to the results described in Section~\ref{subsec:results}, since we are considering underdense regions, we computed the UMF employing the normalized local rescaling~(\ref{eq:locrescnorm}), with normalization parameter $\alpha=1.25$. For every case, we fitted both functions~(\ref{eq:fitRBP}) and~(\ref{eq:fit}) against the bias, as a function of $\deltal$. We show some examples in Figure~\ref{fig:fits}; the use of the additional parameter clearly improves the goodness of the fit, which is capable of reproducing the bias with a relative error in $\sim$ 1--2\,\%. This result is useful for computing the halo abundance as a function of mass in underdense regions, without implementing the full computation of the local CMF rescaling: by computing the bias with~(\ref{eq:fit}) and the integrated UMF with~(\ref{eq:intumf}), their product already yields the integrated CMF, which can be derived to obtain the proper conditional mass function.

The fitting parameters $A$, $b$ and $c$ inherit the dependence on redshift, on the condition size and on the object mass; in addition, they are dependent on the underlying cosmology. Our goal was to give a fitting formula which is capable of reproducing the bias computed with any chosen cosmology, and any scale and redshift. To this aim, we further fitted the dependence of the parameters on these quantities; the full procedure is described in Appendix~\ref{sec:fitparams}. We first provide fitting formulae for computing the parameters $A$, $b$ and $c$ for a reference cosmology, for any choice of the variables $m$ and $Q$. These parameters can then be scaled to a different cosmology. Since the dependences on redshift and on $\sigma_8$ are correlated via the linear growth factor, we first fitted jointly the dependence on these two variables. This way, once the values of $z$, $\sigma_8$, $Q$ and $m$ are chosen, it is possible to compute the fit to the bias using the procedure described in Appendix~\ref{sec:fitparams} and using the common values from the tables presented there; the final accuracy at reproducing the bias is always below $\sim 8.5\,\%$, but it most cases is around a few percents. We attempted also at allowing the cosmological parameters $h$ and $\Omega_{\rm m}$ to vary, including their effect in a rescaling of the mass definition; in this case, however, the quality of the fit degrades and the relative error can grow as large as $40\,\%$ in the worst cases.

\section{Conclusions}
\label{sec:conclusions}

We have compared the predicted CMF of dark matter halos from two theoretical
prescriptions against numerical N-body simulations: the standard rescaling of
the UMF, and a rescaling locally defined inside the condition (RBP08). Both
prescriptions use as a reference the T08 parametrization for the UMF.
Drawing from the results of RBP08, we reminded that the locally rescaled mass
function requires the introduction of an additional parameter $\alpha$ in order
to satisfy the normalization condition, i.e., the integral of the CMF over the
weighted density contrast should yield the initial UMF. The problem of
normalization thus turns into the problem of determining a proper value for
$\alpha$ that satisfies the normalization condition over the whole mass range in
which the mass function is to be computed. We repeated the analysis described in
RBP08, and found that the standard rescaling also requires the normalization to
be implemented explicitly via the parameter $\alpha$. For the first time, we
obtained the value of $\alpha$ needed to normalize the T08 mass function, and
studied its dependence on the size of the condition. We found that this
parameter is well constrained by small regions ($Q \sim 8\,h^{-1}\,\text{Mpc}$),
yielding the estimate $\alpha=1.25$ for both the standard and the local
rescalings; large conditions ($Q \sim 25\,h^{-1}\,\text{Mpc}$), instead, are
much less effective in constraining its value, leaving $\alpha$ undetermined.

In order to test the validity of these CMFs in predicting the abundance of
collapsed objects inside conditions, we tested the results from the theoretical
rescalings against halo counts extracted from numerical simulations. We employed
two different dark matter-only simulations, one at relatively small scales
($250\,h^{-1}\,\text{Mpc}$ box size), the other at larger scales
($1500\,h^{-1}\,\text{Mpc}$ box size). Jointly, the two simulations allowed us
to explore the mass range $[10^{10.5},10^{15.5}]\,h^{-1}\,\text{Mpc}$. The
correspondent halo catalogues were obtained using a FoF algorithm with $b=0.2$,
which corresponds to the $\Delta=200$ spherical overdensity mass function
definition used in T08. As a preliminary check, we compared the halo densities
as a function of mass, extracted from both simulations, with the prediction
computed using the T08 UMF. We found that the simulated counts reproduce the
mass dependence of the theoretical halo abundances but are affected by
systematic effects at the level of $\sim 10\,\%$ relative error. These offsets
are due to the use of a single simulation for each box and have been accounted
for in our analyses.

We then placed a large number of spheres over the simulations, with random
positions and radii, in order to simulate the conditioning regions. We grouped
the spheres in bins of size and density contrast, and obtained the correspondent
conditioned halo density as a function of mass. These distributions were then
compared with the predictions from the different CMF recipes. For every radius
bin, we joined together all spheres with common size and all possible density
contrasts, in order to test a new normalization condition. This is an important
test since by construction we are using the exact probability distribution of
$\deltal$ when averaging over different spheres. We found that the new
normalization condition is satisfied within a few percent accuracy by the
abundances from simulations, when compared to the UMF from the complete halo
catalogue. The theoretical recipes for computing the CMF also achieve that
precision in the normalization when compared to the T08 UMF prediction,
provided the normalization is explicitly applied with the parameter
$\alpha=1.25$. Normalization is no longer achieved in the high mass end, where
errors from the simulation grow large and the formalism for computing the CMF
begins to fail for getting too close to the condition mass.

The following test was checking the CMF prediction for the halo abundance inside
regions with a definite density contrast. We found that the explicitly
normalized CMF with $\alpha=1.25$ is the most accurate recipe in reproducing
simulated halo counts inside underdense regions, both at small and large scales;
the CMF can be computed either with the standard or the local rescalings, both
providing very similar results, although the latter provides a slight better fit. 
 In the case of overdense regions, the use of the same normalization parameter results in an 
improvement of the agreement with simulations at large scales, but in a slight overcorrection 
at small scales. Apart from this case, the use of the locally rescaled CMF normalised 
with $\alpha=1.25$ is the recipe that best works in most cases.

Finally, we presented an analytical fit for computing the CMF in underdense
regions, in terms of the bias with respect to the UMF. By multiplying the
fitting function times the integrated UMF, the integrated CMF can be obtained,
avoiding the time-consuming computation required to implement the normalized
local rescaling in underdense regions. In RBP08, a possible analytical expression
for this fitting function was proposed, which could be applied to extremely
underdense regions (voids); we checked that such a fit is actually inadequate at
reproducing the bias dependence on the density contrast when $\deltal$ gets
closer to zero. We proposed a different fitting function by adding an additional
parameter, that clearly improves the agreement with the CMF bias, yielding
final errors below $\sim 2\,\%$. The dependence of the fitting parameters on the
condition size, the halo mass and the cosmology has also been fitted; it is
possible to use the same fitting parameters for computing the bias with
different values of the mass, the condition radius, the redshift and $\sigma_8$,
with a relative error below $\sim 8\,\%$. It is also possible to apply the same
fitting parameters moving $h$ and $\Omega_{\rm m}$ from their reference values,
computing the bias within a $40\,\%$ error. We present the whole set of fitting
parameters in Appendix~\ref{sec:fitparams}.

\section*{Acknowledgements}
\label{sec:acknowledgements}
JARM acknowledges financial support from the Spanish Ministry of Economy and
Competitiveness (MINECO) under the 2011 Severo Ochoa Program MINECO
SEV-2011-0187. This work has been partially funded by the Spanish Ministry of
Economy and Competitiveness (MINECO) under the projects ESP2013-48362-C2-1-P and
AYA2014-60438-P.

CDV acknowledges financial support from the Spanish Ministry of Economy and
Competitiveness (MINECO) under the 2011 and 2015 Severo Ochoa Programs
SEV-2011-0187 and SEV-2015-0548, and grants AYA2013-46886 and AYA2014-58308.

Simulations have been performed on the Hydra cluster at the Max Planck
Rechenzentrum Garching (RZG) and the COSMA-4 supercomputing facility of the
Institute for Computational Cosmology (ICC) of Durham.

This work used the DiRAC Data Centric system at Durham University, operated by
the Institute for Computational Cosmology on behalf of the STFC DiRAC HPC
Facility (www.dirac.ac.uk). This equipment was funded by BIS National
E-infrastructure capital grant ST/K00042X/1, STFC capital grant ST/H008519/1,
and STFC DiRAC Operations grant ST/K003267/1 and Durham University. DiRAC is
part of the National E-Infrastructure.




\bibliographystyle{mnras}
\bibliography{manuscript} 


\appendix

\section{A derivation of the $\alpha$ parameter based on the analytical normalization condition for the CMF}
\label{sec:normpar}

Following RBP08, we can use the following normalization condition
\begin{equation}
\label{eq:normcond}
\frac{\text{d}n_{\rm u}}{\text{d}m}(m) = \int_{-\infty}^{X}  \text{d}\deltal\; \frac{\text{d}n_{\rm c}}{\text{d}m}(m;Q,\deltal)\;G(\deltal;\sigma_1), \qquad m < m(Q),
\end{equation}
to derive the optimal value of the $\alpha$ parameter. In this equation,
$G(\deltal;\sigma_1)$ stands for a Gaussian with zero mean and variance
$\sigma_1$. Note that by construction, the expression must be evaluated at
masses no larger than the condition mass $m(Q)$. The upper integration limit $X$
is equal to the critical density $\deltac$ for mass functions based on the
physics of spherical collapse \citep{PS74}, and to the value of the moving
barrier shape when the ellipsoidal collapse is allowed \citep{ST02}.

We note that this normalization condition has been previously discussed in the
literature, mainly in connection to the PS mass function.  For example, several
authors \citep{ShethLemson99, Musso12, Neyrinck14} have shown that for the
(unphysical) case of an unbound upper limit of the integral (i.e. $X=+\infty$),
the normalization is satisfied for the PS mass function. Therefore, when the
integration is truncated to the corresponding upper limit, a modification of the
normalization is required in order to verify the equation.

In RBP08, this condition was applied to different prescriptions for the unconditional mass function. The method proposed in RBP08 in order to solve the normalization problem is the introduction of a normalization parameter $\alpha$ in the rescaling for the critical density:  
\begin{equation}
\label{eq:locresc2}
\deltac \rightarrow \deltac' = \deltac\left[ 1 - D(q,Q_1,Q_2) \frac{\delta_1}{\alpha} \right].
\end{equation}
For the T08 parametrization, in which $\deltac$ does not appear explicitly, the new rescaling is to be applied in the form of eq.~(\ref{eq:locrescnorm}).

\begin{figure*}
\includegraphics[trim = 60mm 130mm 30mm 20mm, scale=0.37]{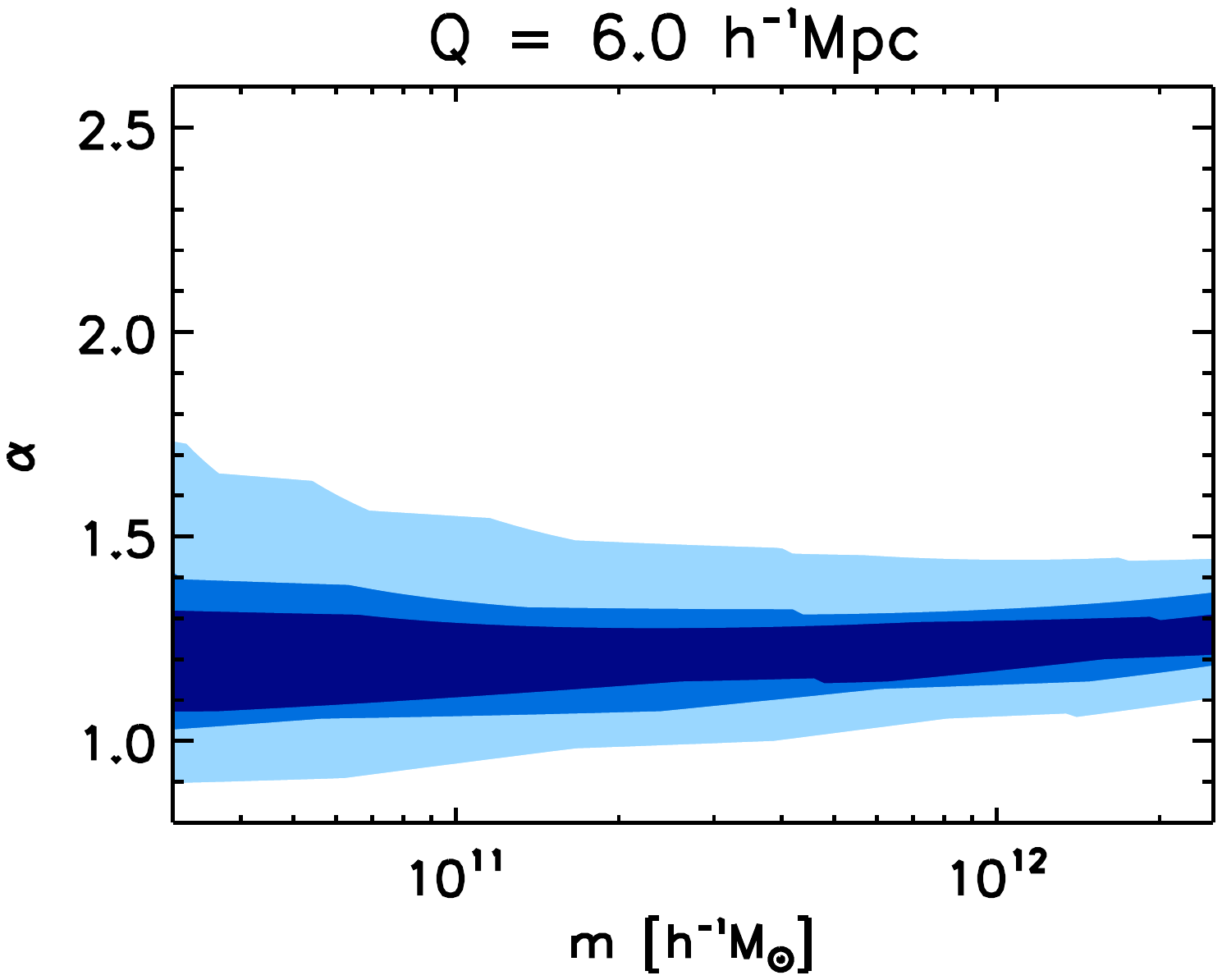} \quad
\includegraphics[trim = 37mm 130mm 30mm 20mm, scale=0.37]{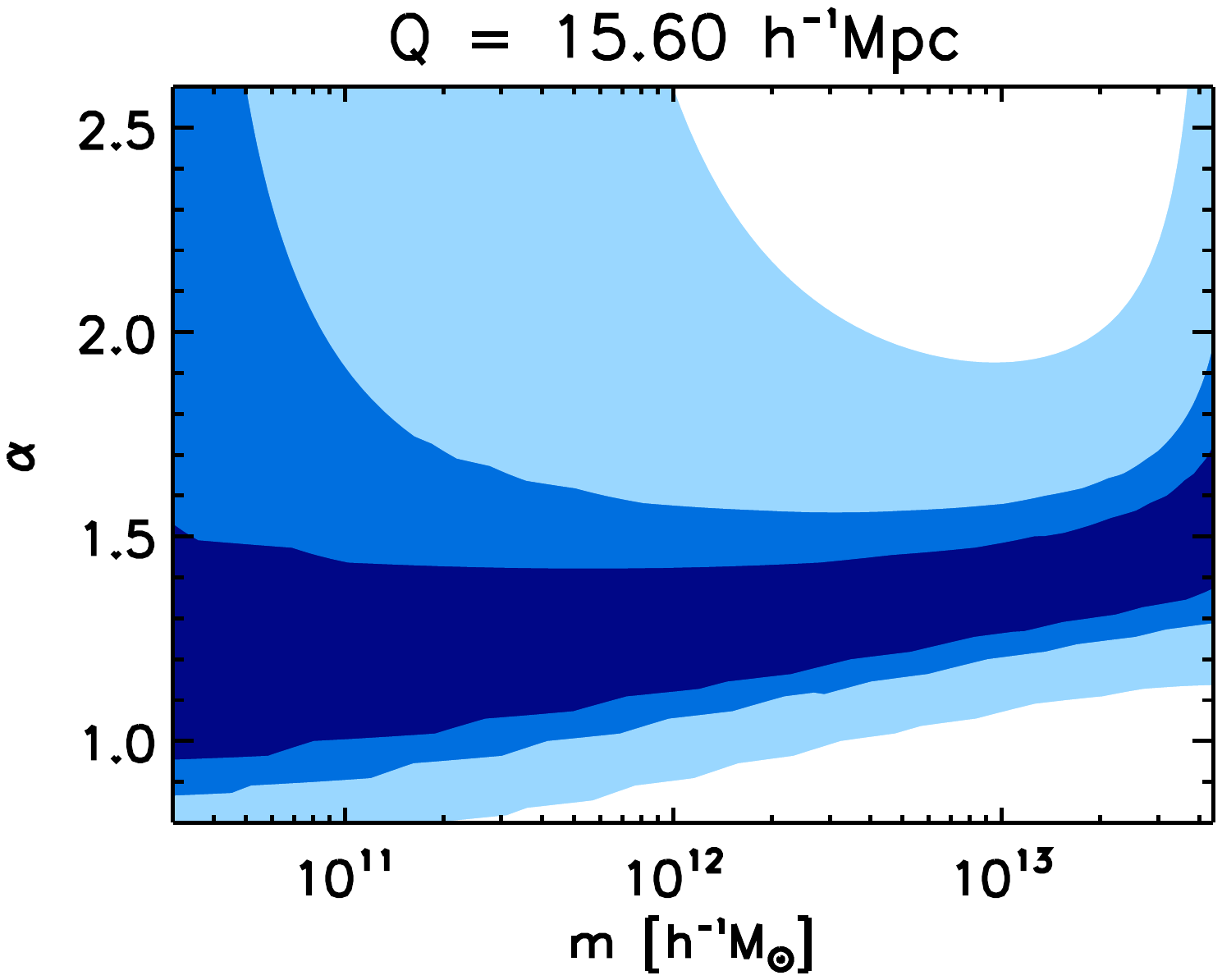} \quad
\includegraphics[trim = 37mm 130mm 30mm 20mm, scale=0.37]{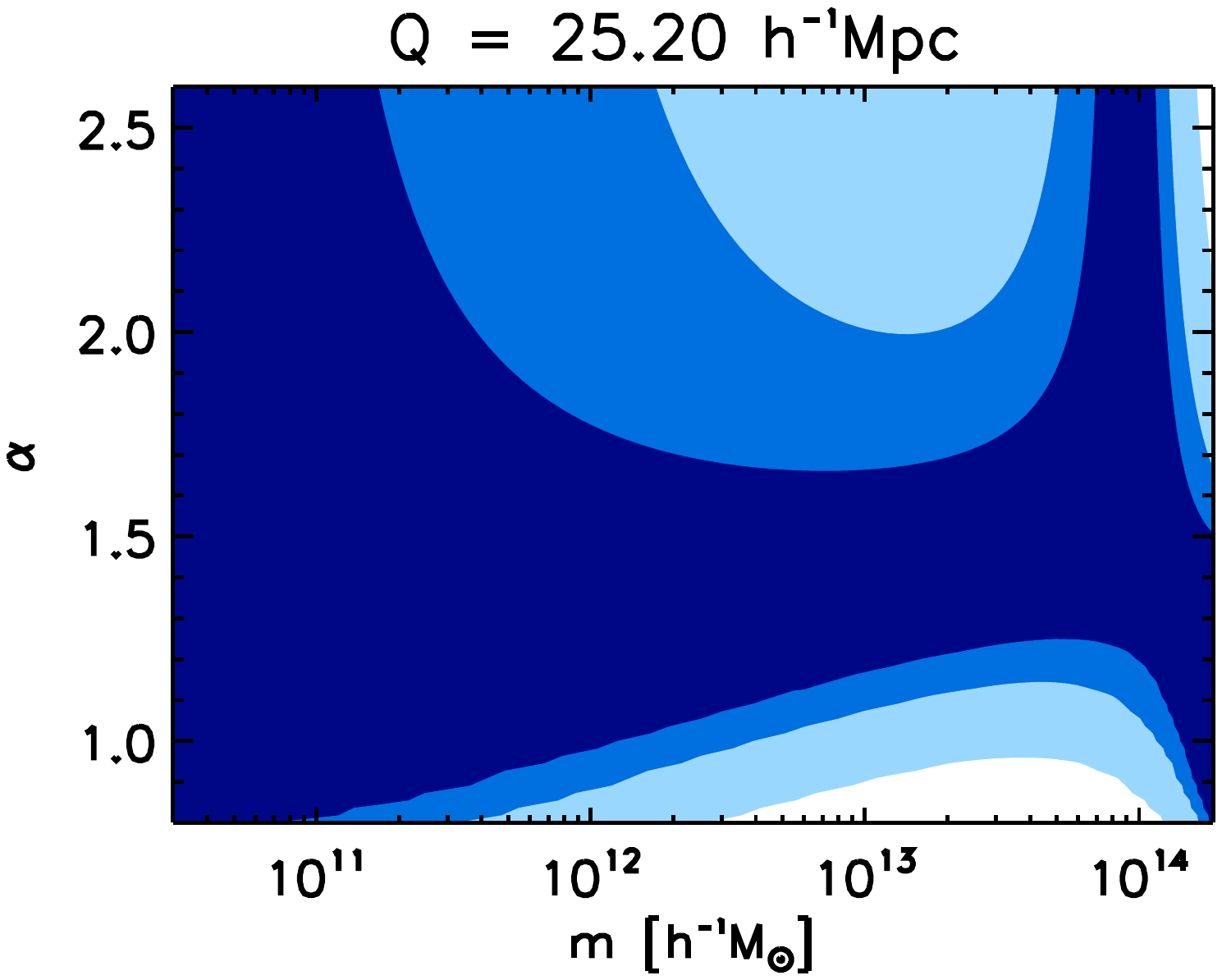} 
\caption{Regions showing the allowed values for the parameter $\alpha$, as a function of the mass, in order to have the normalization condition satisfied with a $10\,\%$, $5\,\%$ and $3\,\%$ accuracy, respectively. The conditional mass function is computed using the T08 parametrization with the local rescaling. The plot shows how the contours evolve when the condition size increases: larger conditions are less effective in determining a unique value for the normalization parameter, ultimately leaving $\alpha$ unconstrained. }
\label{fig:normcontours}
\end{figure*}

\begin{figure*}
\includegraphics[trim = 60mm 130mm 30mm 20mm, scale=0.37]{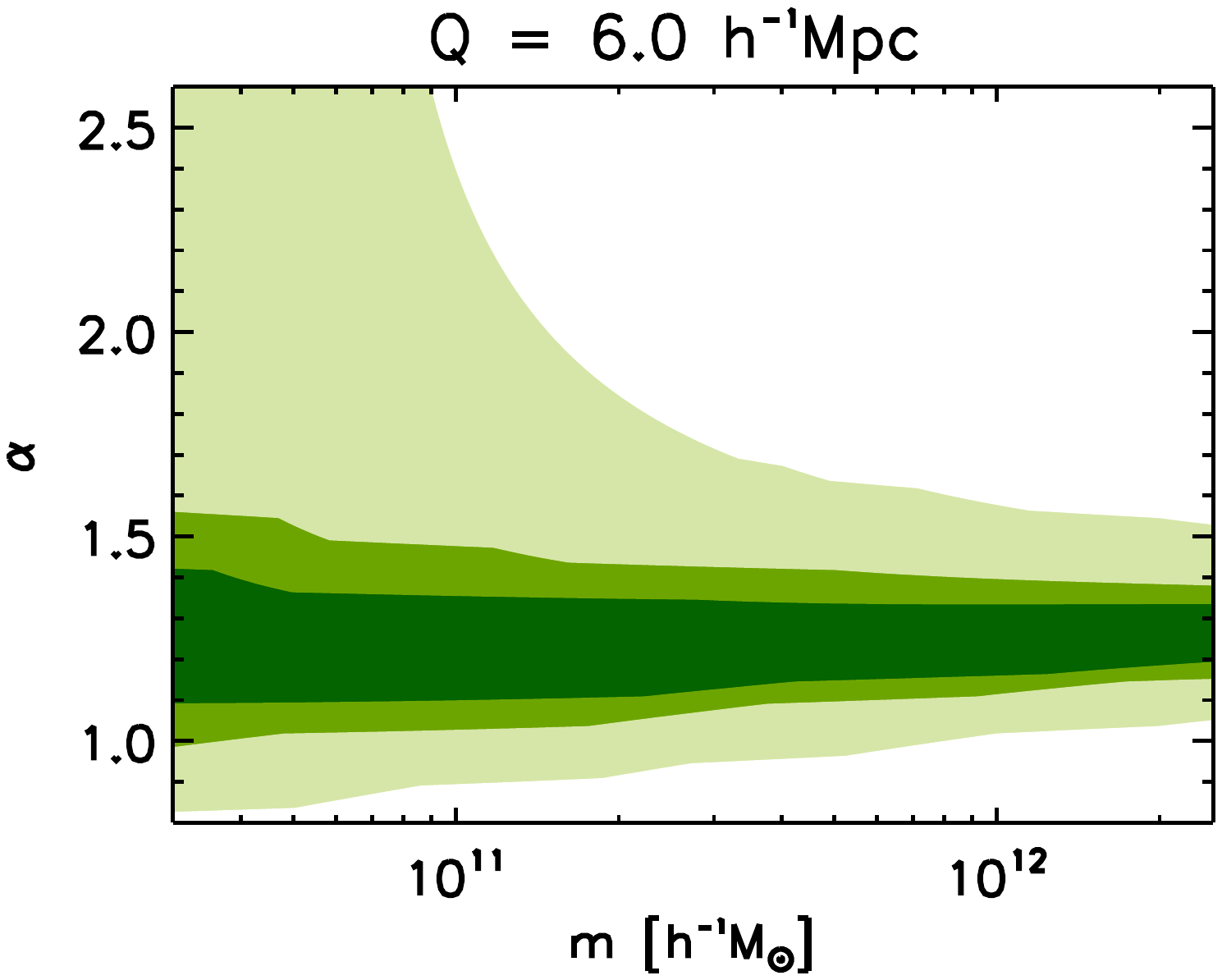} \quad
\includegraphics[trim = 37mm 130mm 30mm 20mm, scale=0.37]{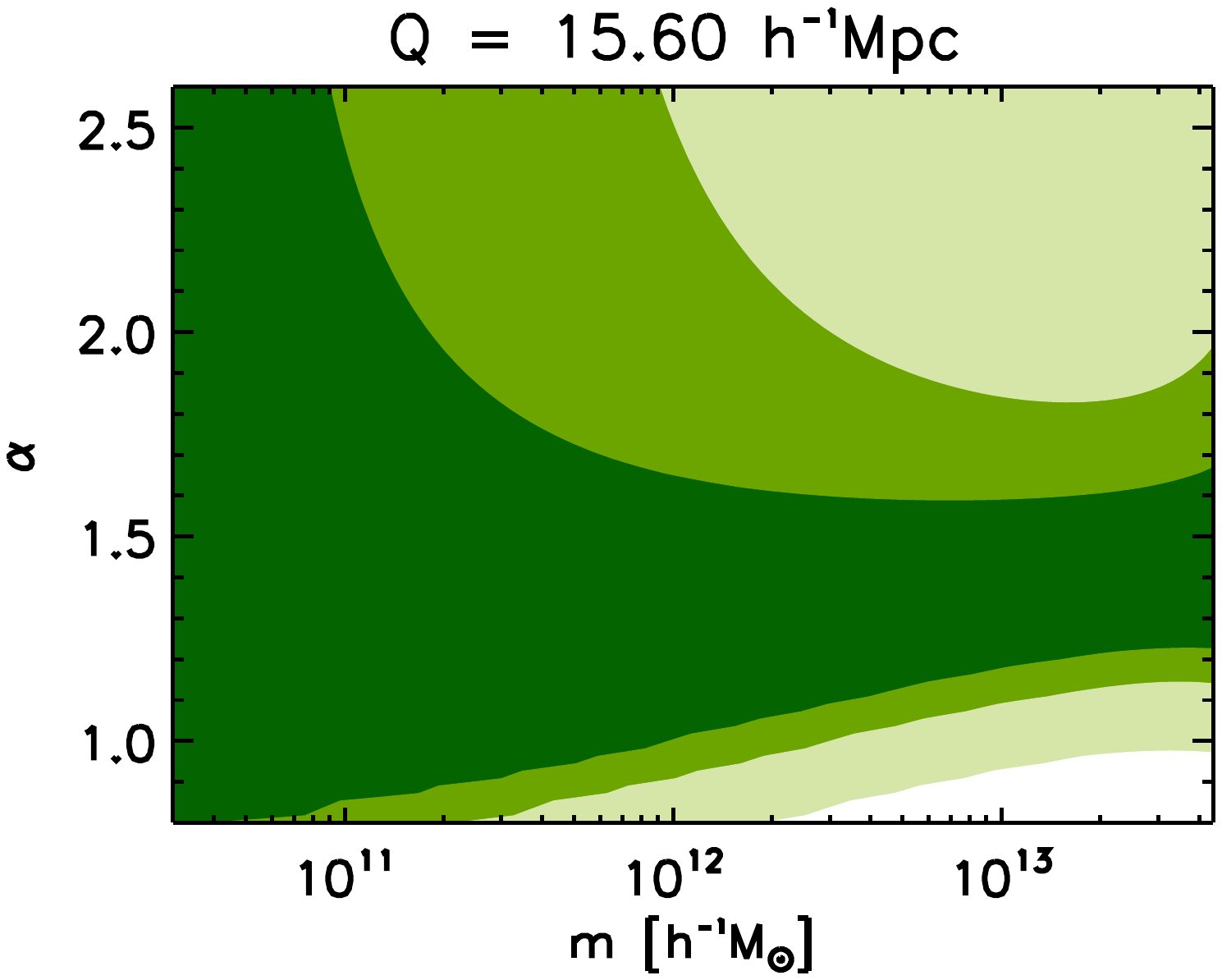} \quad
\includegraphics[trim = 37mm 130mm 30mm 20mm, scale=0.37]{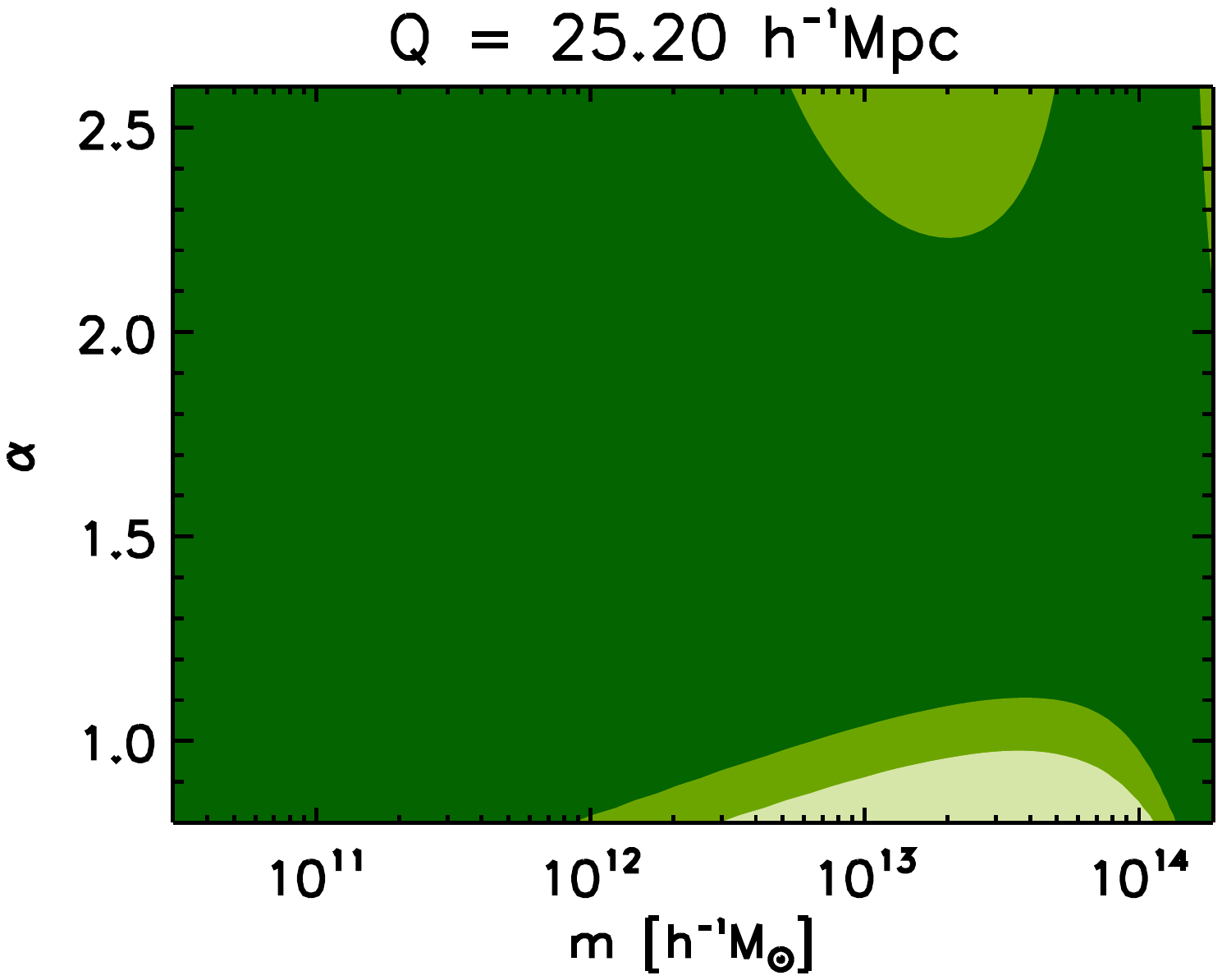} 
\caption{Same as Figure~\ref{fig:normcontours}, but computing the mass function with the standard rescaling.}
\label{fig:normcontours_std}
\end{figure*}
Our goal was to find a proper estimate for the normalization parameter and to explore its dependence on $Q$, both for the standard and the local rescaling. We followed the same method described in RBP08. In order to determine the precision with which the normalization is satisfied by the CMF we computed the ratio: 
\begin{equation}
R(m,Q) \equiv \left( \frac{\text{d}n_{\rm u}}{\text{d}m}(m) \right)^{-1} \int_{-\infty}^{X}  \text{d}\deltal\, \frac{\text{d}n_{\rm c}}{\text{d}m}(m;Q,\deltal)\,G(\deltal;\sigma_1),
\label{eq:normal}
\end{equation}
that quantifies the error in the normalization (how much the integrated conditional mass function differs from the expected unconditional one). In~(\ref{eq:normal}) the CMF is to be computed with the new rescaling eq.~(\ref{eq:locresc2}) or~(\ref{eq:locrescnorm}), depending on the chosen parametrization for the initial UMF. By varying the parameter $\alpha$ we end up with a three-variable function $R(m,Q,\alpha)$. For different values of the condition radius we computed the quantity $(1-R(m,Q,\alpha))\times100$ and considered the regions in the $(m,\alpha)$ plane where it is at most $10\,\%$, $5\,\%$ and $3\,\%$. The result is plotted in Fig.~\ref{fig:normcontours} for the case of local rescaling. We see that it is possible to choose a unique $\alpha$ value for all masses; in the case of $Q = 6\,h^{-1}\,\text{Mpc}$, the value $\alpha=1.25$ satisfies normalization with a $3\,\%$ accuracy. This is the value we chose for the comparison of the locally rescaled CMF with simulations. However, the precision contours largely widens when growing the condition, meaning that larger conditions are not effective in constraining the normalization problem. In fact, we see that for $Q \gtrsim 20\,h^{-1}\,\text{Mpc}$ the usual value $\deltac=1.686$ is allowed for the whole mass range. Setting  $\alpha=\deltac$ is equivalent to using the original form of the local rescaling, eqs.~(\ref{eq:locresc}) and~(\ref{eq:locrescsigma}). We repeated the same analysis also for the standard rescaling, and present the normalization contours in Figure~\ref{fig:normcontours_std}; we found that the contours evolve in a similar way, constraining the value $\alpha=1.25$ in the smaller conditions.

The conclusion, as anticipated in Section~\ref{subsec:normalization}, is that the normalization condition is effective in constraining the value of $\alpha$ in the case of small conditions, with a radius comparable to the ones used in RBP08, but is not enough when dealing with sufficiently large conditions ($Q \gtrsim 20\,h^{-1}\,\text{Mpc}$).

\section{Fitting formulae for the CMF (or halo bias function) in underdense regions}
\label{sec:fitparams}
We present in the following the full procedure we employed to parametrize the residual dependence of the bias fitting parameters, and the algorithm for computing them for any choice of cosmology, mass and condition size.

\subsection{Description}
\begin{table}
\centering
\caption{Complete list of fitting terms for computing the parameters $A$, $b$ and $c$ in the reference cosmology, using eq.~(\ref{eq:refcosmo}). Index $i$ runs over lines, index $j$ runs over columns.}
\label{tab:table1}
\begin{tabular}{c|cccc}
\hline
$a_{ij}$& 0 & 1 & 2 & 3 \\ 
\hline
 0 &       -0.91  &       -0.15  &        0.69  &     -0.13 \\ 
 1 &        1.08  &        0.42  &       -0.59  &      0.10 \\ 
 2 &       -0.12  &       -0.20  &        0.17  &     -0.03 \\ 
 3 &       -0.00  &        0.02  &       -0.02  &      0.00 \\ 
\hline
\end{tabular}

\begin{tabular}{c|cccc}
\hline
$b_{ij}$& 0 & 1 & 2 & 3 \\ 
\hline
 0 &        5.38  &       -2.75  &        0.23  &      0.03 \\ 
 1 &       -4.69  &        2.88  &       -0.42  &     -0.00 \\ 
 2 &        1.48  &       -0.99  &        0.18  &     -0.01 \\ 
 3 &       -0.15  &        0.11  &       -0.02  &      0.00 \\ 
\hline
\end{tabular}

\begin{tabular}{c|cccc}
\hline
$c_{ij}$& 0 & 1 & 2 & 3 \\ 
\hline
 0 &        0.89  &        1.70  &       -0.85  &      0.09 \\ 
 1 &       -0.25  &       -1.84  &        0.85  &     -0.09 \\ 
 2 &        0.09  &        0.55  &       -0.26  &      0.03 \\ 
 3 &       -0.02  &       -0.04  &        0.02  &     -0.00 \\ 
\hline
\end{tabular}

\end{table}

The fitting parameters $A$, $b$ and $c$ defined in~(\ref{eq:bias}) inherit from the bias the dependence on the object mass $m$, the condition size $Q$, the redshift $z$ and the cosmological parameters. It is convenient to split the dependence on the variables $Q$ and $m$ from the cosmological variables (including $z$). 

For a given cosmology the dependence on the condition can be fitted by a polynomial function in powers of $Q$ and $\ln{(m)}$. More precisely, the fitting functions we employed are: 
\begin{align}
\label{eq:refcosmo}
A(m,Q) = \sum_{i=0}^3 \left(\sum_{j=0}^3 a_{ij}Q'^j \right)\ln{(m')}^i, \nonumber \\
b(m,Q) = \sum_{i=0}^3 \left(\sum_{j=0}^3 b_{ij}Q'^j \right)\ln{(m')}^i, \nonumber \\
c(m,Q) = \sum_{i=0}^3 \left(\sum_{j=0}^3 c_{ij}Q'^j \right)\ln{(m')}^i,  
\end{align}
where $m'=m/3.51\times10^{11}\,h^{-1}\,\text{M}_{\odot}$ and $Q'= Q/8\,h^{-1}\,\text{Mpc}$. Polynomials are third degree in both radius and mass, resulting in 16 $a_{ij}$ fitting terms for $A$, 16 $b_{ij}$ fitting terms for $b$ and 16 $c_{ij}$ fitting terms for $c$. For our reference cosmology, these terms are listed in table~\ref{tab:table1}. They were obtained allowing the Lagrangian radius to vary in the range $[18.0,34.0]\,h^{-1}\,\text{Mpc}$, and the halo mass in the range $[10^3,6\times10^3]\,h^{-1}\,\text{Mpc}$; the intervals also fix the limits in which the parameters in table~\ref{tab:table1} can be applied. Using these tabulated values and eq.~(\ref{eq:refcosmo}), it is possible to quickly compute the bias with a typical error $\sim$ 1--2\,\%

The dependence on cosmology is better fitted first in terms of the combination $D(z)\sigma_8$, with $D(z)$ the growth factor of linear perturbation at redshift $z$, normalized to unity at $z=0$. Indeed, other combinations of the variables $z$ and $\sigma_8$ produce quite a large scatter in the points, making this 1-d fit impossible. This can be understood since the mass function does not depend separately on $z$ and $\sigma_8$. In order to fit this dependence, we introduced the variable 
\begin{equation}
\mathcal{C} \equiv D(z)\sigma_8/\tilde{\sigma}_8,
\end{equation}
where $\tilde{\sigma}_8$ is the value for the reference cosmology. In our case, our reference choice was $\tilde{z}=0$, $\tilde{\sigma}_8=0.8$, to which corresponds $\mathcal{C} = 1$. We considered redshift values in $[0,1]$ and $\sigma_8$ values in $[0.6,1.0]$. Now, given the parameters $A$, $b$ and $c$ for the reference cosmology, we found that a convenient function to fit the parameters change when moving the value of $z$ or $\sigma_8$ is: 
\begin{align}
\label{eq:cosmofit}
&\ln{\left[\dfrac{A(m,Q;z,\sigma_8)}{A(m,Q;\tilde{z},\tilde{\sigma}_8)}\right]} = \nonumber \\
&= \ln{\mathcal{C}} \left[\xi_{\rm A}(m,Q) + \eta_{\rm A}(m,Q)\mathcal{C} + \zeta_{\rm A}(m,Q)\mathcal{C}^2 + \omega_{\rm A}(m,Q)\mathcal{C}^3\right], \nonumber \\
&\ln{\left[\dfrac{b(m,Q;z,\sigma_8)}{b(m,Q;\tilde{z},\tilde{\sigma}_8)}\right]} = \nonumber \\
&= \ln{\mathcal{C}} \left[\xi_{\rm b}(m,Q) + \eta_{\rm b}(m,Q)\mathcal{C} + \zeta_{\rm b}(m,Q)\mathcal{C}^2 + \omega_{\rm b}(m,Q)\mathcal{C}^3\right], \nonumber \\
&\ln{\left[\dfrac{c(m,Q;z,\sigma_8)}{c(m,Q;\tilde{z},\tilde{\sigma}_8)}\right]} = \nonumber \\
&= \ln{\mathcal{C}} \left[\xi_{\rm c}(m,Q) + \eta_{\rm c}(m,Q)\mathcal{C}^2 + \zeta_{\rm c}(m,Q)\mathcal{C}^2 + \omega_{\rm c}(m,Q)\mathcal{C}^3\right]. \nonumber \\
\end{align}
The fitting function is the product of a logarithm of $\mathcal{C}$ and a polynomial in $\mathcal{C}$ up to the third power. Note that with this definition of the fitting function, when $\mathcal{C}=1$ the reference values for the parameters are recovered exactly. The fit for each parameter now depends on 4 coefficients, $\xi$, $\eta$, $\zeta$ and $\omega$, which in turn still depend on $m$ and $Q$. 

\begin{table}
\centering
\caption{Complete list of fitting terms to transform the $A$ parameter to the chosen values of $z$ and $\sigma_8$, using eqs.~(\ref{eq:lowlev}) and~(\ref{eq:cosmofit}). Index $i$ selects a line, index $j$ selects a column.}
\label{tab:table2}
\begin{tabular}{c|ccccc}
\hline
$\xi^{\rm A}_{ij}$& 0 & 1 & 2 & 3 & 4 \\ 
\hline
 0 &     -601.87  &      515.84  &     -142.68  &       11.74  &      0.28 \\ 
 1 &      480.91  &     -349.49  &       59.17  &        6.42  &     -1.73 \\ 
 2 &     -151.77  &       89.94  &       -0.73  &       -7.22  &      1.04 \\ 
 3 &       24.26  &      -12.77  &       -1.32  &        1.58  &     -0.21 \\ 
 4 &       -1.67  &        0.95  &        0.03  &       -0.09  &      0.01 \\ 
\hline
\end{tabular}
 \qquad
\begin{tabular}{c|ccccc}
\hline
$\eta^{\rm A}_{ij}$& 0 & 1 & 2 & 3 & 4 \\ 
\hline
 0 &     1314.85  &     -996.58  &      184.93  &       13.28  &     -4.48 \\ 
 1 &     -942.80  &      479.62  &       78.01  &      -71.26  &      9.16 \\ 
 2 &      274.40  &      -62.62  &      -92.19  &       41.22  &     -4.65 \\ 
 3 &      -45.12  &        6.77  &       18.44  &       -7.77  &      0.86 \\ 
 4 &        3.51  &       -1.16  &       -0.88  &        0.45  &     -0.05 \\ 
\hline
\end{tabular}
 \\
\begin{tabular}{c|ccccc}
\hline
$\zeta^{\rm A}_{ij}$& 0 & 1 & 2 & 3 & 4 \\ 
\hline
 0 &    -1357.21  &      951.22  &     -109.53  &      -40.10  &      7.36 \\ 
 1 &      957.06  &     -396.99  &     -172.61  &      102.48  &    -12.41 \\ 
 2 &     -280.42  &       32.45  &      127.75  &      -53.09  &      5.89 \\ 
 3 &       47.90  &       -4.19  &      -23.22  &        9.53  &     -1.05 \\ 
 4 &       -3.86  &        1.25  &        1.05  &       -0.53  &      0.06 \\ 
\hline
\end{tabular}
 \qquad
\begin{tabular}{c|ccccc}
\hline
$\omega^{\rm A}_{ij}$& 0 & 1 & 2 & 3 & 4 \\ 
\hline
 0 &      646.58  &     -516.54  &      107.32  &        3.35  &     -2.02 \\ 
 1 &     -485.83  &      297.08  &        3.29  &      -27.64  &      3.98 \\ 
 2 &      149.69  &      -64.78  &      -25.95  &       16.04  &     -1.97 \\ 
 3 &      -24.88  &        9.90  &        5.18  &       -2.94  &      0.36 \\ 
 4 &        1.84  &       -0.96  &       -0.18  &        0.16  &     -0.02 \\ 
\hline
\end{tabular}
\end{table}

\begin{table}
\centering
\caption{Complete list of fitting terms to transform the $b$ parameter to the chosen values of $z$ and $\sigma_8$, using eqs.~(\ref{eq:lowlev}) and~(\ref{eq:cosmofit}). Index $i$ selects a line, index $j$ selects a column.}
\label{tab:table3}
\begin{tabular}{c|ccccc}
\hline
$\xi^{\rm b}_{ij}$& 0 & 1 & 2 & 3 & 4 \\ 
\hline
 0 &    -5098.88  &     3004.96  &      -36.68  &     -248.98  &     37.40 \\ 
 1 &     5948.68  &    -4065.13  &      526.96  &      148.63  &    -29.76 \\ 
 2 &    -2635.03  &     2035.06  &     -435.59  &       -6.77  &      7.41 \\ 
 3 &      517.88  &     -440.23  &      120.16  &       -8.72  &     -0.48 \\ 
 4 &      -37.68  &       34.46  &      -10.82  &        1.24  &     -0.02 \\ 
\hline
\end{tabular}
 \qquad
\begin{tabular}{c|ccccc}
\hline
$\eta^{\rm b}_{ij}$& 0 & 1 & 2 & 3 & 4 \\ 
\hline
 0 &       -7.31  &    15936.67  &   -13941.07  &     4094.47  &   -400.37 \\ 
 1 &    -6188.86  &    -8012.52  &    10187.80  &    -3312.39  &    339.99 \\ 
 2 &     4944.70  &     -371.62  &    -2216.28  &      907.10  &   -101.58 \\ 
 3 &    -1294.53  &      673.68  &       76.99  &      -89.84  &     12.17 \\ 
 4 &      110.78  &      -81.52  &       14.21  &        1.60  &     -0.45 \\ 
\hline
\end{tabular}
 \\
\begin{tabular}{c|ccccc}
\hline
$\zeta^{\rm b}_{ij}$& 0 & 1 & 2 & 3 & 4 \\ 
\hline
 0 &   -16937.67  &     1881.75  &     7188.21  &    -3034.33  &    344.50 \\ 
 1 &    24667.19  &   -11863.01  &    -2414.89  &     2031.29  &   -266.51 \\ 
 2 &   -12435.60  &     8586.89  &    -1080.03  &     -343.33  &     67.33 \\ 
 3 &     2627.79  &    -2160.41  &      532.22  &      -17.38  &     -5.38 \\ 
 4 &     -198.39  &      180.62  &      -55.52  &        5.84  &     -0.04 \\ 
\hline
\end{tabular}
 \qquad
\begin{tabular}{c|ccccc}
\hline
$\omega^{\rm b}_{ij}$& 0 & 1 & 2 & 3 & 4 \\ 
\hline
 0 &     2523.95  &     4337.38  &    -5179.53  &     1679.53  &   -173.02 \\ 
 1 &    -6083.56  &      279.16  &     2962.50  &    -1212.21  &    136.61 \\ 
 2 &     3702.39  &    -1958.01  &     -217.79  &      265.72  &    -36.47 \\ 
 3 &     -858.07  &      643.84  &     -117.70  &      -11.54  &      3.51 \\ 
 4 &       68.11  &      -59.61  &       16.85  &       -1.31  &     -0.06 \\ 
\hline
\end{tabular}
\end{table}

\begin{table}
\centering
\caption{Complete list of fitting terms to transform the $c$ parameter to the chosen values of $z$ and $\sigma_8$, using eqs.~(\ref{eq:lowlev}) and~(\ref{eq:cosmofit}). Index $i$ selects a line, index $j$ selects a column.}
\label{tab:table4}
\begin{tabular}{c|ccccc}
\hline
$\xi^{\rm c}_{ij}$& 0 & 1 & 2 & 3 & 4 \\ 
\hline
 0 &      748.68  &      -72.63  &     -320.20  &      131.82  &    -14.59 \\ 
 1 &    -1142.51  &      618.11  &       48.90  &      -73.50  &      9.99 \\ 
 2 &      574.04  &     -434.57  &       82.90  &        5.72  &     -2.02 \\ 
 3 &     -118.26  &      104.21  &      -29.84  &        2.53  &      0.06 \\ 
 4 &        8.61  &       -8.27  &        2.77  &       -0.36  &      0.01 \\ 
\hline
\end{tabular}
 \qquad
\begin{tabular}{c|ccccc}
\hline
$\eta^{\rm c}_{ij}$& 0 & 1 & 2 & 3 & 4 \\ 
\hline
 0 &    -1221.81  &    -1707.65  &     2094.52  &     -672.53  &     67.91 \\ 
 1 &     2894.40  &     -639.47  &     -924.38  &      420.64  &    -48.04 \\ 
 2 &    -1695.05  &     1094.91  &      -81.54  &      -65.14  &     10.68 \\ 
 3 &      374.61  &     -313.11  &       79.72  &       -3.58  &     -0.63 \\ 
 4 &      -28.23  &       26.57  &       -8.59  &        1.02  &     -0.03 \\ 
\hline
\end{tabular}
 \\
\begin{tabular}{c|ccccc}
\hline
$\zeta^{\rm c}_{ij}$& 0 & 1 & 2 & 3 & 4 \\ 
\hline
 0 &     -484.44  &     4245.93  &    -3390.87  &      958.13  &    -90.91 \\ 
 1 &    -1646.29  &    -1238.83  &     1889.74  &     -634.97  &     65.44 \\ 
 2 &     1381.05  &     -612.86  &     -168.43  &      121.24  &    -15.28 \\ 
 3 &     -343.50  &      263.65  &      -53.79  &       -2.33  &      1.12 \\ 
 4 &       27.28  &      -24.96  &        7.74  &       -0.83  &      0.01 \\ 
\hline
\end{tabular}
 \qquad
\begin{tabular}{c|ccccc}
\hline
$\omega^{\rm c}_{ij}$& 0 & 1 & 2 & 3 & 4 \\ 
\hline
 0 &     -272.51  &     -999.77  &      956.32  &     -286.49  &     27.94 \\ 
 1 &     1016.88  &      -84.11  &     -409.65  &      168.12  &    -18.58 \\ 
 2 &     -646.65  &      415.56  &      -40.68  &      -20.78  &      3.61 \\ 
 3 &      146.42  &     -125.71  &       35.66  &       -2.88  &     -0.09 \\ 
 4 &      -11.07  &       10.74  &       -3.73  &        0.52  &     -0.02 \\ 
\hline
\end{tabular}
\end{table}

The next step is to fit the residual dependence of these coefficients. We use a fitting function similar to the one in eq.~(\ref{eq:refcosmo}), but since the dependence of these coefficients on mass and radius is more complex than it was for $A$, $b$ and $c$, we employed a 4-degree polynomial in both $m$ and $Q$. In formulae:
\begin{align}
\label{eq:lowlev}
\xi(m,Q) = \sum_{i=0}^4 \left(\sum_{j=0}^4 \xi_{ij}Q'^j \right)\ln{(m')}^i, \nonumber \\
\eta(m,Q) = \sum_{i=0}^4 \left(\sum_{j=0}^4 \eta_{ij}Q'^j \right)\ln{(m')}^i,  \nonumber \\
\zeta(m,Q) = \sum_{i=0}^4 \left(\sum_{j=0}^4 \zeta_{ij}Q'^j \right)\ln{(m')}^i, \nonumber \\
\omega(m,Q) = \sum_{i=0}^4 \left(\sum_{j=0}^4 \omega_{ij}Q'^j \right)\ln{(m')}^i.
\end{align}
In this case we have 25 terms $\xi_{i,j}$ for $\xi$, 25 terms $\eta_{i,j}$ for $\eta$ and so on. The whole set must be repeated three times (for $A$, $b$ and $c$), ending up with 300 parameters for this last fit. Tables~\ref{tab:table2},~\ref{tab:table3} and~\ref{tab:table4} list these parameters for $A$, $b$ and $c$ respectively. This set of fitting parameters allows the computation of the bias for any value of $Q$, $m$, $z$ and $\sigma_8$ in the specified ranges, with a final error below $\sim 8\,\%$.

The last step is the dependence of fitting parameters $A$, $b$ and $c$ on the matter density parameter $\Omega_{\rm m}$ and the Hubble parameter $h$. We found that the same parameters obtained for the reference cosmology can still be employed, provided the mass is scaled as:
\begin{equation}
\label{eq:massscale}
 m \rightarrow  m \,\left(\dfrac{\Omega_{\rm m}\,h^4}{\tilde{\Omega}_{\rm m}\,\tilde{h}^4}\right)^{f(h,\Omega_{\rm m})},
\end{equation}
where $\tilde{h}$ and $\tilde{\Omega}_{\rm m}$ are the values for the reference cosmology, and
\begin{equation}
\label{eq:exponent}
f(h,\Omega_{\rm m}) = \sum_{i=0}^4 \left(\sum_{j=0}^4 f_{ij}\Omega_{\rm m}^j \right)h^i.
\end{equation}
The terms $f_{ij}$ are listed in table~\ref{tab:table5}. We computed them allowing $h$ to vary in $[0.65,0.75]$, and $\Omega_{\rm m}$ in $[0.25,0.35]$. With this recipe it is possible to compute the bias using the same fitting parameters listed in tables~\ref{tab:table1} to~\ref{tab:table4}. In this case, the final relative error with respect to the bias computed analytically is usually in the range $\sim$ 10--20\,\%, but in some cases it can grow up to $40\,\%$. 
\begin{table}
\centering
\caption{Complete list of fitting terms for computing the exponent $f(h,\Omega_m)$ using eq.~(\ref{eq:exponent}), entering the mass scaling~(\ref{eq:massscale}). Index $i$ selects a line, index $j$ selects a column.}
\label{tab:table5}
\begin{tabular}{c|ccccc}
\hline
$f_{ij}$& 0 & 1 & 2 & 3 & 4 \\ 
\hline
 0 &     7824.37  &    25237.25  &     3138.79  &  -323056.91  & 351919.38 \\ 
 1 &    55357.30  &    15986.33  &  -244637.95  &  -256695.08  & 598394.06 \\ 
 2 &    72768.05  &   -31325.23  &  -342372.66  &    74755.55  & 391854.50 \\ 
 3 &    20716.17  &     6647.99  &  -158499.00  &    43559.96  & 161916.81 \\ 
 4 &    -1549.50  &     4181.00  &     6828.17  &   -51471.20  &  55152.71 \\ 
\hline
\end{tabular}

\end{table}

\subsection{Implementation}
To sum up, in order to determine the value of the bias, given as input a chosen cosmology and the values of $Q$, $m$, $z$, and $\deltal$, one should:
\begin{enumerate}
\item for the chosen values of the Hubble parameter and the matter density parameter, compute the exponent $f(h,\Omega_{\rm m})$ using eq.~(\ref{eq:exponent}) with the parameters from table~\ref{tab:table5};
\item scale the mass according to eq.~(\ref{eq:massscale}), with the value of $f$ just computed. From now on this is the new mass that has to be used;
\item compute the parameters $A(m,Q;\tilde{z},\tilde{\sigma_8})$, $b(m,Q;\tilde{z},\tilde{\sigma_8})$ and $c(m,Q;\tilde{z},\tilde{\sigma_8})$ for the reference cosmology, using eq.~(\ref{eq:refcosmo}) and the parameters listed in table~\ref{tab:table1};
\item compute the parameters $(\xi_{\rm A},\eta_{\rm A},\zeta_{\rm A},\omega_{\rm A})$,  $(\xi_{\rm b},\eta_{\rm b},\zeta_{\rm b},\omega_{\rm b})$ and  $(\xi_{\rm c},\eta_{\rm c},\zeta_{\rm c},\omega_{\rm c})$, using the fit described in eq.~(\ref{eq:lowlev}) with parameters from tables~\ref{tab:table2},~\ref{tab:table3} and~\ref{tab:table4};
\item evolve $A$, $b$ and $c$ to the chosen values of $z$ and $\sigma_8$ using eq.~(\ref{eq:cosmofit});
\item compute the final bias with eq.~(\ref{eq:fit}).
\end{enumerate}
This computation, though requiring storing a large number of fitting parameters, is much faster than the implementation of the full expression~(\ref{eq:bias}). Reading the parameters from files and computing polynomials is a task that can easily be implemented in a numerical code, and is efficiently handled by a processor.

\bsp	
\label{eq:lastpage}
\end{document}